\documentclass[useAMS,usenatbib]{mn2e}
\usepackage{longtable,lscape}
\usepackage{multirow}

\usepackage{graphicx}
\usepackage{aas_macros}
\usepackage{amssymb}
\usepackage[draft]{hyperref}

\newcommand{\about}{$\sim\!\!$~}
\newcommand{\kms}{km~s$^{-1}$}

\newcommand{\be}{\begin{displaymath}}
\newcommand{\ee}{\end{displaymath}}

\def\lsim{\hbox{\rlap{\raise 0.425ex\hbox{$<$}}\lower 0.65ex\hbox{$\sim$}}}
\def\gsim{\hbox{\rlap{\raise 0.425ex\hbox{$>$}}\lower 0.65ex\hbox{$\sim$}}}

\newcommand{\ion}[2]{#1$\;${\small{#2}}\relax}

\newcommand{\cair}{CaIR3}
\newcommand{\cahk}{CaHK}
\newcommand{\siii}{\ion{Si}{II} $\lambda$6355}

\graphicspath{{/Users/jsilv/BSNIP/high_vel/plots/}}

\title[High-Velocity Features in SNe Ia]{High-Velocity Features of
  Calcium and Silicon in the Spectra of Type~Ia Supernovae}

\author[Silverman, et al.]{Jeffrey M. Silverman,$^{1,2,3}$ J\'{o}zsef Vink\'{o},$^{1,4}$ G. H. Marion,$^{1,5}$
\newauthor
J. Craig Wheeler,$^{1}$ Barnab\'{a}s Barna,$^{4}$ Tam\'{a}s Szalai,$^{4}$
\newauthor
Brian W. Mulligan,$^{1}$ Alexei V. Filippenko$^{6}$ \\
$^{1}$Department of Astronomy, University of Texas at Austin, Austin, TX 78712, USA \\
$^{2}$NSF Astronomy and Astrophysics Postdoctoral Fellow \\
$^{3}$email: jsilverman@astro.as.utexas.edu \\
$^{4}$Department of Optics and Quantum Electronics, University of Szeged, D\'{o}m t\'{e}r 9, 6720 Szeged, Hungary \\
$^{5}$Harvard-Smithsonian Center for Astrophysics, Cambridge, MA 02138, USA \\
$^{6}$Department of Astronomy, University of California, Berkeley, CA 94720-3411, USA }

\begin{document}
\date{Accepted  . Received   ; in original form  }
\pagerange{\pageref{firstpage}--\pageref{lastpage}} \pubyear{2015}
\maketitle
\label{firstpage}

\begin{abstract}
``High-velocity features'' (HVFs) are spectral features in Type~Ia
supernovae (SNe~Ia) that have minima indicating significantly higher
(by greater than about 6000~\kms) velocities than typical ``photospheric-velocity
features'' (PVFs). The PVFs are absorption features with minima
indicating typical photospheric (i.e., bulk ejecta) velocities (usually
\about9000--15,000~\kms\ near $B$-band maximum brightness). In this
work we undertake the most in-depth study of HVFs ever performed. The
dataset used herein consists of 445 low-resolution optical and near-infrared
(NIR) spectra (at epochs up to 5~d past maximum brightness) of 210
low-redshift SNe~Ia that follow the ``Phillips relation.'' A series of
Gaussian functions is fit to the data in order to characterise
possible HVFs of \ion{Ca}{II}~H\&K, \siii, and the \ion{Ca}{II}
NIR triplet.  The temporal evolution of
the velocities and strengths of the PVFs and HVFs of these three
spectral features is investigated, as are possible correlations with
other SN~Ia observables. We find that while HVFs of \ion{Ca}{II} are
regularly observed (except in underluminous SNe~Ia, where they are 
never found), HVFs of \siii\ are significantly rarer, and they tend to exist
at the earliest epochs and mostly in objects with large photospheric
velocities. It is also shown that stronger HVFs of \siii\ are found in
objects that lack \ion{C}{II} absorption at early times and that have
red ultraviolet/optical colours near maximum brightness. These results 
lead to a self-consistent connection between the presence and strength of HVFs
of \siii\ and many other mutually correlated SN~Ia observables,
including photospheric velocity.
\end{abstract}

\begin{keywords}
{methods: data analysis -- techniques: spectroscopic -- supernovae: general}
\end{keywords}


\section{Introduction}\label{s:intro}

Observations of Type~Ia supernovae (SNe~Ia) led to the discovery of
the accelerating expansion of the Universe
\citep{Riess98:lambda,Perlmutter99} and have been extremely useful as
a way to accurately measure cosmological parameters
\citep[e.g.,][]{Suzuki12,Betoule14,Rest14}. The cosmological
utility of SNe~Ia as precise distance indicators relies on the fact
that their luminosity can be standardised. \citet{Phillips93} was
the first to convincingly 
show that the light-curve decline rate of most SNe~Ia
is well correlated with luminosity at peak brightness, a connection
now known as the ``Phillips relation.''

SNe~Ia arise from the thermonuclear explosion of C/O white dwarfs
\citep[WDs; e.g.,][]{Hoyle60,Colgate69,Nomoto84,Nugent11,Bloom12}, but
beyond that basic statement, we still lack a detailed understanding of
the progenitor systems and explosion mechanisms of SNe~Ia (see
\citealt{Howell11} and \citealt{Maoz14} for further information). In
general, the two 
leading progenitor scenarios are the single-degenerate (SD) channel,
when the WD accretes matter from a nondegenerate companion star
\citep[e.g.,][]{Whelan73}, and the double-degenerate (DD) channel,
which is the result of the merger of two WDs
\citep[e.g.,][]{Iben84,Webbink84}.

Detailed spectroscopic studies of large collections of low-redshift
SNe~Ia have been undertaken in the past
\citep[e.g.,][]{Barbon90,Branch93,Nugent95,Hatano00,Folatelli04,Benetti05,Bongard06,Hachinger06,Bronder08,Foley08:uv,Branch09,Wang09,Walker11,Nordin11a,Blondin11,Konishi11,Foley11:vel,Silverman12:BSNIPII},
and have focused mainly on ``photospheric-velocity features'' (PVFs),
which are absorption features with minima indicating typical
photospheric (i.e., bulk ejecta) velocities (usually
\about9000--15,000~\kms\ near $B$-band maximum brightness). These
features are formed at the outer edge of the optically--thick portion
of the ejecta; thus, most absorption features in the spectra of
SNe~Ia should be PVFs. However,
some recent work has focused on carefully identifying and
characterising so-called ``high-velocity features'' (HVFs), which are
spectral features that have minima indicating significantly higher
velocities than typical photospheric velocities \citep[i.e.,
6000--13,000~\kms\ larger than PVFs,
e.g.,][]{Mazzali05,Maguire12,Folatelli13,Childress14,Maguire14}.

In addition to these extensive samples, many studies of individual SNe~Ia
have presented evidence for HVFs
\citep[e.g.,][]{Hatano99:hv,Li01:00cx,Gerardy04,Thomas04,Wang09:05cf,Foley12,Parrent12,Silverman12:12cg,Childress13:12fr,Marion13:09ig,Maund13,Pereira13,Silverman13:13bh}  
and have shown that they appear strongest in early-time spectra
and weaken with time (as the PVFs strengthen). Previous work
has also shown that HVFs are most often seen in the \ion{Ca}{II}~H\&K
(hereafter \cahk), \siii, and \ion{Ca}{II} NIR triplet (hereafter
\cair) features, though they are sometimes also present in other features as well 
\citep[e.g.,][]{Parrent12,Marion13:09ig}. Furthermore, the
line-forming regions of the PVFs and HVFs appear to be physically
distinct and substantially asymmetric, based in part on numerous
spectropolarimetric observations
\citep[e.g.,][]{Leonard05,Wang03,Wang06,Chornock08,Patat09,Maund13}. 

It has been suggested that the velocity of the \cahk\ feature is
correlated with light-curve width \citep[e.g.,][]{Maguire12} and that
HVFs are responsible for this relationship. However, \citet{Foley13}
claims that \ion{Si}{II} $\lambda$3858 usually dominates the \cahk\
profile and is actually the cause of the observed
correlation. Recently, \citet{Childress14} examined HVFs of \cair\ in
58 low-redshift SNe~Ia with spectra within 5~d of $B$-band maximum
brightness 
and found that the existence and strength of HVFs is (positively)
correlated with light-curve width and uncorrelated with SN
colour. They also find that the existence and strength of the \cair\
HVFs are anticorrelated with \siii\ (photospheric) velocity. These
results are confirmed by \citet{Maguire14}, who 
studied a different dataset, consisting of 258 low-redshift SNe~Ia
with spectra earlier than 5~d after maximum brightness. This study
finds that \about80~per~cent (60--70~per~cent) of SNe~Ia at epochs
earlier than 5~d before (after) maximum show evidence for HVFs of
\cair, and that these features have velocities that are
\about7000~\kms\ faster than the PVFs seen in the same spectra. 

Despite the recent interest in HVFs, an explanation of
the physical origin of these features and how they might be related to
SN~Ia progenitors and their environments is still lacking. Interaction
with circumstellar material (CSM) is one of the leading proposed
causes of HVFs, which could arise from the SN ejecta sweeping up (or
otherwise interacting with) a clumpy CSM, or a torus or shell of CSM
\citep[e.g.,][]{Kasen03,Wang03,Gerardy04,Mazzali05,Tanaka06,Patat09}. Alternatively,
HVFs could arise naturally from the SN~Ia explosion mechanism itself,
such as from helium detonations in WD envelopes
\citep[e.g.,][]{Shen14}. No matter what the origin of HVFs, it seems
likely that an abundance or density enhancement at high velocity
(i.e., large radius in homologously expanding SN~Ia ejecta) must be
present \citep[e.g.,][]{Mazzali05,Tanaka08}, though perhaps ionisation
effects play a role as well \citep{Blondin13}. 

In this work, we explore a large dataset of low-redshift ($z < 0.1$),
low-resolution, optical and NIR 
SN~Ia spectra observed earlier than 5~d before maximum brightness
(described in Section~\ref{s:data}), a subset of which was studied by 
\citet{Childress14}. In these data we carefully search for and
measure the profiles of HVFs and PVFs of \cahk, \siii, and \cair\
(discussed in detail in Section~\ref{s:procedure}). The temporal
evolution of these features, and how their velocities and strengths
correlate with each other and other observables, are described in
Section~\ref{s:results}. We summarise our conclusions in
Section~\ref{s:conclusions}.


\section{Dataset}\label{s:data}

The majority of the SN~Ia spectra used in this study come from the
Berkeley SN~Ia Program (BSNIP) and have been published in BSNIP~I
\citep{Silverman12:BSNIPI}. Most of these data were obtained with the
Shane 3~m telescope at Lick Observatory using the Kast double
spectrograph \citep{Miller93}. The typical wavelength coverage of
3300--10,400~\AA\ (with resolutions of \about11 and \about6~\AA\ on
the red and blue sides, respectively) allows us to observe the \cahk\
and \cair\ features simultaneously. All objects have $z < 0.1$ with a
median redshift of 0.02.

We require that each SN~Ia have a
well-determined date of maximum brightness so that we can assign an
age to each spectrum. In this work, we only investigate spectra
obtained earlier than 5~d after maximum brightness. Note that this is
a superset of what was studied by \citet{Childress14}, who only used
BSNIP spectra {\it within} 5~d of maximum.
We removed objects which do not follow the ``Phillips relation'' {\it
  a priori}, including the extremely peculiar SN~2000cx
\citep[e.g.,][]{Li01:00cx}, SNe~Iax
\citep[e.g.,][]{Li03:02cx,Jha06:02cx,Foley13:Iax}, and
super-Chandrasekhar-mass objects
\citep[e.g.,][]{Howell06,Scalzo10,Silverman11}. A handful of the
remaining spectra had signal-to-noise ratios (S/N) that were too low
to reliably measure any spectral features or did not cover the
wavelengths any of the three features under investigation (\cahk,
\siii, and \cair). After all of these cuts, 226 spectra of 169 SNe~Ia
remained.

To this sample, we added low-resolution optical spectra obtained using
the Marcario Low-Resolution Spectrograph \citep[LRS;][]{Hill98} on the
9.2~m Hobby-Eberly Telescope (HET) at McDonald Observatory and the
Robert Stobie Spectrograph \citep[RSS;][]{Nordsieck01} on the 11.1~m
by 9.8~m Southern African Large Telescope (SALT), and low-resolution
NIR spectra obtained using SpeX \citep{Rayner03} on the NASA
Infrared Telescope Facility (IRTF). Applying the same cuts as for the
BSNIP sample, this yielded 128 spectra of 48 SNe~Ia that covered at
least the \cair\ feature. Most of these data are unpublished and will
appear in upcoming works (e.g., Marion et~al., in preparation),
although a handful of these spectra have appeared in previous
publications \citep[e.g.,][]{Quimby06,Marion09:IR,Parrent11}. There
are 11 SNe~Ia with spectra in both this sample and BSNIP.

We also include in the current study 91 published spectra of 5
extremely well-observed SNe~Ia: SNe~2009ig \citep{Marion13:09ig},
2011by\footnote{Note that there are also spectra of SN~2011by in the
  aforementioned HET/SALT/IRTF sample.} \citep{Silverman13:late},
2011fe \citep{Vinko12,Parrent12}, 2012cg
\citep{Silverman12:12cg,Marion12}, and 2012fr
\citep{Childress13:12fr,Zhang14}. This yields a total of 445 spectra
of 210 
SNe~Ia that we analyse herein. Table~\ref{t:objects} lists the names
and phases of the spectra for each object. Note that all
results discussed in Section~\ref{s:results} are consistent with what
is found when using just the BSNIP sample alone. Thus, adding the
other spectra into the current study does not bias any of our
findings, yet it adds statistical weight and significance to the
results.

To better characterise the objects in our sample, we attempt to
classify each SN~Ia using a variety of classification schemes. We
consider an object ``spectroscopically normal'' if it is classified 
as ``Ia-norm'' by the SuperNova IDentification code
\citep[SNID;][]{Blondin07} as implemented in BSNIP~I
\citep{Silverman12:BSNIPI}. Other ``SNID Types'' used in this work
include ``Ia-91bg'' \citep[e.g.,][]{Filippenko92:91bg,Leibundgut93},
which represent typically underluminous SNe~Ia, and ``Ia-91T''
\citep[e.g.,][]{Filippenko92:91T,Phillips92} and ``Ia-99aa''
\citep[e.g.,][]{Li01:pec,Strolger02,Garavini04}, which together
represent typically overluminous SNe~Ia. 

Using the expansion velocity of the \siii\ feature, \citet{Wang09}
classified spectroscopically ``normal'' SNe~Ia within 5~d of maximum
brightness as either ``normal velocity'' (N) or ``high  velocity''
(HV), with a velocity cutoff of 11,800~\kms\ at maximum brightness. While a
sharp distinction between the two ``Wang Types'' may not exist
\citep[e.g.,][]{Silverman12:BSNIPII}, we nonetheless 
utilise this classification scheme for illustrative purposes. Note
that an individual SN~Ia can be classified as N or HV, and each of its
spectra may have PVFs, HVFs, or both. In other words, the Wang Type is
used to classify a SN~Ia, while the presence or absence of PVFs and
HVFs is determined for each spectrum.

Another spectral classification scheme often used in SN~Ia research
was first introduced by \citet{Branch06}. Using the pseudo-equivalent
widths (pEWs) of \siii\ and \ion{Si}{II} $\lambda$5750 in spectra near
maximum brightness, they divide their spectral sample into four
different groups: core normal (CN), broad line (BL), cool (CL), and
shallow silicon (SS). This classification scheme is not used in the
current work because it is effectively equivalent to a combination of
SNID Types and Wang Types (CN = N, BL = HV, CL = Ia-91bg, SS =
Ia-91T/99aa).

\citet{Benetti05} used the rate of decrease of the \siii\ expansion
velocity before and near maximum brightness to define the velocity
gradient, $\dot{v}$. Adopting these values, they separated their SN~Ia
sample into three subclasses, or ``Benetti Types.'' High velocity
gradient (HVG) and low velocity gradient (LVG) objects are
normal-luminosity or overluminous SNe~Ia with $\dot{v} \ge
70$~\kms~d$^{-1}$ and $\dot{v} <
70$~\kms~d$^{-1}$, respectively. The third subclass (FAINT) have
moderately large velocity gradients, but are underluminous ($\Delta
m_{15}(B) \ga 1.6$~mag). All three of the aforementioned
classifications are listed for each object in Table~\ref{t:objects}. 

Photometric information was obtained from published sources, when
available. This includes the date of $B$-band maximum
for each object, as well as light-curve width (characterised by
$\Delta m_{15}(B)$) and $\left(B-V\right)_0$ colour (the observed
$B-V$ colour of the SN at $B$-band maximum brightness). For the BSNIP
data, this information came from \citet{Jha06}, \citet{Hicken09}, and
\citet{Ganeshalingam10:phot_paper}. Photometric information for the
HET, SALT, and IRTF data were obtained from a variety of sources
\citep{Quimby06,Ganeshalingam10:phot_paper,Stritzinger11,Maguire12,Hicken12,Silverman13:late}. As
for the five well-studied objects, their spectroscopic and photometric
references are listed above. About two-thirds of the objects in this
study have published $\Delta m_{15}(B)$ and $\left(B-V\right)_0$
values, and these are also presented in Table~\ref{t:objects}.

\section{Measurement Procedure}\label{s:procedure}

The measurement procedure used in this study is implemented in {\tt
  IDL} and based in part on the one utilised extensively in BSNIP~II
\citep{Silverman12:BSNIPII}. It is briefly described by
\citet{Silverman13:13bh}, but the description of our procedure herein
is more in-depth. Each spectrum is first deredshifted (adopting the
redshift listed in NED) and corrected for Milky Way (MW) reddening
using values from \citet{Schlegel98}. Each of the three features
measured (\cahk, \siii, and \cair) is then investigated individually.

For each feature, a local minimum in the spectrum is found, and
the first local, relatively broad maxima are recorded to the left and 
right of this minimum. Note that the local maximum to the right of the
minimum often corresponds to the peak of the P-Cygni profile. A
concave downward quadratic function is fit to these local maxima, and
the peaks of these parabolas are considered the endpoints of the
spectral feature. These endpoints were visually inspected for every
feature measured, and in about one-third of cases one or both of the
endpoints were clearly incorrect, either still within the feature
profile or very far from it. In these cases, the endpoints were chosen
manually.

\begin{figure*}
\centering$
\begin{array}{cc}
\includegraphics[width=3.5in]{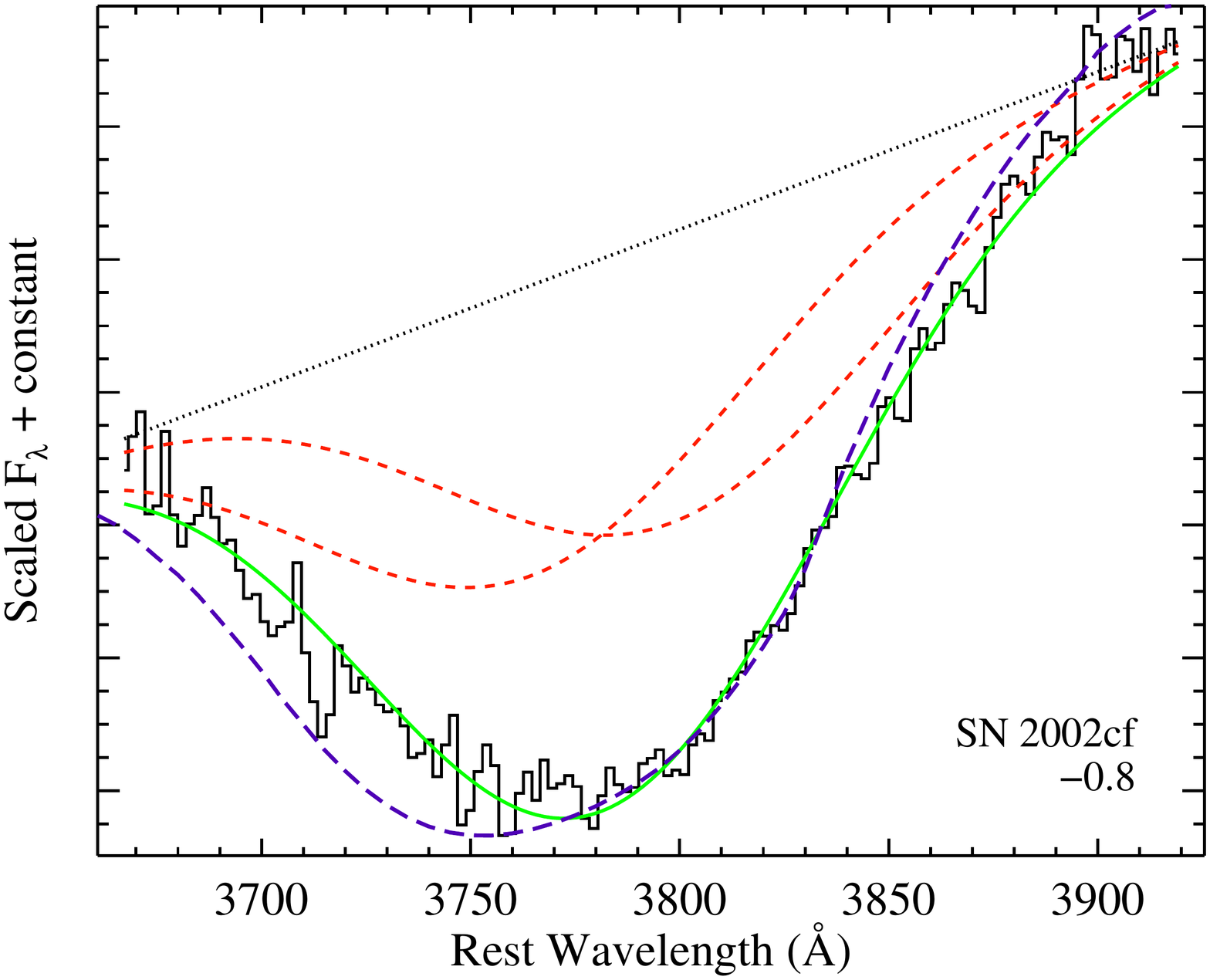} &
\includegraphics[width=3.5in]{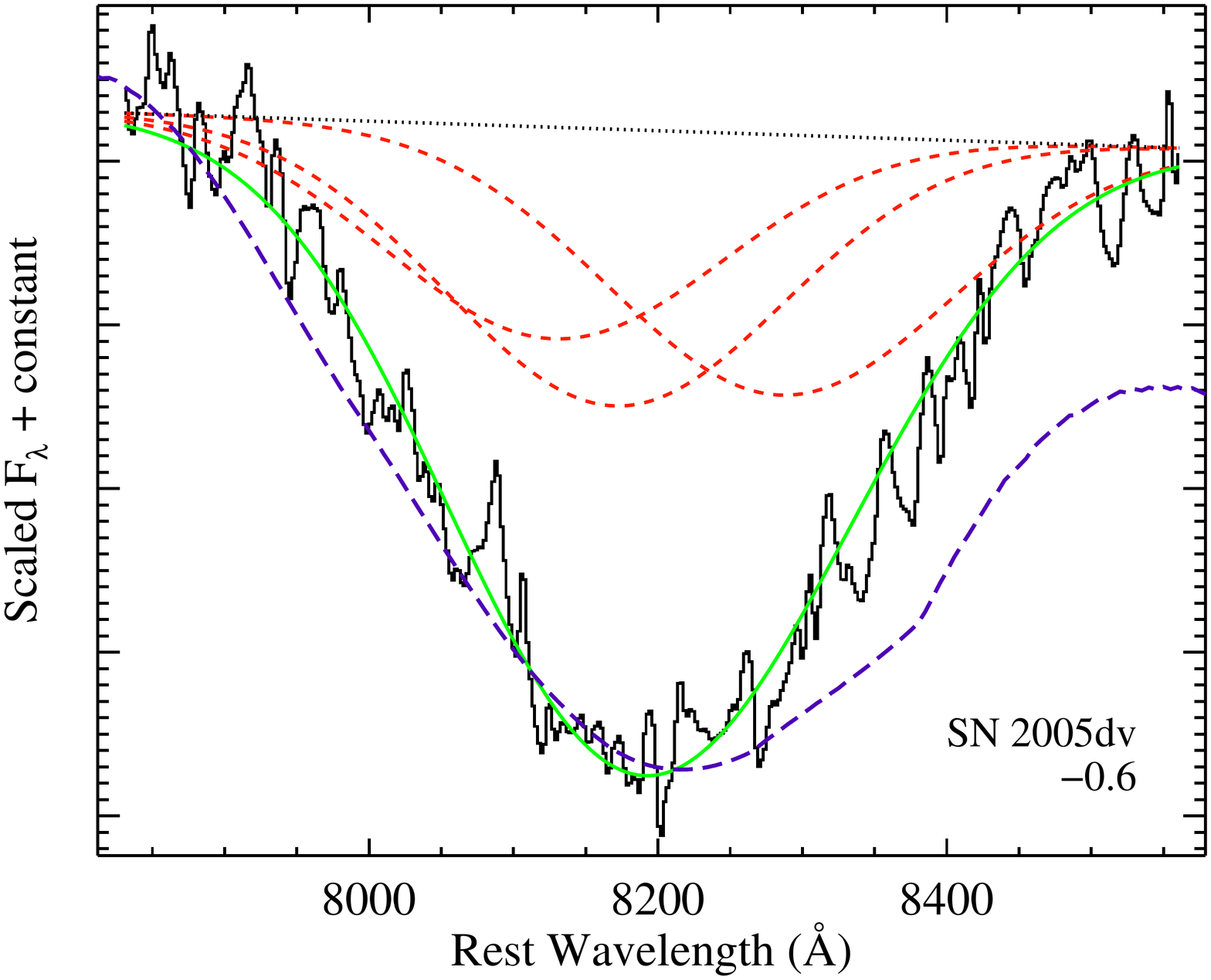} \\
\includegraphics[width=3.5in]{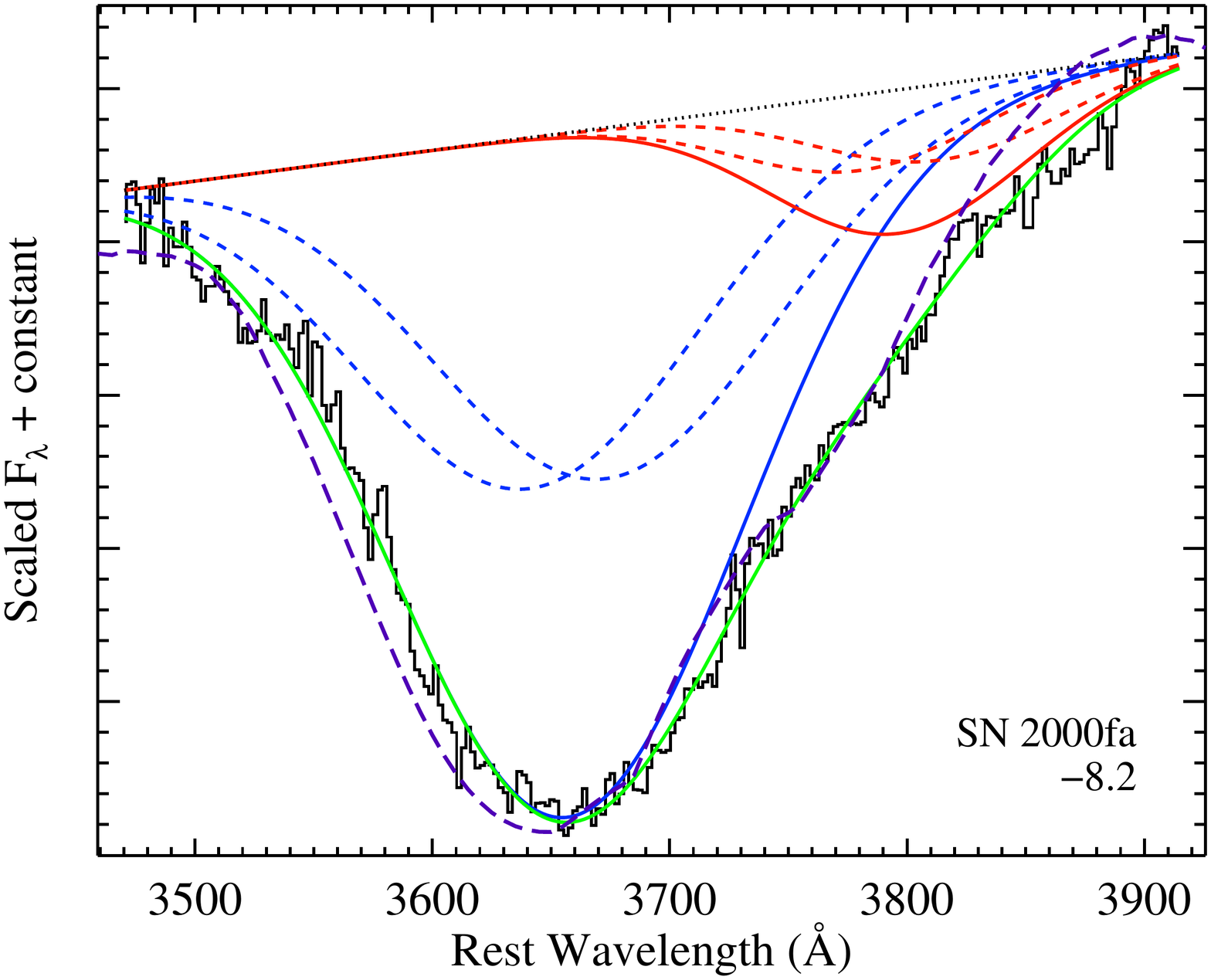} &
\includegraphics[width=3.5in]{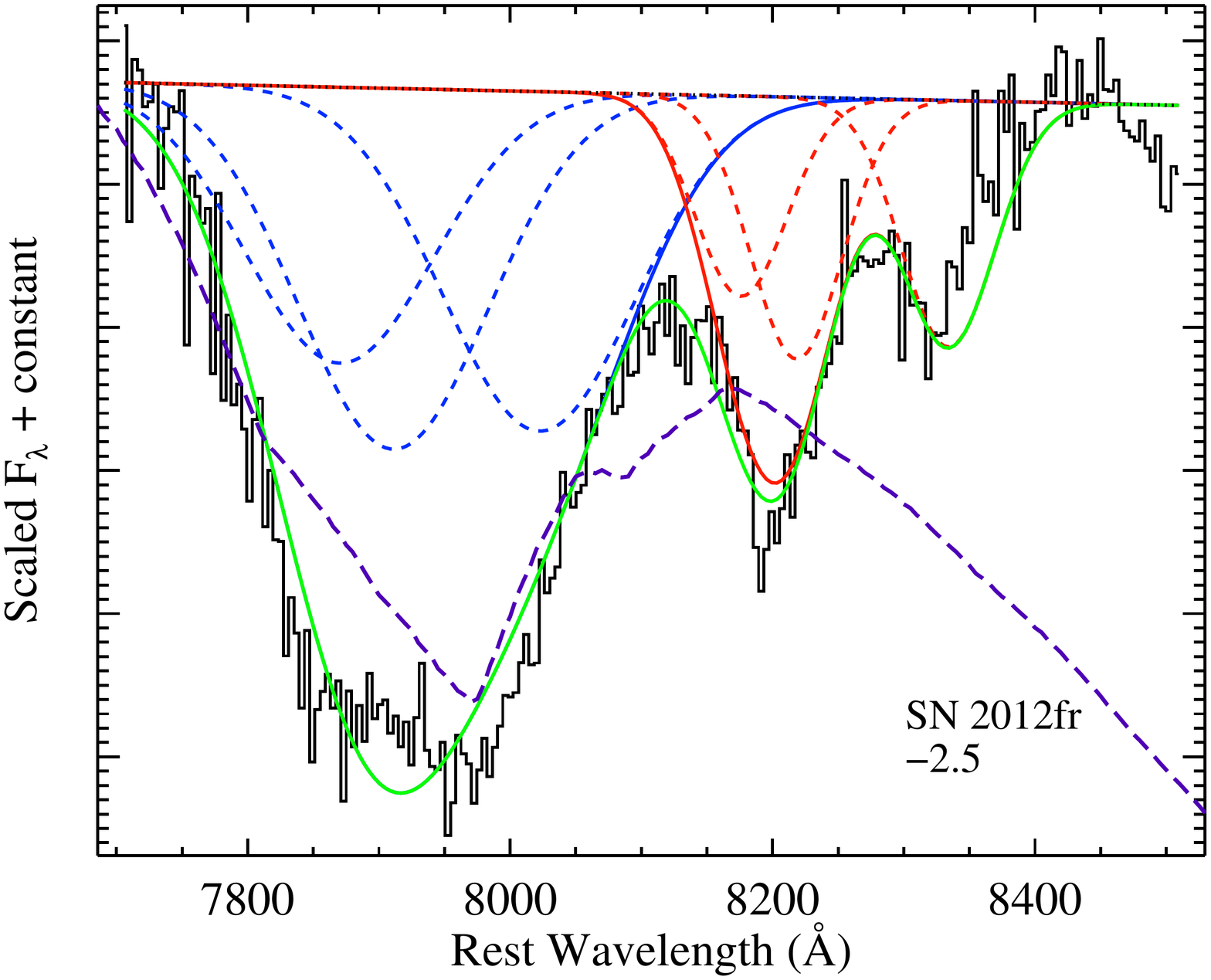} \\
\includegraphics[width=3.5in]{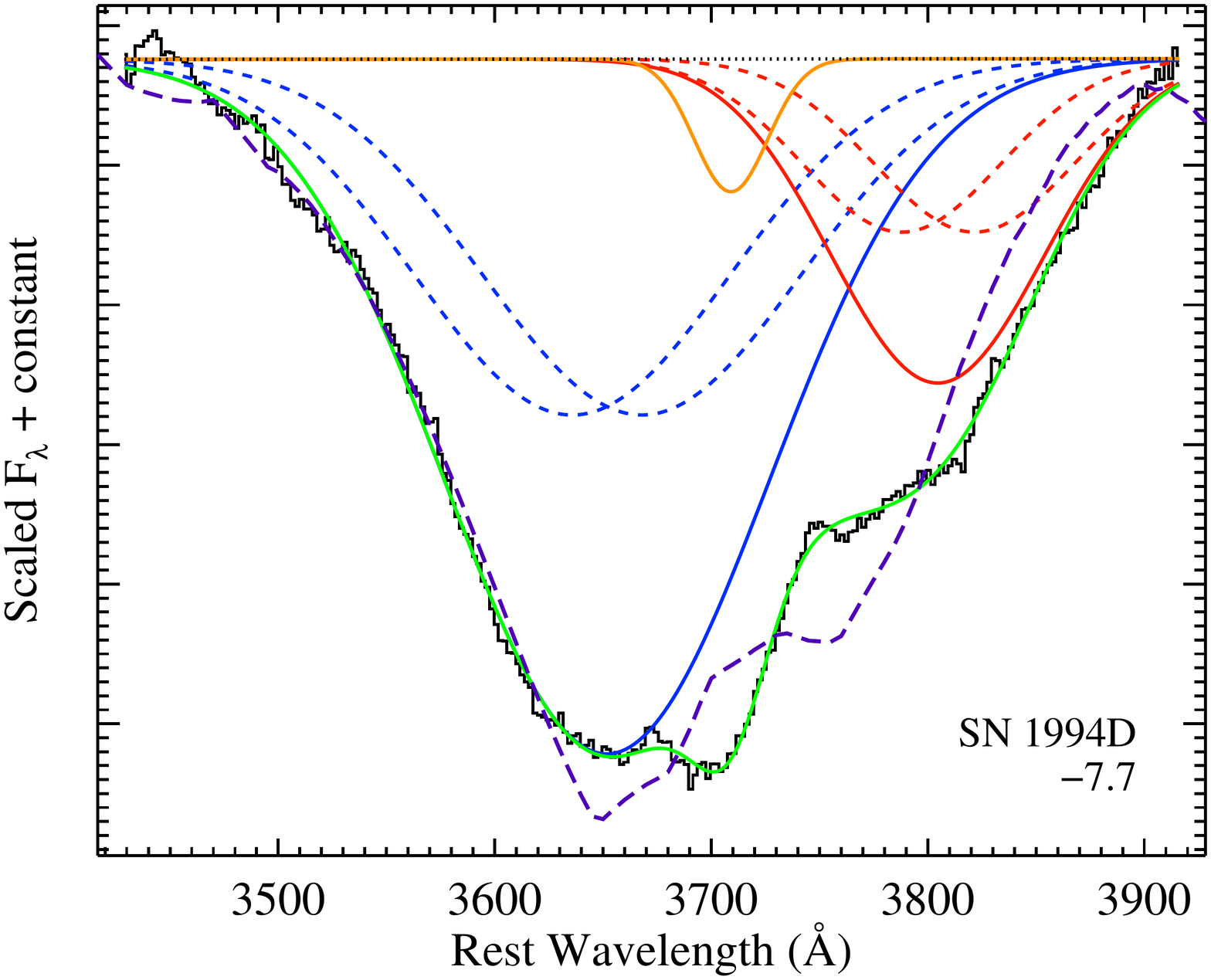} &
\includegraphics[width=3.5in]{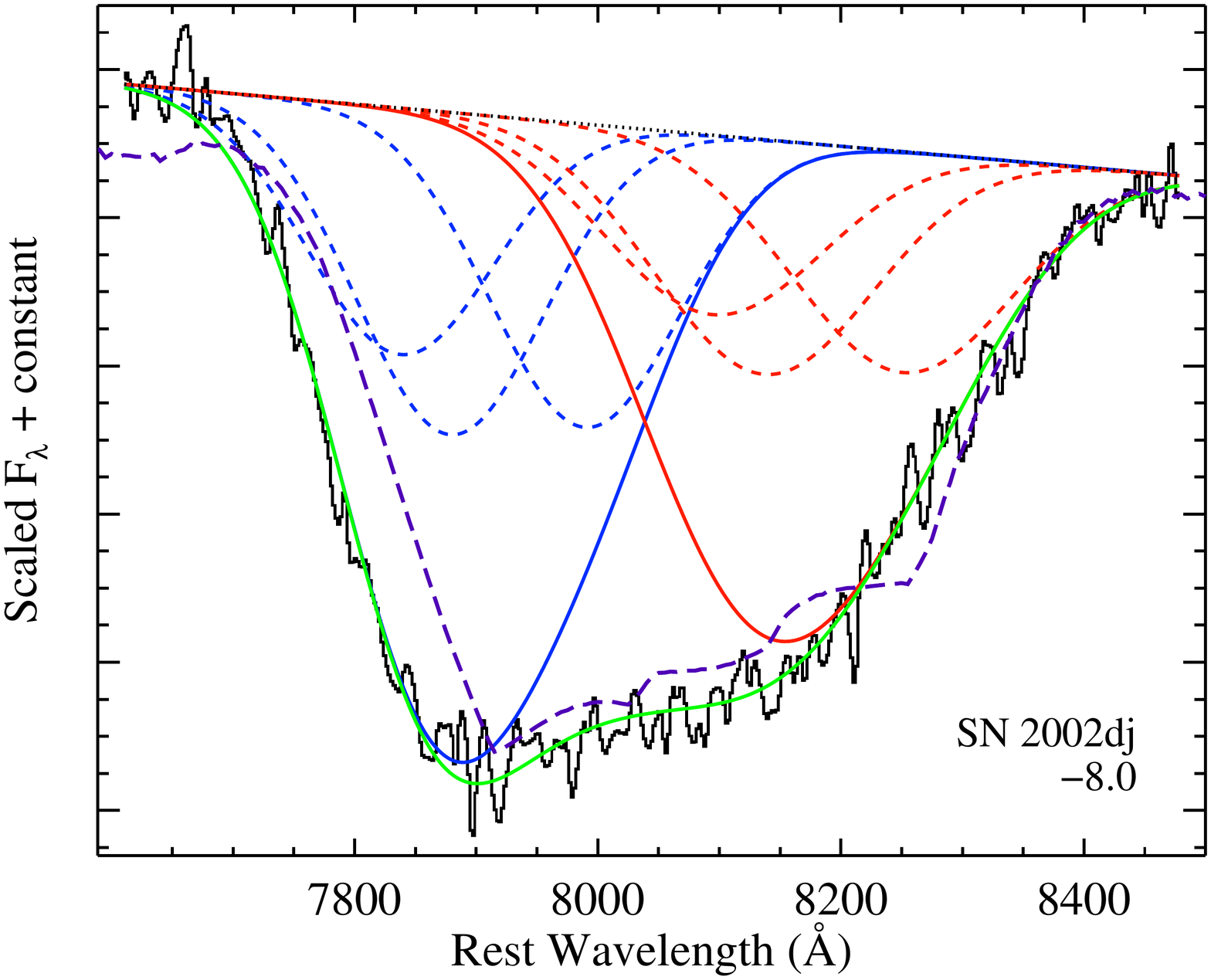} \\
\end{array}$
\caption{Fits to \cahk\ ({\it left column}) and \cair\ ({\it right
    column}) showing PVFs ({\it red}) and HVFs ({\it blue}, where
  required). Individual Gaussian components are {\it short-dashed} and
  their sum is {\it solid}. Also shown are the sum of the total fit
  ({\it green}), the data ({\it black, solid}), the linear continuum ({\it
    black, dotted}), and a  {\tt SYNAPPS} \citep{Thomas11:synapps} fit ({\it
    purple, long-dashed}). \ion{Si}{II} $\lambda$3858 is also required
  in the {\it bottom-left} fit ({\it orange, solid}). Each spectrum is
  labeled with its object name and age relative to $B$-band maximum 
  brightness.}\label{f:fits}  
\end{figure*}

The two endpoints for each feature are then connected with a straight
line, and this becomes the pseudo-continuum (black, dotted lines in
Fig.~\ref{f:fits}). The continuum flux at each pixel is then
subtracted from the observed flux, yielding the background-subtracted
spectrum used in the procedure described below. This step is sometimes
referred to as flattening the spectra and was used previously in
BSNIP~II \citep{Silverman12:BSNIPII}. One might instead {\it divide}
the observed flux at each pixel by the continuum flux, though our
tests indicate that this alternative approach does not significantly 
change the derived fit parameters; the values differ only at the
few-percent level.

For each feature, the local minimum is found (when fitting only one
velocity component, either PVF {\it or} HVF) or two separate local
minima are found (when fitting two velocity components simultaneously,
i.e., a PVF {\it and} HVF). These minima are then used as initial
estimates in a nonlinear least-squares fitting routine 
that fits the entire profile between the two endpoints with either one
or two velocity components. Each component consists of one
(for \siii), two (for \cahk, left column of Figure~\ref{f:fits}), or three
(for \cair, right column of Figure~\ref{f:fits}) Gaussian
functions and contains three free parameters. The
relative separations of the \ion{Ca}{II} lines come from their 
rest wavelengths, while their relative strengths come from their 
$gf$-weights\footnote{{\tt\url{http://www.nist.gov/pml/data/asd.cfm}}.}. We 
are thus operating in the optically thin limit, an assumption which
has been shown in previous work using similar spectral feature
fitting methods to not strongly affect the results
\citep{Childress14,Maguire14,Pan15}. 

For \cair\ and \siii, two fits were attempted for each profile: a 
one-component fit (PVF {\it or} HVF) and a two-component fit (both HVF
{\it and} PVF). For \cahk, however, the possible presence of \ion{Si}{II}
$\lambda$3858 \citep[e.g.,][]{Foley13} complicates matters. Thus, we
attempt four fits for each \cahk\ profile: a one-component fit (PVF
{\it or} HVF), a two-component fit (both HVF {\it and} PVF), a
different two-component fit (PVF {\it or} HVF, but with a single
component of \ion{Si}{II} $\lambda$3858 included), and a three-component 
fit (both HVF {\it and} PVF, but with a single component of
\ion{Si}{II} $\lambda$3858 included). Each fit is then 
visually inspected, and in extreme cases where the fits do not match
the data well, the initial estimates of local minima are changed and
the fit is redone. As mentioned above, some spectra, mostly ones with
low S/N, did not yield any acceptable fit.

While the {\it relative} separations and strengths of the spectral
features are fixed as mentioned above, we do not impose any other
constraints on the fit parameters. This differs from what was done by 
\citet{Maguire14}, who required that the \cair\ PVF velocities be
within 25~per~cent of the \siii\ velocity and that the \cair\ HVF
velocities be at least 2000~\kms\ faster than the \siii\ velocity.

To decide which combination of fit components best represents the
data, a variety of methods were used. All fits of a given spectral
profile were visually inspected and the best fit was chosen via
``$\chi$-by-eye.'' This choice was then compared to the reduced-$\chi^2$ 
value and the Bayesian information criterion (BIC) value for
each fit. In the vast majority of cases, all three methods agreed
unanimously. In the few cases where there was serious disagreement, we
erred on the side of trusting fits with fewer parameters. 

Once a best fit was chosen for each profile, the Gaussian fit
parameters were used to calculate a velocity using the relativistic
Doppler equation and a pEW
\citep[e.g.,][]{Garavini07,Silverman12:BSNIPII} for each component of
the 
fit. These values (and their uncertainties) for \cahk, \siii, and
\cair\ are listed in Tables~\ref{t:cahk}, \ref{t:siii}, \ref{t:cair},
respectively. The formal uncertainty of the Gaussian fits indicates
that the typical velocity error is \about60~\kms. The minima of the
spectral fits, however, are only accurate to a few \AA, which implies
a velocity uncertainty more like \about200--400~\kms. This measurement
uncertainty increases for weaker features. A few examples of
\ion{Ca}{II} ``best fits'' are displayed in Figure~\ref{f:fits}.

\subsection{Ambiguous \cahk\ Fits}\label{ss:ambig_hk}

As mentioned above, the \cahk\ feature overlaps with the \ion{Si}{II}
$\lambda$3858 feature, which can affect the observed spectral profile
\citep[e.g.,][]{Foley13}. This was seen in our data, as many
spectra were fit equally well (in a reduced-$\chi^2$ sense) by both a
HVF and a PVF of \cahk, and \ion{Si}{II} $\lambda$3858 and a PVF of
\cahk. To break this degeneracy, we exploited the fact that the
majority of the spectra studied herein include both the \cahk\ and
\cair\ features in the same observation. We assumed that if a spectrum
showed a HVF of \cair\ (based on the method outlined above), then it
should also have a HVF of \cahk\ (and vice versa). In the two
ambiguous cases where the spectra did not cover the \cair\ feature, we
found that the inferred \ion{Si}{II} $\lambda$3858 velocity was
significantly larger than the \siii\ velocity in the same
observations, and thus we identify those profiles as containing HVFs
of \cahk\ (instead of \ion{Si}{II} $\lambda$3858).

To test our assumption that a HVF of \cair\ implies a HVF of \cahk, we
temporarily changed all of our \ion{Si}{II} $\lambda$3858
identifications to HVFs of \cahk. This led to large differences in the
velocities of the HVFs of \cahk\ and the HVFs of \cair\ in the same
spectrum (\about5000~\kms, as opposed to the more typical value of
\about500~\kms; see below). It also led to relatively small
differences between the velocities of the HVFs and PVFs of \cahk\ in
the same spectrum (\about5500~\kms, as opposed to the average of
\about9000~\kms; again, see below). Therefore, it seems that our
identifications of \ion{Si}{II} $\lambda$3858 are
correct. Furthermore, we also compare the velocity of \ion{Si}{II}
$\lambda$3858 (when we detect it) with that of \siii\ in the same
spectra and find the typical difference to be \about600~\kms,
consistent with previous work on velocities of various \ion{Si}{II}
spectral features \citep[e.g.,][]{Silverman12:BSNIPII}. 

The opposite test to the one described above was also
performed. Namely, we temporarily changed all of our HVFs of \cahk\ to
\ion{Si}{II} $\lambda$3858. After doing this, the average \ion{Si}{II}
$\lambda$3858 velocity was found to be \about15,000~\kms, and on
average about 2600~\kms\ faster than the \siii\ velocity in the same
observation. Thus, these inferred \ion{Si}{II} $\lambda$3858
velocities are too high to be real and so our HVF \cahk\
identifications appear to be correct.

\begin{figure}
\centering$
\begin{array}{c}
\includegraphics[width=3.5in]{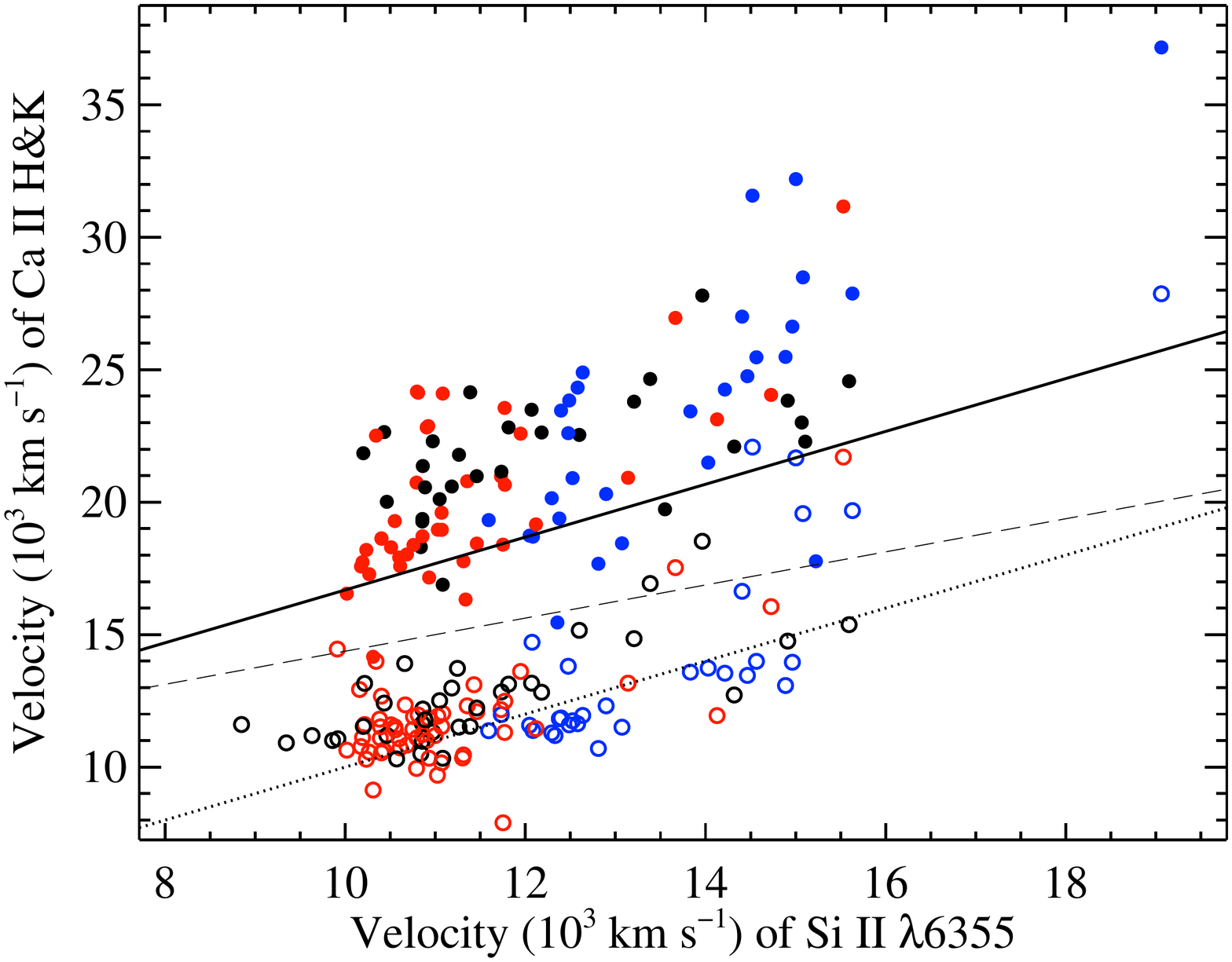} \\
\includegraphics[width=3.5in]{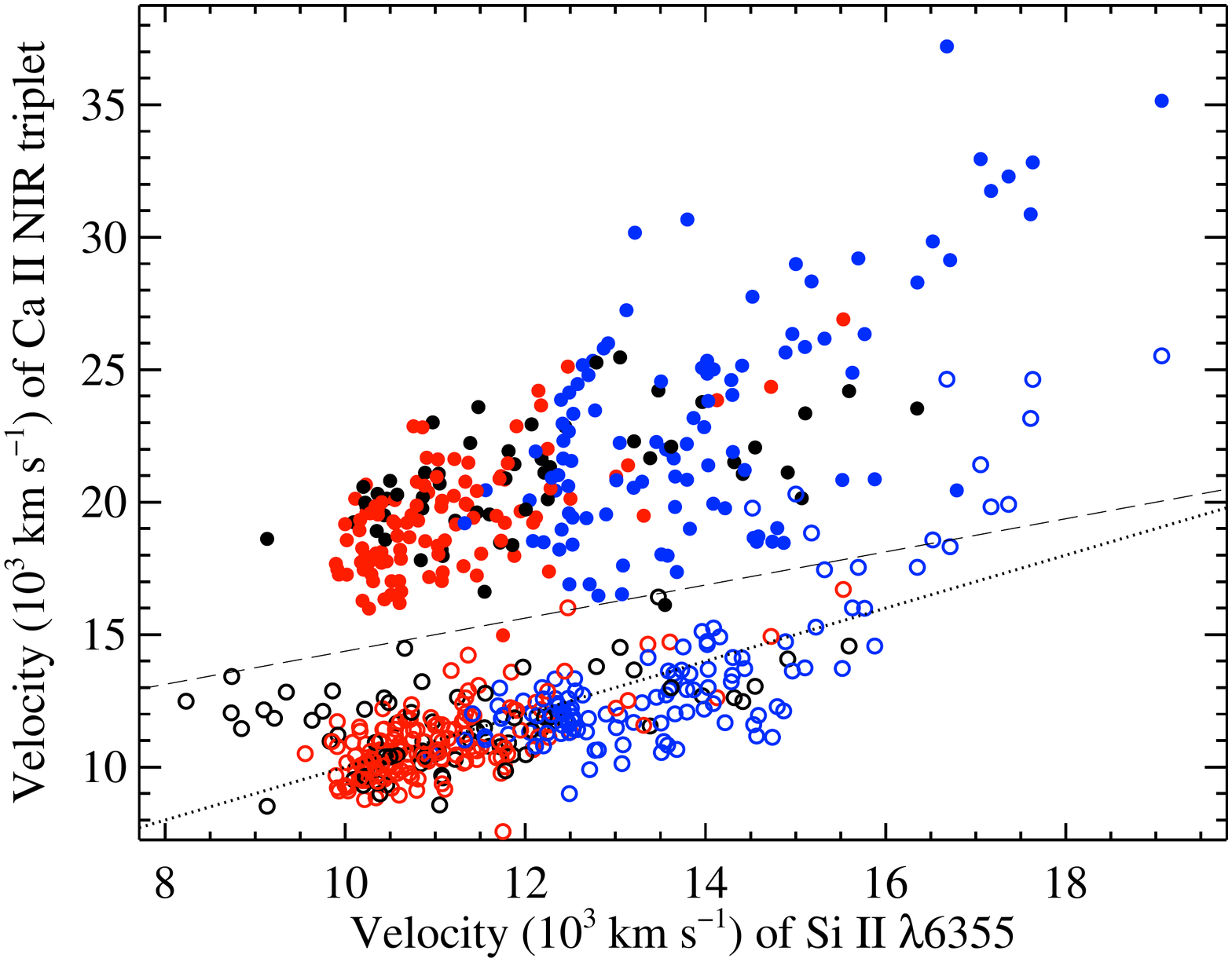} \\
\end{array}$
\caption{The velocity of \cahk\ ({\it top}) and \cair\ ({\it bottom})
  versus the velocity of \siii. Open points are PVFs of \ion{Ca}{II}
  and filled points are HVFs of \ion{Ca}{II}. Blue points
  are HV objects, red points are N objects, and black points are
  objects without a Wang Type. The dotted line is the
  one-to-one line; the dashed line is the cutoff between HVFs and PVFs
  used by \citet{Foley13}. The solid line in the top panel represents
  \ion{Si}{II} $\lambda$3858 at the same velocity as \siii, if our
  HVFs of \cahk\ were actually misidentified \ion{Si}{II}. Since most
  of the filled points in the top panel lie above the solid line, it 
  is unlikely that they are actually \ion{Si}{II} $\lambda$3858, and
  thus they are probably HVFs of \cahk, as our assumption
  implies.}\label{f:cahk_si_vels}
\end{figure}

Another way to visualise this is shown in
Figure~\ref{f:cahk_si_vels}. There we plot the velocity of \siii\
versus the velocity of \cahk. The open points are PVFs of \cahk\ while
the filled points are HVFs of \cahk, as determined using our method
described above. The dotted line is the one-to-one line and shows that
PVFs of \cahk\ are slightly faster than \siii\ at low \ion{Si}{II}
velocities and comparable at higher \ion{Si}{II} velocities. The
dashed line is the cutoff between HVFs and PVFs used by
\citet{Foley13}; the classifications from his study mostly match those 
in this work. Finally, the solid line represents \ion{Si}{II}
$\lambda$3858 at the same velocity as \siii, if our HVFs of \cahk\
were actually misidentified \ion{Si}{II}. Thus, if a solid point fell
directly on this line, the velocity 
of our measured HVF of \cahk\ would match that of \siii\ if it were
actually \ion{Si}{II} $\lambda$3858.

The fact that most of the filled points lie above this line implies
that our identification of HVFs of \cahk\ is correct and that if those
features were actually \ion{Si}{II} $\lambda$3858, then their
velocities would be significantly higher than that of \siii\ in the
same spectra (as discussed above). Finally, we note that our inferred
velocities of the HVFs of \cahk\ and \cair\ are highly correlated, as
are the velocities of the PVFs of \cahk\ and \cair\, as well as the
velocities of \ion{Si}{II} $\lambda$3858 and \siii. This once again
supports our spectral identifications.

\subsection{Ambiguous \siii\ Fits}

When applying the aforementioned fitting algorithm to the \siii\
feature, we discovered that a single Gaussian (plus linear background)
fits most spectral profiles quite well. However, as has been seen
previously \citep[e.g.,][]{Silverman12:BSNIPII}, the stronger \siii\
profiles appear non-Gaussian and look more Lorentzian in shape
\citep[though these are mostly at greater than 3~d past maximum
brightness, well 
after HVFs of \siii\ usually disappear; e.g.,][]{Marion13:09ig}. In
addition, two Gaussian profiles (i.e., both a HVF and a PVF) fit nearly
every observation very well, both via visual inspection as well as in
a reduced-$\chi^2$ sense. Thus, to decide whether one or two
components were present in a given profile, other factors must be
considered.

Some of the HVF+PVF fits to \siii\ had the difference in velocity
between the two components less than 4500~\kms. This is significantly smaller
than the smallest difference between \ion{Ca}{II} HVFs and PVFs (i.e.,
\about6000~\kms; see Section~\ref{ss:v_t_ca}) and our fitting
algorithm is not capable of reliably distinguishing between two 
components that are so close to each other in velocity space (see
Section~\ref{sss:synthetic}). Thus, we are unable to say with
confidence that two components are present and in these cases we
prefer the one-component fit. Other HVF+PVF fits to \siii\ indicated a
velocity 
of the PVF of $\lesssim9000$~\kms, which is never seen at these epochs in
the ``relatively normal'' SNe~Ia used herein 
\citep[e.g.,][]{Silverman12:BSNIPII}. Therefore, we regard these fits
as unreliable as well, and we instead use the one-component fit for these
data.

After removing the unphysical two-component fits mentioned above, we
find that there are nearly no reliable two-component fits where the
measured HVF velocity is less than 16,500~\kms. Thus, our analysis indicates
that HVFs of \siii\ always remain above \about16,500~\kms,
consistent with previous work
\citep[e.g.,][]{Marion13:09ig}. Hence, we make the assumption that 
for a two-component (i.e., HVF+PVF) fit of \siii\ to be preferred over
a one-component fit, the inferred HVF velocity must be larger than
16,500~\kms.

\begin{figure}
\centering
\includegraphics[width=3.5in]{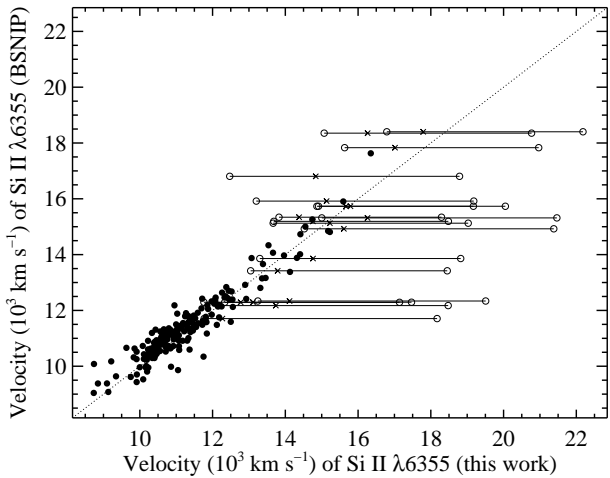}
\caption{\siii\ velocities for the 201 spectra in the current study
  (on the abscissa) that were also analysed in BSNIP~II \citep[][on the
  ordinate]{Silverman12:BSNIPII}. Filled points are spectra with one 
  velocity component. Open points connected with a horizontal line are
  spectra with two velocity components; the left endpoint is the PVF,
  the right endpoint is the HVF, and the ``$\times$'' is the
  pEW-weighted mean of the \siii\ PVF and HVF velocities for that
  spectrum. The dotted line is the one-to-one
  line.}\label{f:bsnip_vsi} 
\end{figure}

Under this assumption, our measured PVF velocities of \siii\ are
consistent with previous measurements of the same data
\citep{Silverman12:BSNIPII,Childress14}. Figure~\ref{f:bsnip_vsi}
shows this by plotting the \siii\ velocities for the 201
spectra in the current study (on the abscissa) that were also analysed
in BSNIP~II \citep[][on the ordinate]{Silverman12:BSNIPII}. Filled
points represent spectra for which only one velocity component is
detected in the current work. Pairs of open points connected with a
horizontal line represent spectra for which both a PVF and a HVF
velocity are measured in this work, with the left endpoint
representing the PVF and the right endpoint representing the HVF. The
``$\times$'' along each line segment represents the pEW-weighted mean
of the \siii\ PVF and HVF velocities for that spectrum. The dotted
line is the one-to-one line.

For velocities less than about 16,000~\kms, only the PVF velocity was
measured in BSNIP~II, whether or not a HVF was actually present in the
\siii\ profile. There are 11 spectra in which we detect two components
herein that fall into this category, and all of them have relatively
weak HVFs (which were simply missed by the fitting algorithm used in
BSNIP~II). For \siii\ velocities that are greater than
\about16,000~\kms, BSNIP~II typically measured the pEW-weighted mean
of the \siii\ PVF and HVF velocities. The 8 
spectra which have both a HVF and PVF component that are in this
category were mostly observed at early times when the PVF velocity was
high and the pEW of the HVF was large. This likely caused the two
components to be severely blended and thus the pEW-weighted mean of
the two velocities was measured in BSNIP~II. Finally, when considering
only spectra with one component, the BSNIP~II velocities and those 
measured in the current work are consistent. 

\subsection{Checks of the Fitting Procedure}\label{ss:checks}

\subsubsection{Synthetic Data}\label{sss:synthetic}

To test the limits of our fitting algorithm, we constructed synthetic
spectral profiles with a variety of Gaussian input parameters. The
first test varied only the separation in velocity/wavelength space
between the two components (i.e., HVFs and PVFs). We started with
representative values of PVF and HVF velocities and widths from our
fit to one of our relatively high-S/N spectra (SN~2002dj, $t =
-8$~d). 
The velocity of the PVF was held constant at its value from the fit to
the actual data (\about14,600~\kms), while the velocity of the HVF
component was varied. Random noise was also added to the Gaussian
functions to more closely resemble real data.

The HVF velocities tested ranged from the original value from the fit
to the data (\about24,200~\kms) down to \about17,700~\kms, in steps of
500~\kms. This allowed us to create 13 synthetic spectra with velocity
differences between the HVF and PVF of 9600--3100~\kms. We then
applied our spectral fitting algorithm as outlined above to these
synthetic data. For all spectra with velocity separations greater than
or equal to 4500~\kms, our fitting procedure preferred a two-component
profile, while spectra with separations in velocity below this value were
better fit by a one-component \cair\ profile. Thus, it seems that our
fitting algorithm is able to ``resolve'' distinct PVF and HVF
components when they are separated by greater than about 4500~\kms. As
we will show below, the smallest velocity difference between HVFs and
PVFs seen in our data is \about5000~\kms. Therefore, it seems unlikely
that nature produces HVFs and PVFs with velocity separations of less
than about 5000~\kms; otherwise, our algorithm would probably have
detected them.

A similar test was performed by varying the depth of the HVF, while
holding the depth of the PVF constant (along with the width and
velocity of both components). The difference in velocity between the
HVF and PVF used in this test was \about9000~\kms, which is typical
for our dataset (see below). The depth of the HVF was varied such that
we tested depth ratios (HVF depth divided by PVF depth) of 1--0.05, in
steps of 0.05. Two-component fits to the \cair\ profile (i.e.,
HVF+PVF) were preferred in spectra where the ratio of depths was
greater than 0.1. Thus, if there is a HVF whose depth is less than
about 10~per~cent that of the 
PVF, or vice versa, we will likely be unable to detect it using our
fitting algorithm. The smallest finite ratio between the strengths of
HVFs and PVFs of \ion{Ca}{II} that we measure in our data is about
0.12 (see below). Therefore, nature {\it may} produce HVFs that are so
weak compared to their PVFs (or vice versa) that our algorithm cannot 
detect them.

\subsubsection{{\tt SYNAPPS} Fits}

As another check to our spectral measurement technique, we use the
spectrum-synthesis code {\tt SYNAPPS} \citep{Thomas11:synapps}. {\tt
  SYNAPPS} (and its modeling kernel {\tt SYN++}) is derived from {\tt
  SYNOW} \citep{Synow}, which can compute spectra of SNe in
the photospheric phase using the Sobolev approximation
\citep{Sobolev60,Castor70,Jeffery89}. By varying many parameters
automatically and simultaneously, {\tt SYNAPPS} can find an optimum
fit to an input spectrum via $\chi^2$-minimisation. {\tt SYNAPPS}
assumes that spectral lines are formed via resonance scattering above
a sharp photosphere. The location of this photosphere (in velocity
space) is defined by the $v_{\rm ph}$ parameter, and the ejecta
are assumed to be in homologous expansion at a photospheric
temperature defined by the $T_{\rm phot}$ parameter.

For each input ion, the minimum and maximum velocity of the
line-forming region are defined by the $v_{\rm min}$ and
$v_{\rm max}$ parameters, respectively. In cases where
$v_{\rm min} \ga ~v_{\rm ph}$, the line-forming region is
considered ``detached'' from the photosphere; by definition, this is
the case for all HVFs. Each input ion also requires a value for the
optical depth of a ``reference line,'' $\tau_{\rm ref}$ (usually
the strongest optical line), an $e$-folding velocity width of the
optical-depth profile above the photosphere, $v_e$, and an excitation
temperature, $T_{\rm exc}$.

{\tt SYNAPPS} was used to fit 11 spectra that were chosen
semirandomly to represent various \cahk\ and \cair\ profile shapes
(i.e., differing relative strengths of HVFs, PVFs, and \ion{Si}{II}
$\lambda$3858). Each fit used ions that are typically found in SNe~Ia
(i.e., \ion{O}{I}, \ion{Ca}{II}, \ion{Mg}{II}, \ion{Si}{II},
\ion{S}{II}, and \ion{Fe}{II}); some fits also included \ion{Si}{III},
\ion{Fe}{III}, or \ion{Ti}{II}. {\tt SYNAPPS} was also allowed to
include a ``detached'' version of \ion{Ca}{II}, representing the
HVFs. Some of these fits can be seen in Figure~\ref{f:fits} as the
purple, long-dashed curves.

In general, the {\tt SYNAPPS} fits match well with our Gaussian fits
to the data in that the spectral shapes agree, and when our fitting
algorithm detects a HVF, it is also required in the {\tt SYNAPPS} fit
(and vice versa). Note that the {\tt SYNAPPS} fits were not intended
to perfectly match the entire spectral range covered by the data; they
were used mainly to identify and disentangle the HVF and PVF
components of the \cahk\ and \cair\ features. The sum of the
Gaussian fits used herein tend to reproduce the details of the profile 
better than the {\tt SYNAPPS} fits, as the latter appear to be
``smoothed out'' and are unable to match the smaller-scale features in
the data. 

Given the uncertainties of the measured velocities using our
fitting algorithm and the degeneracy in the {\tt
  SYNAPPS} fits between input velocity and $\tau_{\rm ref}$, the
velocities derived using the two methods are consistent within
\about2$\sigma$ (i.e., \about1200~\kms). The largest disagreement in
the velocities comes when the PVFs are weak. In these cases, there are
large uncertainties in the {\tt SYNAPPS} parameters, and one can change
the value of $\tau_{\rm ref}$ a small amount to force the velocity
of the PVFs to match what is measured using our fitting algorithm. 

As for the measured strengths of the features as characterised by
their pEWs, the {\tt SYNAPPS} fits and the measurements from
our fitting algorithm are roughly consistent, but not as close to each 
other as the velocities of the features. This is possibly caused by the
uncertainty associated with the $\tau_{\rm ref}$ parameter in the
{\tt SYNAPPS} fits. Furthermore, {\tt SYNAPPS} indicates that the
\ion{Si}{II} $\lambda$3858 feature is always present in the
observations, but is only noticeably strong in spectra where our
fitting algorithm required it to be included in the Gaussian fits. In
conclusion, while both {\tt SYNAPPS} and our spectral feature fitting
algorithm have their distinct pros and cons, their general agreement
(especially regarding the existence of HVFs) is encouraging.

\subsection{Comparisons to Previous Measurements}

\citet{Marion13:09ig} present a detailed study of the well-observed
Type~Ia SN~2009ig, specifically focusing on HVFs of various ions in
the pre- and near-maximum-brightness spectra. This study is one of the
very few that seriously investigates HVFs of \siii\ (as we also do in
this work). To compare and contrast with SN~2009ig, they also discuss
HVFs in a handful of other objects. To derive the velocities,
\citet{Marion13:09ig} fit Gaussians to the cores of the features
without removing the continuum. They inspect the positions of the
derived minima visually and then calculate the velocity and
uncertainty of HVFs and PVFs in \cahk, \siii, and \cair.

The dataset used herein includes 16 of their spectra of SN~2009ig, as
well as 16 spectra of 9 other SNe~Ia studied by
\citet{Marion13:09ig}. For the 7 (27) spectra present in both samples
that include the \cahk\ (\cair) feature, we find that both the HVF and
PVF velocities are consistent at the 2--3$\sigma$ level, with a nearly
constant offset of \about1400~\kms\ (\about900~\kms) between the two
studies. Similar results 
are found for the \siii\ feature, where the average offset in the HVF
velocity is \about1100~\kms\ (for 8 spectra) and in the PVF velocity
is \about400~\kms\ (for 14 spectra). We detect about half the number
of HVFs of \siii\ as \citet{Marion13:09ig}, likely owing to the
different fitting algorithms employed in the two studies. Finally, we
note that the offsets are such that the HVF velocities measured herein
tend to be higher than those of \citet{Marion13:09ig}, while the PVF
velocities tend to be lower, thus leading to the current study finding
larger velocity differences between the two components.

\citet{Childress14}, as mentioned above, studied HVFs of \cair\ in a
relatively large sample of SNe~Ia spectra near maximum brightness,
which represents a subset of the sample used herein. They used two 
Gaussians to fit each \cair\ profile and assumed equal strength in
each of the triplet components. \citet{Childress14} also forced a
minimum velocity difference between the HVFs and the PVFs in a given
spectrum of 2000~\kms, and required that the velocity and width of the
PVFs be within 10~per~cent of those of the \siii\ feature in the same
spectrum. As described in Section~\ref{s:procedure}, our fitting
algorithm does not impose such strict limits on the fit parameters. 

The measured \siii\ PVF (\cair\ HVF and PVF) velocities of the 56
spectra that are in both datasets are consistent at the
1--2$\sigma$ level, with typical offsets of \about300~\kms\
(\about500~\kms). These offsets are such that the velocities measured
herein tend to be larger than those reported by
\citet{Childress14}. They also measure pEWs for the \siii\ PVFs and
the \cair\ PVFs and HVFs. These values are consistent with what we
measure at the 2--3$\sigma$ level (offsets of \about9 \AA), and once
again our values tend to be larger than those of
\citet{Childress14}. These relatively minor differences are likely
caused by the assumption of optically thin (this work, see above)
versus optically thick \citep{Childress14} spectral features.


\section{Results \& Analysis}\label{s:results}

\subsection{The Existence of HVFs in  \cahk\ and \cair}\label{ss:ca}

Using the aforementioned algorithm, we calculate the pEW and expansion
velocities of HVFs and PVFs for the \cahk\ and \cair\ features; these
are listed in Tables~\ref{t:cahk} and \ref{t:cair}, respectively. For
\cahk, we fit a total of 126 spectra of 84 SNe~Ia; 5 of these spectra
have HVFs only, 12 have PVFs only, 79 have both HVFs and PVFs present,
15 have PVFs and a \ion{Si}{II} $\lambda$3858 feature, and 15 have
both HVFs and PVFs, in addition to \ion{Si}{II} $\lambda$3858. On the
other hand, we fit the \cair\ feature in a total of 382 spectra of 192
SNe~Ia; 16 of these spectra have HVFs only, 105 have PVFs only, and
261 have both HVFs and PVFs present.

There are eight SNe~Ia in the sample that exhibit only HVFs in their
earliest spectra; most of these observations are earlier than 7~d
before maximum brightness. We have multiple spectra of three of these
objects, and all three eventually develop
PVFs. \citet{Childress13:12fr} found evidence for HVFs, but not PVFs,
in their earliest spectra of SN~2012fr, consistent with what
is found herein using the same observations. On the other hand,
\citet{Marion13:09ig} detected HVFs, but {\it not} PVFs, in early-time
spectra of SN~2009ig, while we {\it do} detect PVFs (as well as HVFs)
in the same data. Note that \citet{Maguire14} found no spectra with
only HVFs in their sample.

SN~Ia spectra tend to evolve from having only HVFs (in \about4~per~cent of
cases), to having both HVFs and PVFs (in the majority of spectra,
i.e., \about65--75~per~cent), to having only PVFs (in \about20--30~per~cent of the
observations). This generic picture of spectra changing with time
(HVFs$\rightarrow$HVFs+PVFs$\rightarrow$PVFs) is consistent with what
has been seen in previous work
\citep[e.g.,][]{Childress13:12fr,Marion13:09ig}. As mentioned above,
spectra with only HVFs are seen almost exclusively at very early
times. Spectra with only PVFs are seen as early as \about12~d before
maximum brightness and as late as 5~d past maximum (which are the
oldest spectra included in the current study). Data that show both
HVFs and PVFs simultaneously are observed at all epochs studied herein
($-16 < t < 5$~d), implying that there is evidence of some SNe~Ia
showing HVFs at epochs as late as 5~d past maximum brightness, though 
most HVFs are gone by about 5~d {\it before} maximum.

When considering the entire dataset studied herein, we find that
\about67~per~cent of all objects show HVFs in at least one spectrum. This is
almost exactly the same percentage that was found by
\citet{Maguire14}. When looking at only early-time observations ($t
\la -4$~d), \about91~per~cent of the objects show evidence of HVFs, which is
consistent with, but slightly higher than, what was found previously
\citep[83~per~cent,][]{Maguire14}.

Of the SNe~Ia for which we fit the \cahk\ or \cair\ features,
\about28~per~cent of them are HV objects, consistent with the 
overall SN~Ia population
\citep[e.g.,][]{Wang09,Silverman12:BSNIPII}. Of the SNe~Ia with a
known Wang Type, 77~per~cent (71~per~cent) of HV objects show HVFs of \cahk\
(\cair), while 70~per~cent (62~per~cent) of N objects shows HVFs of \cahk\
(\cair). Similarly, (SNID Type) Ia-norm objects contain HVFs of \cahk\
78~per~cent of the 
time and HVFs of \cair\ 70~per~cent of the time, and all 10
Ia-91T/99aa objects in our 
dataset show HVFs of \ion{Ca}{II}. Given the number of SNe~Ia in each
category, these percentages are all mutually consistent. On the other
hand, only 1 out of 17 Ia-91bg objects show HVFs of \ion{Ca}{II}. This
significant dearth of HVFs in underluminous SNe~Ia (i.e., Ia-91bg 
objects) has been noticed in previous work as well
\citep{Maguire12,Childress14,Maguire14}.

The entire BSNIP dataset averages \about2 spectra per object
\citep{Silverman12:BSNIPI}, and since these data make up the bulk of
the sample used herein, there are not many objects for which we
have multiple spectra. Thus, we are only able to determine a Benetti
Type for a handful of the objects studied in this work. For those
SNe~Ia with a Benetti Type, 62~per~cent (72~per~cent) of HVG objects show HVFs of
\cahk\ (\cair), while 85~per~cent (82~per~cent) of LVG objects shows HVFs of \cahk\
(\cair). Only 1 of 8 SNe~Ia with a Benetti Type of FAINT (i.e.,
underluminous objects) contained HVFs. These numbers are consistent
with what was found above using Wang and SNID Types, given the
association of HV/N/Ia-91bg objects with HVG/LVG/FAINT objects
\citep[e.g.,][]{Silverman12:BSNIPII}.

A possible bias leading to the above result is that we do not have any
spectra of Ia-91bg (or FAINT) objects at epochs earlier than 3 days
before maximum brightness. Thus, perhaps, Ia-91bg/FAINT objects have
HVFs, but they disappear earlier than in other SN~Ia subtypes. We
reject this idea in part because at $t = -3$~d, about half of all
SNe~Ia show HVFs, and this is also the same epoch when HVFs and PVFs
tend to be about equal in strength (see
Section~\ref{ss:ew_t_ca}). Conversely, 6 of the 9 objects that show
only PVFs at $t < -3$~d are spectroscopically somewhat similar to
SN~1991bg or have relatively narrow light curves and thus appear to be
border cases between Ia-norm and Ia-91bg. The remaining 3 objects in
this category all have their earliest spectra at $t \approx -6$~d, so
HVFs could have been present at earlier times, but have faded by the
time our spectra were obtained. 

We further investigate whether the apparent lack of HVFs in
Ia-91bg/FAINT objects is an observational bias by determining the
typical epoch at which HVFs ``disappear.'' This was done by taking
each object with more than 1 spectrum in our dataset and fitting a
line to the strength of the HVF relative to the PVF (see
Section~\ref{ss:ew_t_ca}) versus time in order to find the epoch
at which the relative strength of the HVF drops below our detection
threshold of 0.1 (see Section~\ref{sss:synthetic}); we refer to this
as the ``epoch of disappearance.'' This epoch is then compared to the
light-curve width (i.e., $\Delta m_{15}(B)$) in order to search for
any relationship between peak luminosity and the epoch of
disappearance.

In the current sample, there were 26 SNe~Ia with known $\Delta
m_{15}(B)$ values and for which we were able to determine an epoch of
disappearance. The latter for these objects is 
about $t = -1$~d to $t = +0.5$~d. When comparing the epoch of
disappearance to $\Delta m_{15}(B)$, we find a large amount of
scatter. The epoch of disappearance may decrease with increasing
$\Delta m_{15}(B)$, but the slope of the linear fit is consistent
with 0. Using our best linear fit to the data, we find the epoch of
disappearance to be about $-1.0$~d for $\Delta m_{15}(B)$ values of
1.4--1.6~mag \citep[e.g.,][typical for Ia-91bg/underluminous
objects]{Ganeshalingam10:phot_paper}. 

There are no objects classified as Ia-91bg in this work with spectra
obtained earlier than 3~d before maximum brightness. However,
according to the above analysis, Ia-91bg spectra obtained earlier than
\about1~d before maximum {\it should} show HVFs. Thus, Ia-91bg objects
(equivalently, Benetti FAINT objects or SNe~Ia with narrow light
curves) seem to {\it never} show HVFs, while all other SN~Ia subtypes
studied herein (HV, N, Ia-norm, Ia-91T/99aa, LVG, and HVG objects)
{\it always} show HVFs (in spectra obtained earlier than \about6~d
before maximum). Owing to there being relatively few Ia-91bg objects
in our sample, however, there is a small possibility that they may
have HVFs at epochs earlier than about 3~d before maximum, but these
features would have to disappear even earlier than one would expect
based on the rest of our dataset.

\subsection{ \ion{Ca}{II} pEWs}\label{ss:ew_t_ca}

\begin{figure*}
\centering$
\begin{array}{c}
\includegraphics[width=3.7in]{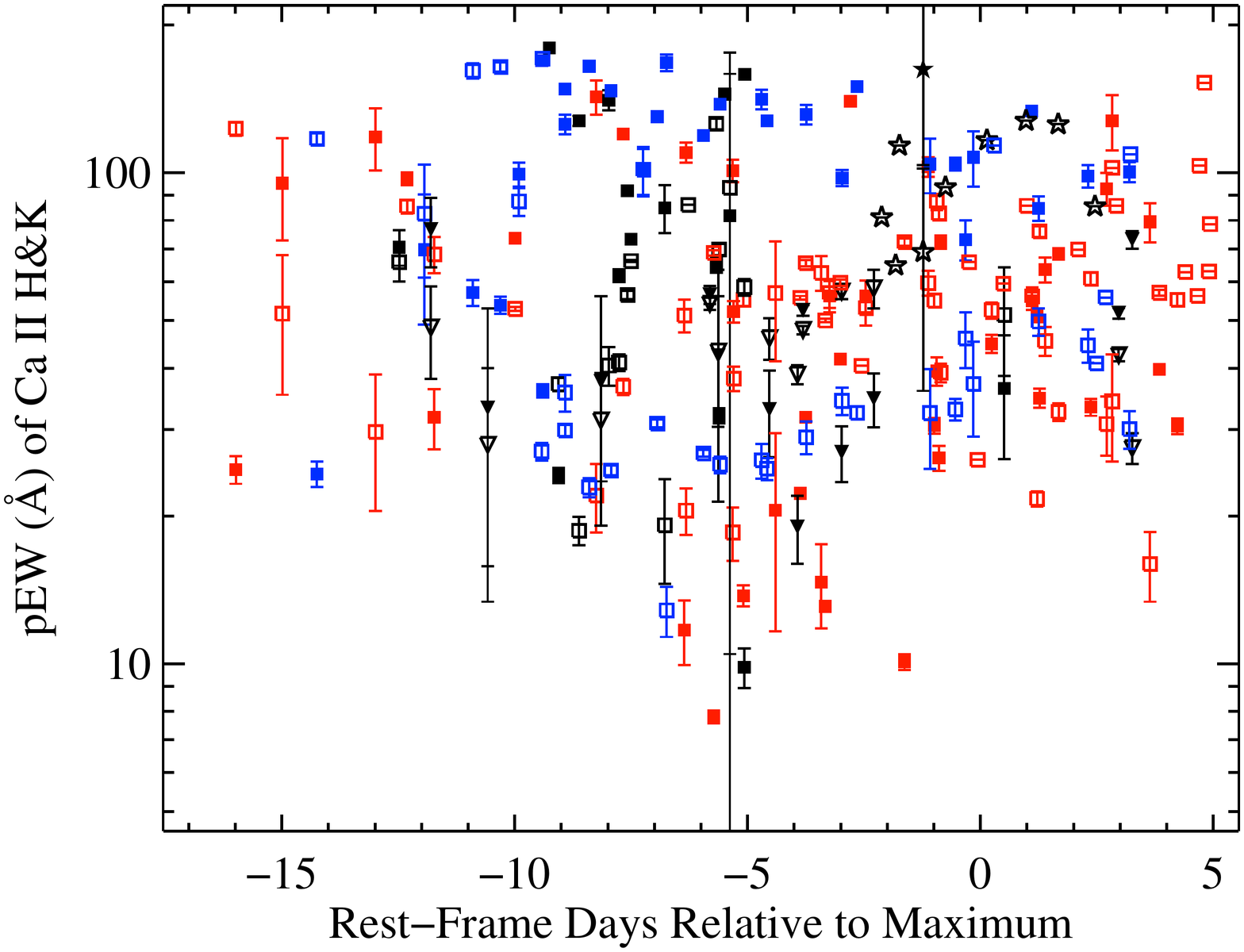} \\
\includegraphics[width=3.7in]{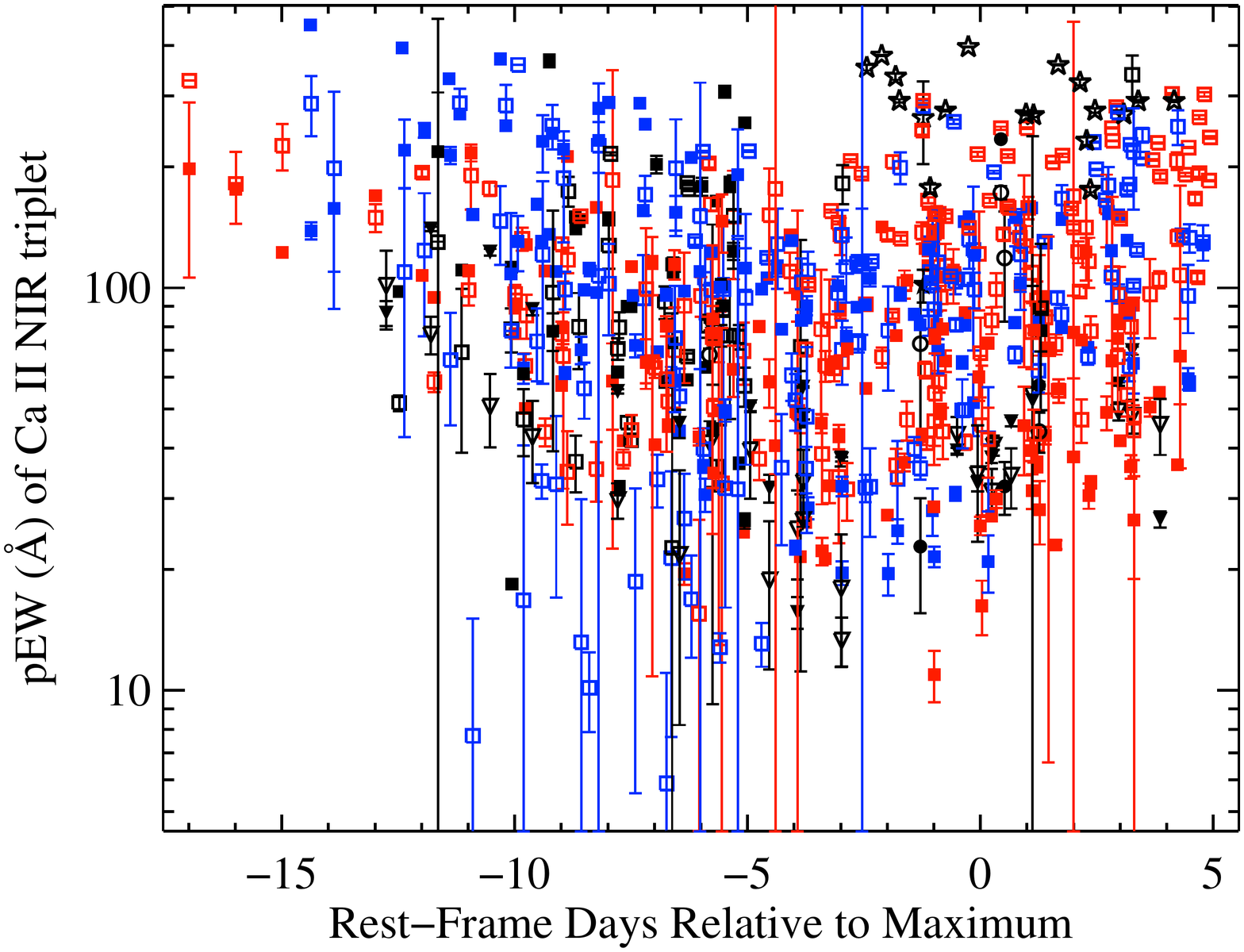} \\
\includegraphics[width=3.7in]{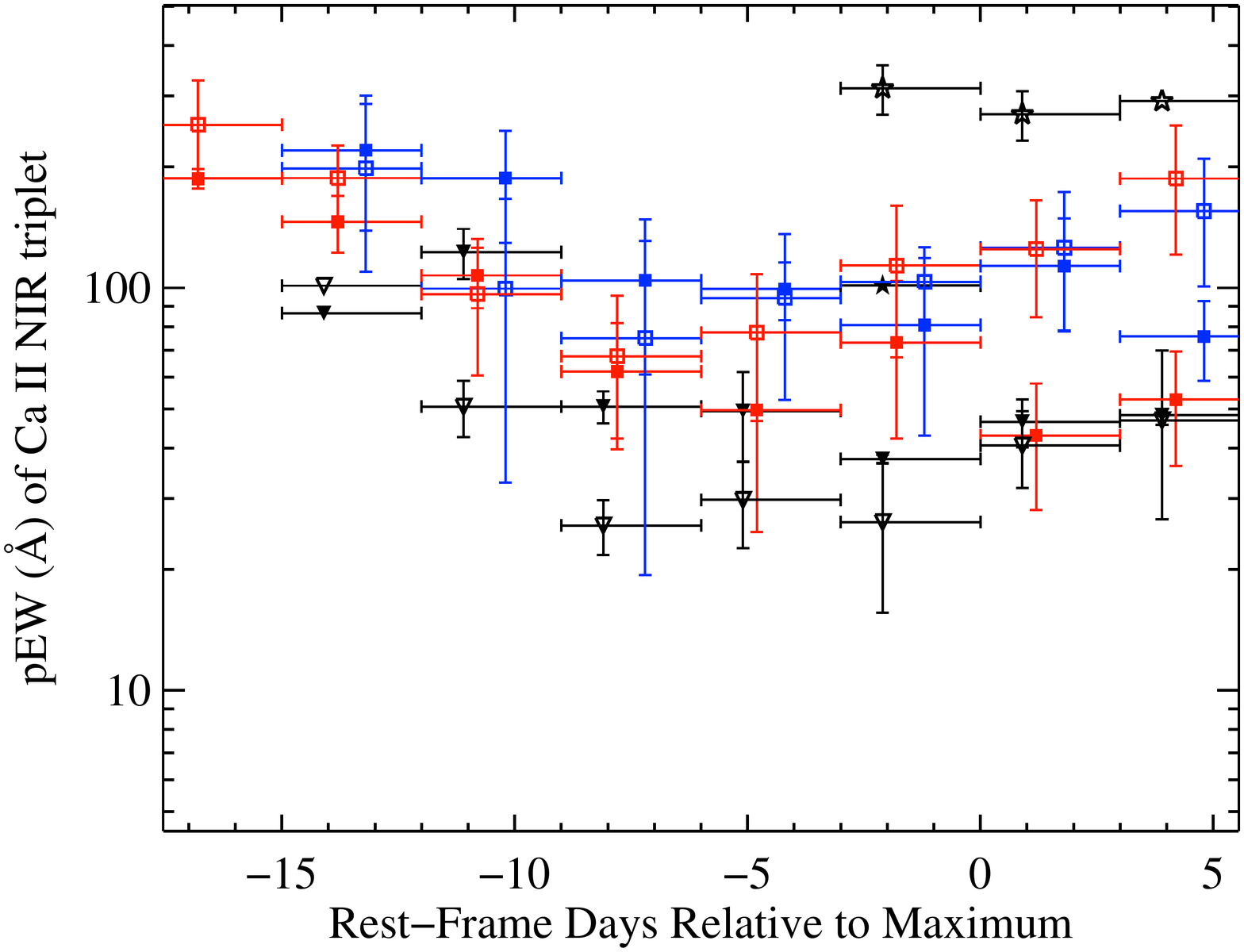} \\
\end{array}$
\caption{The \cahk\ ({\it top}) and \cair\ ({\it middle} and {\it
    bottom}) pEWs versus time. The bottom panel shows the median pEW
  in time bins of 3~d for objects that are classified as Ia-91bg,
  Ia-91T/99aa, HV, or N (shifted slightly from bin centre for 
  clarity). The horizontal error bars represent 
  the width of each bin while the vertical error bars are the median
  absolute deviation in each bin. Open symbols are PVFs; filled
  symbols are HVFs. Blue points are HV objects, red points are N
  objects, and black points are objects without a Wang Type. Squares
  are Ia-norm, stars are Ia-91bg, triangles are Ia-91T/99aa, and
  circles are objects without a SNID Type. There is large scatter in
  the pEWs of HVFs and PVFs for both \ion{Ca}{II} features at all
  epochs, though the pEWs of HVFs (PVFs) tend to decrease (increase)
  with time.}\label{f:ew_t_ca}
\end{figure*}

The pEWs of \cahk\ and \cair\ for both HVFs and PVFs are listed in
Tables~\ref{t:cahk} and \ref{t:cair},
respectively. The temporal evolution of these pEWs is displayed in
Figure~\ref{f:ew_t_ca}. Open symbols represent PVFs while filled
symbols represent HVFs. Blue points are high-velocity (HV) objects,
red points are normal-velocity (N) objects, and black points are
objects for which we could not determine a Wang Type. Squares are
Ia-norm objects, stars are Ia-91bg objects, triangles are Ia-91T/99aa
objects, and circles are objects which do not have a SNID Type.

At all epochs there is large scatter in the pEWs of HVFs and
PVFs for both \ion{Ca}{II} features. For $t \ga -9$~d, the pEWs of the
HVFs tend to decrease with time while those of the PVFs tend to
increase with time, as expected. One specific counterexample to this
is SN~2009ig, which has stronger PVFs than HVFs at the very earliest
times, but quickly evolves to have the HVFs dominate the profile
\citep[until $t \approx -6$~d; see also][]{Marion13:09ig}. The typical
epoch at which the strengths of 
the HVFs and PVFs of the \ion{Ca}{II} features are equal is about 4~d
before maximum brightness. That being said, individual SNe~Ia achieve
equal HVF and PVF strength at a range of epochs ($-8 < t < 2$~d),
which matches what has been found previously \citep{Marion13:09ig}. 

HV objects tend to have strong HVFs at the earliest times, but they
decrease in strength relatively quickly, leading to somewhat weak HVFs
in HV objects near maximum brightness. The latter part of this result
has been seen previously \citep{Childress14,Maguire14}, but at a much
stronger level than what is found in the current study. We attribute
this difference not only to the epochs studied (the previous works
only used spectra within 5~d of maximum brightness), but also to the 
fact that these prior studies contained too few HV objects
\citep[\about13~per~cent of their sample, versus 28~per~cent
herein;][]{Childress14,Maguire14}. This difference is discussed in
more detail in Section~\ref{ss:comp_ca}. 

While Ia-91bg objects never show HVFs, they do exhibit some of the
largest pEWs of PVFs. On the other hand, Ia-91T/99aa objects always
show HVFs, but the pEWs of their PVFs and HVFs are some of the lowest
values seen in Figure~\ref{f:ew_t_ca}. These results have been found
previously and are relatively unsurprising since strong (weak)
absorption features are a defining characteristic of Ia-91bg
(Ia-91T/99aa) objects
\citep{Silverman12:BSNIPII,Folatelli13,Childress14}.

To investigate the {\it relative} strength of HVFs to PVFs,
\citet{Childress14} defined $R_\mathrm{HVF}$ as the ratio of pEW of
the HVF of \cair\ to the pEW of the PVF of \cair. In this work, we
define $R_\mathrm{\cahk}$ and $R_\mathrm{\cair}$ as the ratios of pEWs
of the HVFs to the PVFs of \cahk\ and \cair, respectively. Note that
spectra with only PVFs have a ratio of identically zero, while spectra
with only HVFs have an undefined ratio. The values of
$R_\mathrm{\cahk}$ and $R_\mathrm{\cair}$ are listed in
Tables~\ref{t:cahk} and \ref{t:cair}, respectively.

The ratios found herein span a range of 0--20, though most are less 
than 4. This is much larger than what was measured by \citet{Childress14}, 
who do not find ratios larger than \about2. The difference is likely
caused by the smaller epoch range studied in \citet{Childress14}; they
only use spectra within 5~d of maximum. When considering only spectra
from these epochs in the current work, we find that most of the ratios
are less than 2.5, with only 4 spectra falling above this value. Thus, our  
$R_\mathrm{\cair}$ values are consistent with those in
\citet{Childress14}. While one might expect $R_\mathrm{\cahk}$ to be
correlated with $R_\mathrm{\cair}$, we find that this is not true. One
explanation is that the \cahk\ and \cair\ absorption strengths depend
on temperature in different ways and the material that is
responsible for the HVFs is likely at a different temperature than the
photospheric material \citep[e.g.,][]{Childress14}. In addition, the
values of $R_\mathrm{\cahk}$ might be skewed slightly by the presence
of weak \ion{Si}{II} $\lambda$3858 absorption, though we find that
this is likely a relatively small contamination (see
Section~\ref{ss:si3858} for more).

\subsection{ \ion{Ca}{II} Velocities}\label{ss:v_t_ca}
 
\begin{figure*}
\centering$
\begin{array}{c}
\includegraphics[width=5.5in]{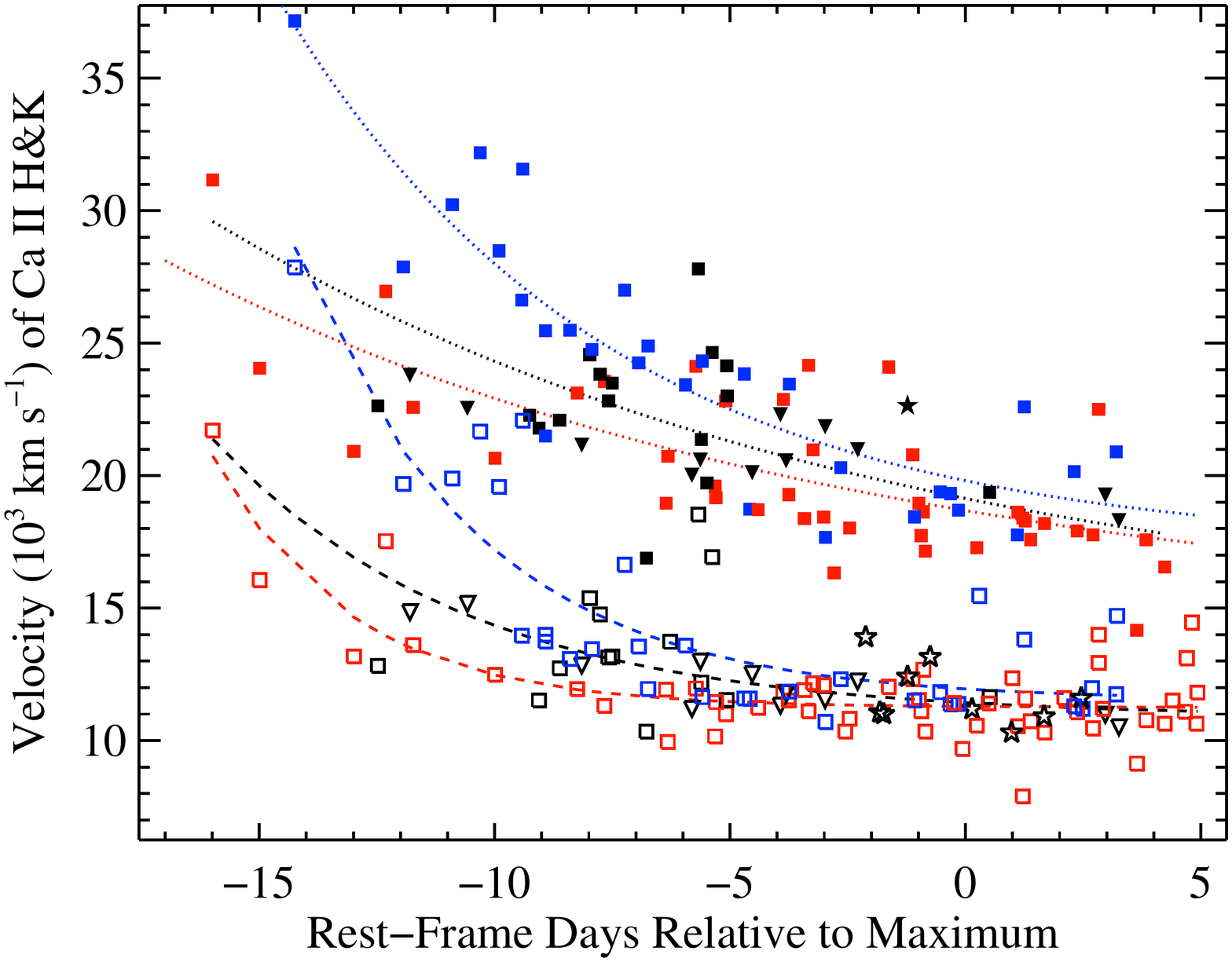} \\
\includegraphics[width=5.5in]{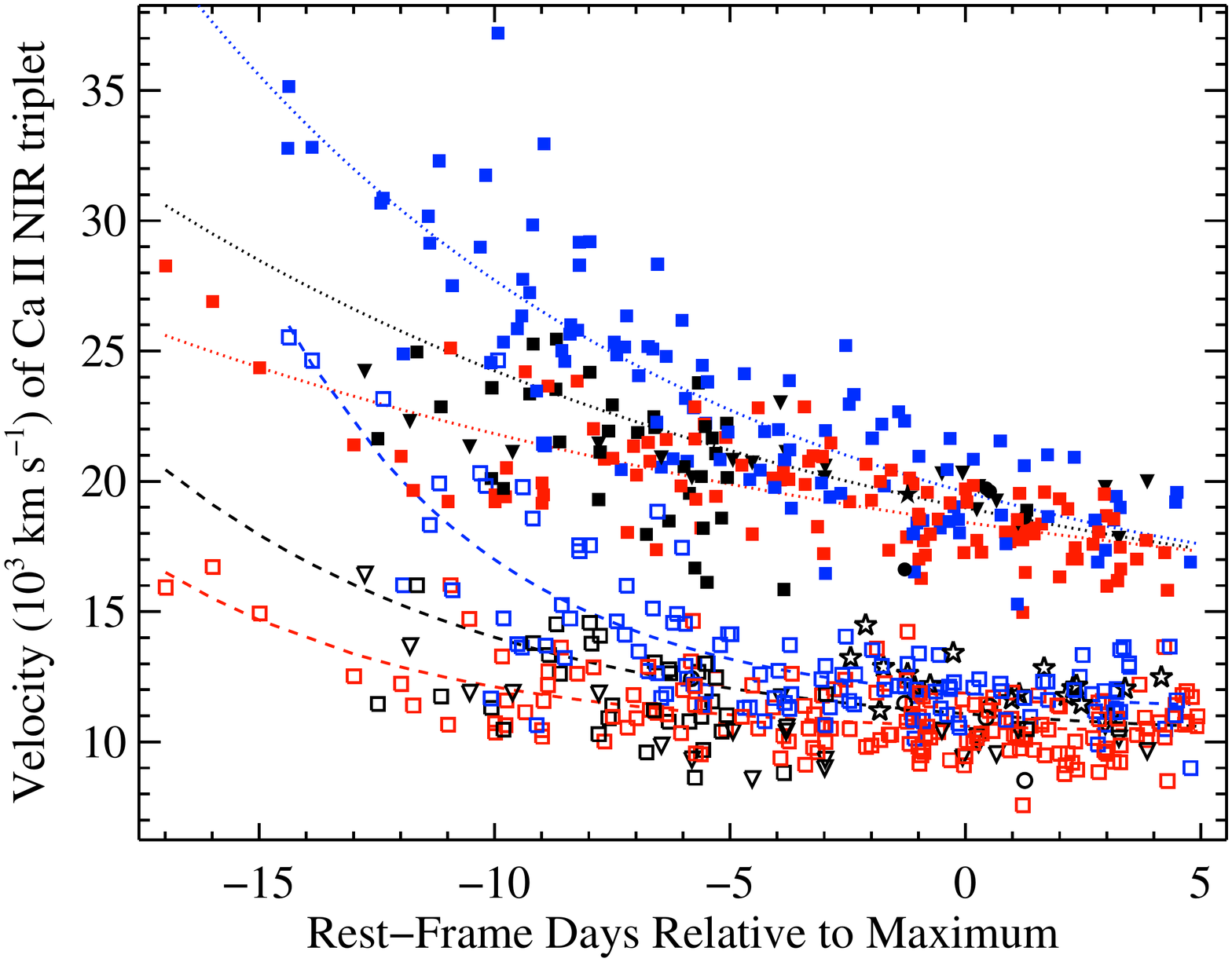} \\
\end{array}$
\caption{The \cahk\ ({\it top}) and \cair\ ({\it bottom}) velocities
  versus time. Colours and shapes of data points are the same as in
  Figure~\ref{f:ew_t_ca}. Measurement uncertainties are comparable to the size of
  the data points. The black, blue, and red dashed lines are natural
  exponential function fits to PVF velocities of all objects, HV
  objects only, and N objects only, respectively. The black, blue, and
  red dotted lines are natural exponential function fits to HVF
  velocities of all objects, HV objects only, and N objects only,
  respectively. Note the gap between the HVF and PVF points,
  especially for $t \ga -5$~d; this minimum difference between HVF and
  PVF velocities appears to be real and not merely a measurement
  artifact.}\label{f:v_t_ca} 
\end{figure*}

The expansion velocities of HVFs and PVFs for the \cahk\ and \cair\
features are listed in Tables~\ref{t:cahk} and \ref{t:cair},
respectively. Figure~\ref{f:v_t_ca} shows the temporal evolution of
the \cahk\ (top) and \cair\ (bottom) velocities. Colours and shapes of
data points are the same as in Figure~\ref{f:ew_t_ca}; measurement uncertainties
are comparable to the size of the data points. The black dashed line
represents the best-fitting natural exponential function to all of
the PVF velocities, while the blue and red dashed lines use only HV
and N objects, respectively. Similarly, the black dotted line is the
best-fitting natural exponential function to all of the HVF
velocities, and the blue and red dotted lines use only HV and N
objects, respectively.

For any given object, all of the measured velocities tend to decrease
with time, as expected and as seen in previous work
\citep[e.g.,][]{Silverman12:BSNIPII}. Furthermore, in a given
spectrum, the difference in velocity between the \cahk\ and \cair\
features (for both PVFs and HVFs) is typically \about500~\kms. The
exponential fits in Figure~\ref{f:v_t_ca} show that in
general, for both \ion{Ca}{II} features, the HVFs (dotted lines) and
PVFs (dashed lines) of HV objects start out with higher velocities
than the N objects, and their velocities decrease more quickly with
time. Consequently, the HV and N objects have similar HVF and PVF
velocities near maximum brightness. This may not be surprising
(i.e., that HV objects have higher velocities), but we note that the
Wang Type classification is based on the near-maximum-brightness
velocity of \siii, and not the \ion{Ca}{II} features.

Furthermore, we find that Ia-norm and Ia-91bg objects have consistent
PVF velocities (recall that only a single Ia-91bg object shows a HVF), 
while Ia-91T/99aa objects have significantly lower HVF and PVF
velocities. The Ia-91T/99aa objects also show a much slower decrease
in their velocities with time, so once again the velocities become
consistent with the rest of the sample by maximum brightness. As for
the Benetti Types, HVG objects tend to start with higher velocities
and decrease their velocities more quickly, as compared to LVG
objects, consistent with the behaviour of the Wang HV and N
objects above.

These results are somewhat different than what was seen in the
early-time \ion{Ca}{II} velocities reported in BSNIP~II
\citep{Silverman12:BSNIPII}, but the studies are consistent for data
closer to maximum brightness. This is likely due to strong HVFs of
\ion{Ca}{II} in early-time spectra being blended with PVFs and biasing
the measurements in BSNIP~II. The velocities presented herein more
accurately reflect the 
actual spectral profiles and expansion velocities present in the
data since we carefully take into account the (possible) presence of
HVFs in each observation.

\begin{figure*}
\centering
\includegraphics[width=5.5in]{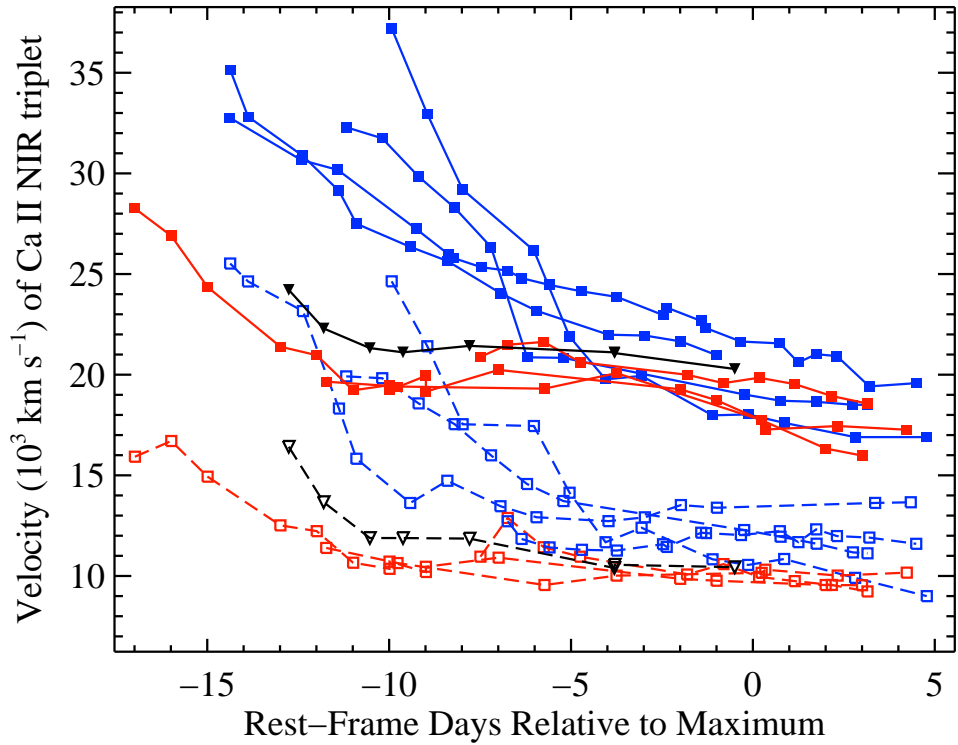}
\caption{The \cair\ velocities versus time for the eight SNe~Ia for which
  we have more than seven spectra (see main text for the list of
  objects). All PVF (HVF) velocities of a given object are connected
  with a dashed (solid) line. Colours and shapes of data points are
  the same as in Figure~\ref{f:ew_t_ca}. Measurement uncertainties are comparable
  to the size of the data points.}\label{f:v_t_ca_many}
\end{figure*}

In order to show how the velocities of a few individual objects evolve
with time, in Figure~\ref{f:v_t_ca_many} we plot a subset of the data
displayed in the bottom panel of
Figure~\ref{f:v_t_ca}. Figure~\ref{f:v_t_ca_many} shows only \cair\
velocities of objects for which we have more 
than seven spectra.\footnote{SN~2009ig is the only object in our dataset
  with more than seven spectra where we are able to fit \cahk. Thus, we
  did not make a plot corresponding to Figure~\ref{f:v_t_ca_many} for
  \cahk.} All of the PVF (HVF) velocities of a given
object are connected with a dashed (solid) line. This sample of eight
objects includes the five extremely well-observed SNe~Ia mentioned above
(SNe~2009ig, 2011by, 2011fe, 2012cg---the lone Ia-91T/99aa object in
the Figure, and 2012fr), in addition to SNe~2006X, 2010kg, and 
2011ao. The above conclusions for the entire sample appear
to also hold for this subset. Namely, HV objects tend to
have faster HVFs and PVFs and their velocities decrease more
quickly with time than N objects. 

\begin{figure}
\centering$
\begin{array}{c}
\includegraphics[width=3.5in]{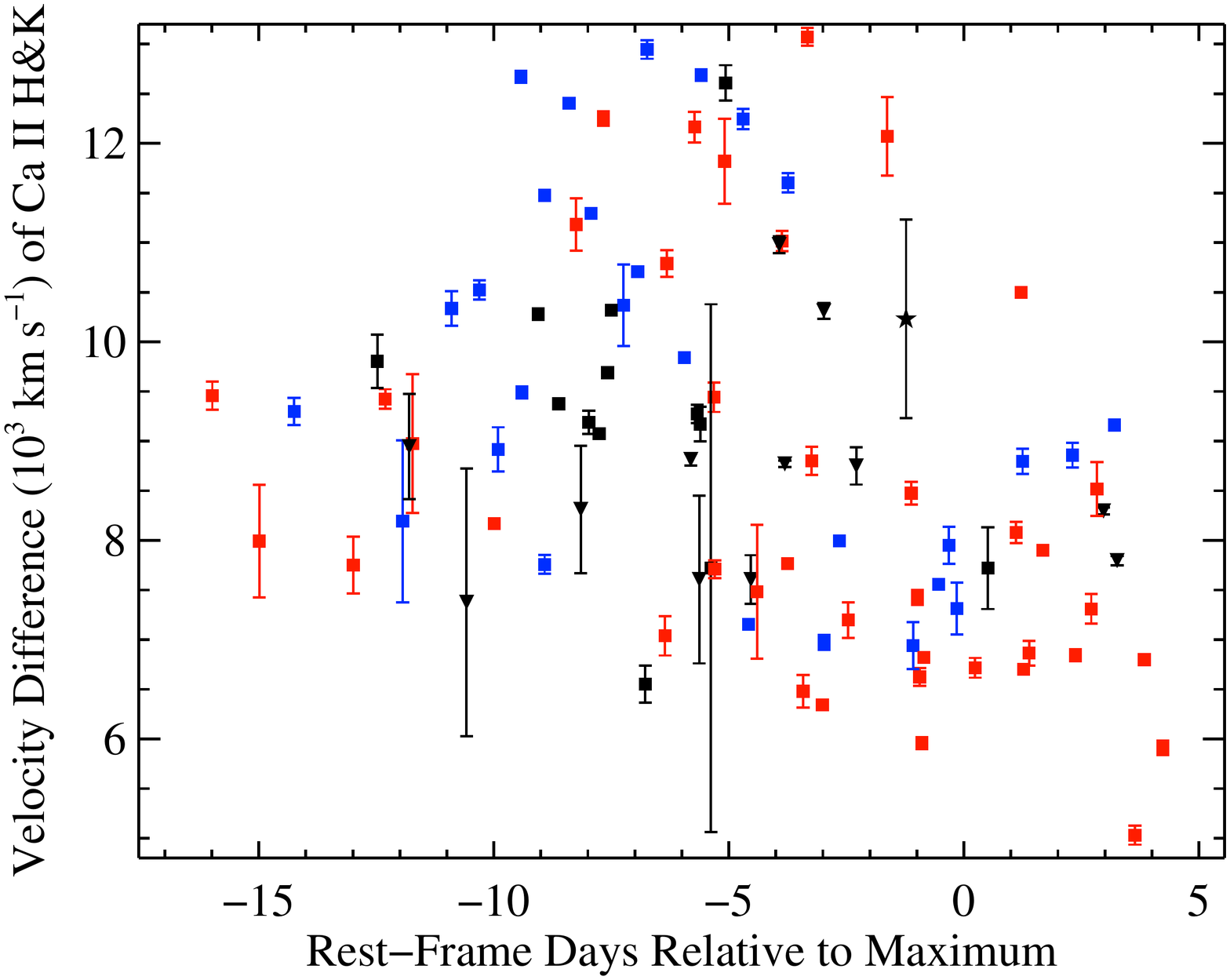} \\
\includegraphics[width=3.5in]{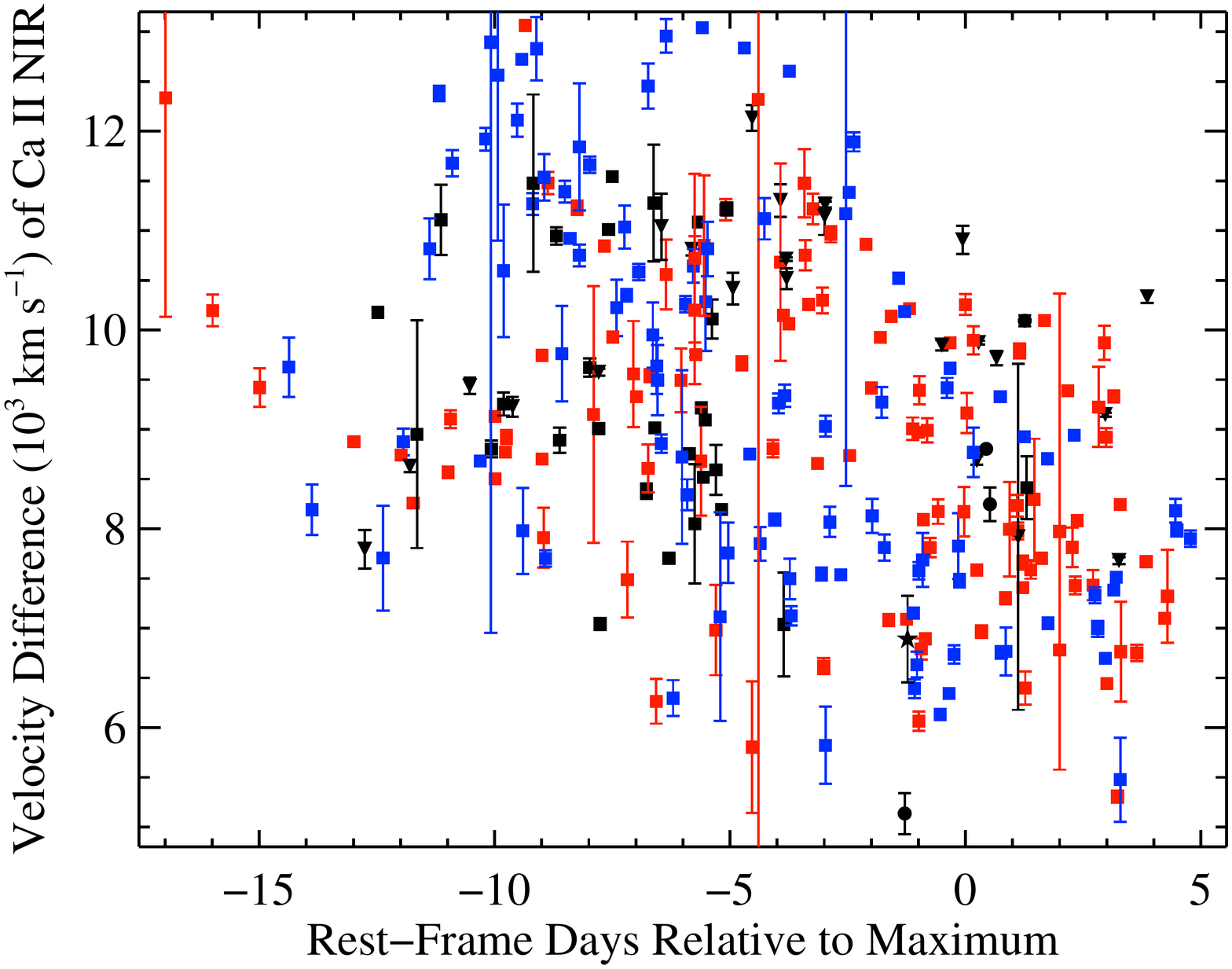} \\
\end{array}$
\caption{The temporal evolution of the difference in velocity between
  HVFs and PVFs of \cahk\ ({\it top}) and \cair\ ({\it
    bottom}). Colours and shapes of data points are the same as in
  Figure~\ref{f:ew_t_ca}. Note that three of these objects have early-time 
  PVFs with higher velocities than their later-time
  HVFs.}\label{f:vdiff_t_ca}   
\end{figure}

In both panels of Figure~\ref{f:v_t_ca} and in
Figure~\ref{f:v_t_ca_many}, there is a noticeable gap between the HVF
and PVF points, especially for $t \ga -5$~d. We further investigate
this gap by calculating the difference in velocity between the HVFs
and PVFs in a given spectrum for all observations where both
components are observed. The temporal evolution of this separation for
\cahk\ (\cair) is shown in top (bottom) panel of
Figure~\ref{f:vdiff_t_ca}.

Both \ion{Ca}{II} features, in all subtypes, show a large range of
values for the 
velocity separation at all epochs, but the difference tends to
decrease with time. In fact, a linear fit to the data indicates a
decrease at the 4$\sigma$ level (7$\sigma$ level) for \cahk\ (\cair)
from \about11,000~\kms\ to \about8000~\kms. The typical velocity
separation for both \ion{Ca}{II} features is \about9000~\kms, slightly
higher than the 7000~\kms\ value found by \citet{Maguire14}, and all
of the SN~Ia subtypes studied herein have consistent typical velocity
differences. No velocity differences are detected in the present study
less than 5000~\kms, consistent with \citet{Marion13:09ig}. In
fact, the vast majority of the velocity differences are greater than
6000~\kms, 
significantly larger than the minimum separation that our
fitting algorithm is able to ``resolve'' (see
Section~\ref{sss:synthetic}). Thus, the gaps between the HVFs and PVFs
in Figures~\ref{f:v_t_ca} and \ref{f:v_t_ca_many} appear to be real.

While Figure~\ref{f:vdiff_t_ca} plots the velocity difference between
HVFs and PVFs in a given spectrum, we also investigated the
velocity separations for a given object at all epochs. To do
this, each object's maximum PVF velocity, usually from the earliest
spectrum of the object in question, was compared to its minimum HVF
velocity (usually from the latest spectrum of the object in
question). The vast majority of objects, \about96~per~cent, have all
of their HVF velocities larger than all of their PVF velocities
(i.e., the minimum HVF velocity is larger than the maximum PVF
velocity).

In contrast, there are five objects with a measured PVF velocity that is
larger than the lowest HVF velocity. Four of these objects (SNe  
2002bo, 2006X, 2009ig, and 2010kg) show some of the fastest
photospheric velocities ever observed in SNe~Ia (e.g.,
\citealt{Benetti04,Wang08,Marion13:09ig}; Silverman et~al., in
preparation, respectively) and are thus all 
classified as HV objects. There are many other HV objects in the
current sample, however, that do not show a PVF velocity larger than
their lowest HVF velocity. Perhaps this is caused by the fact that we do
not have sufficiently early spectra for these other HV objects to show such
a fast PVF. The fifth object in this category is SN~2011fe, which was
spectroscopically observed at extremely early epochs
\citep[e.g.,][]{Parrent12}. It is interesting to note that all five of
these objects also show evidence for a HVF of \siii\ in their earliest
epochs. Although, once again, a handful of other objects show HVFs of
\siii\ but do not have any PVF velocities that are larger than their
lowest HVF velocity (see Section~\ref{ss:si}).

\subsection{HVFs of \ion{Ca}{II} Compared to Other Observables}\label{ss:comp_ca}

In order to connect our analysis of HVFs to possible SN~Ia progenitors
and environments, we compare the absolute strengths (pEWs), relative
strengths ($R_\mathrm{\cahk}$ and $R_\mathrm{\cair}$), and velocities
measured herein to other observables. Using the photometric
information discussed at the end of Section~\ref{s:data}, we find no
correlation between $\left(B-V\right)_0$ and the pEWs of the HVFs or
the PVFs of \cahk\ and \cair\ at any epoch. The latter was also seen
by \citet{Childress14} for their low-reddening ($-0.15 <
\left(B-V\right)_0 < 0.15$~mag), near maximum-brightness (within 5~d
of maximum) sample. There is also no significant correlation between
$\left(B-V\right)_0$ and $R_\mathrm{\cahk}$ or $R_\mathrm{\cair}$ at
any epoch, again consistent with \citet{Childress14}.

\begin{figure*}
\centering$
\begin{array}{c}
\includegraphics[width=5.5in]{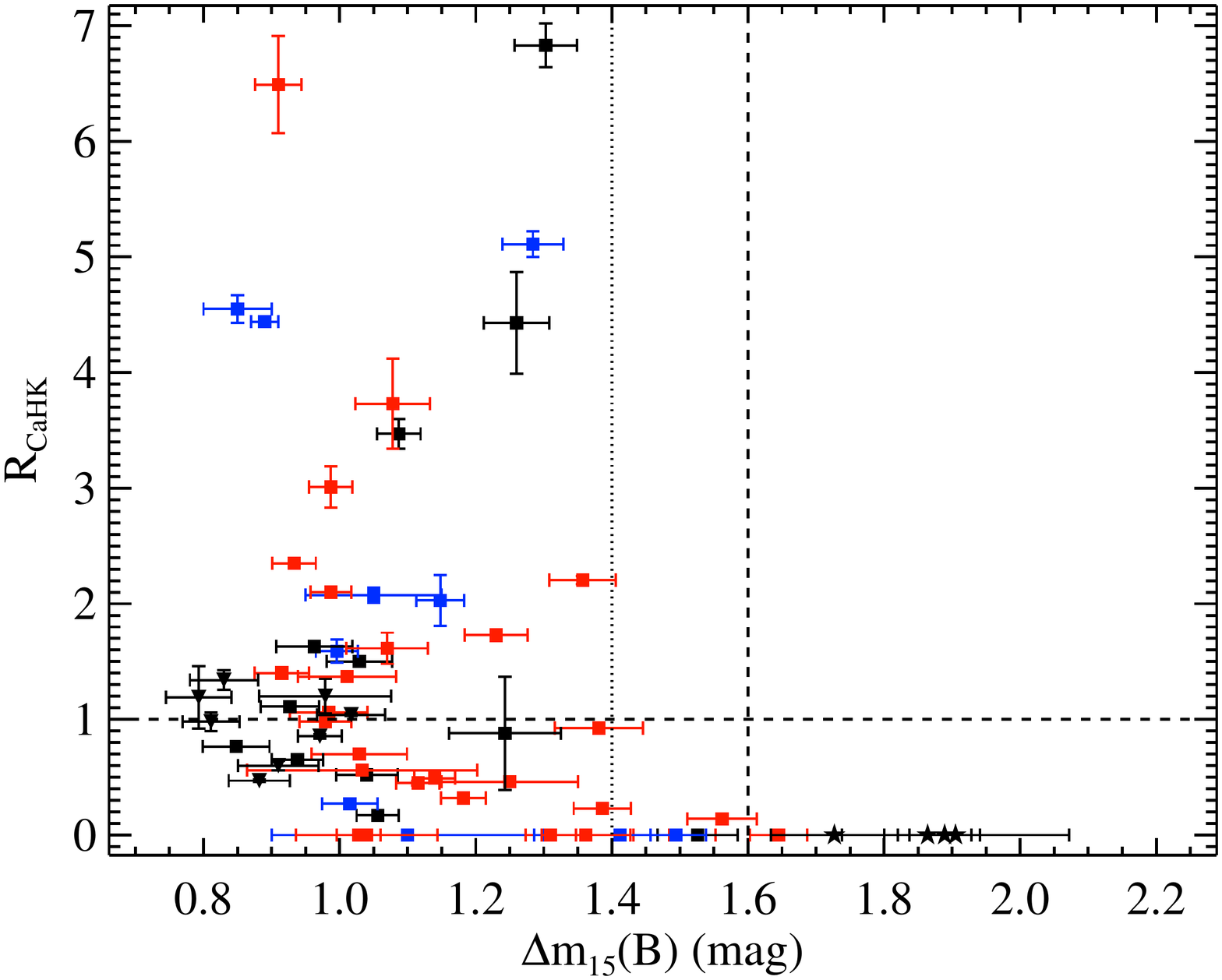} \\
\includegraphics[width=5.5in]{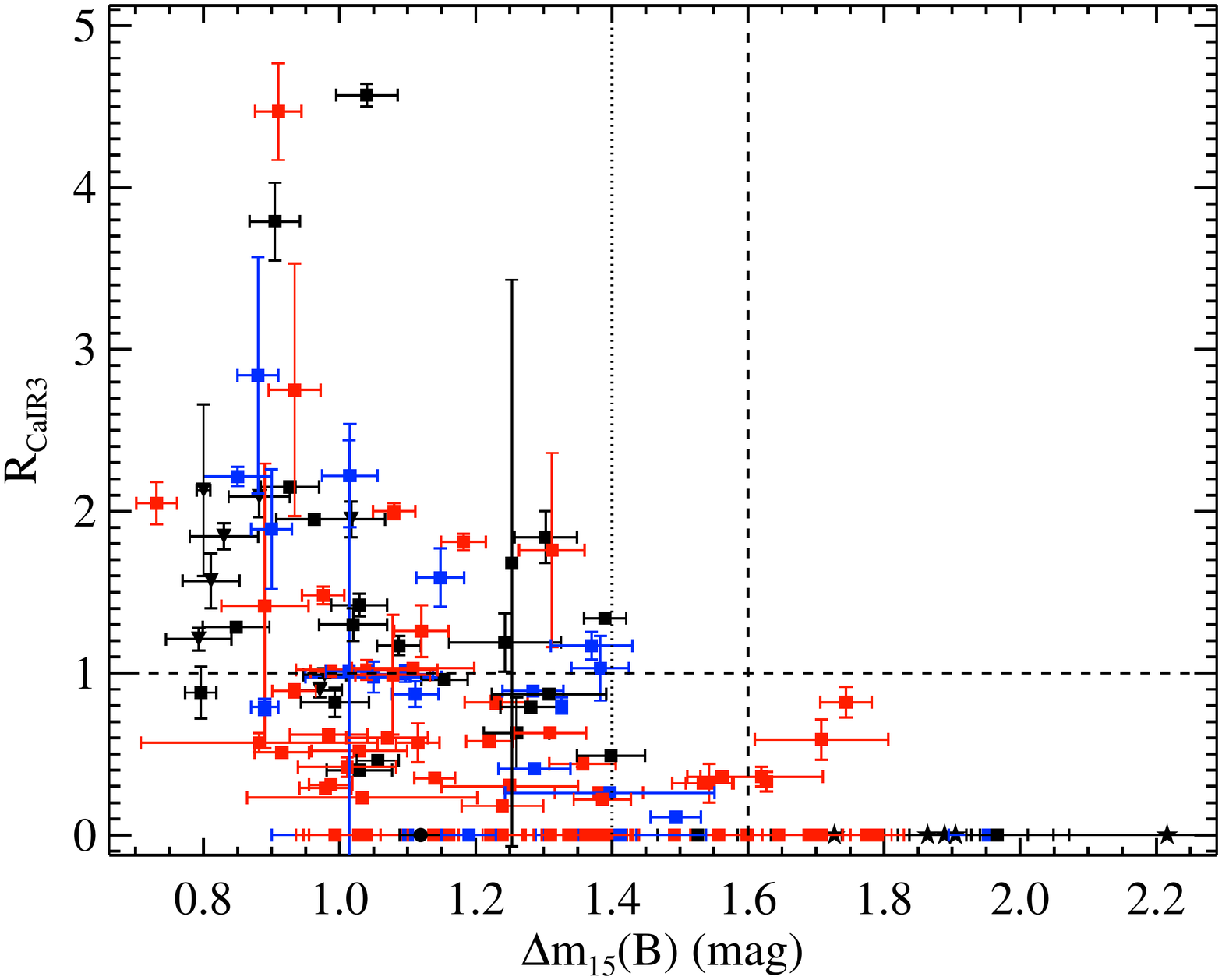} \\
\end{array}$
\caption{$R_\mathrm{\cahk}$ ({\it top}) and $R_\mathrm{\cair}$ ({\it
    bottom}) versus light-curve decline rate ($\Delta m_{15}(B)$). The
  median $R$ value of a given object is used for objects with multiple
  spectra. The 
  dashed vertical line is a typical cutoff between normal- and
  fast-declining objects; the dotted vertical line is a more
  conservative cutoff. The horizontal dashed line is where the pEWs of
  the HVFs and PVFs are equal. Colours and shapes of data points are
  the same as in Figure~\ref{f:ew_t_ca}.}\label{f:r_dm15_ca}
\end{figure*}

The so-called ``Phillips relation'' correlates the light-curve decline
rate of SNe~Ia with their luminosity at peak brightness
\citep{Phillips93}. Faster-declining SNe~Ia tend to be underluminous
and are also often spectroscopically Ia-91bg objects. In contrast,
slow-declining objects are usually overluminous and are of the
Ia-91T/99aa subtype. Figure~\ref{f:r_dm15_ca} compares
$R_\mathrm{\cahk}$ (top) and $R_\mathrm{\cair}$ (bottom) to the
light-curve decline rate, characterised by the $\Delta m_{15}(B)$
parameter. For objects with multiple spectra, the median $R$ value for
a given object is plotted in the figure.\footnote{Here, and elsewhere,
  when using the median $R$ value, we note that the results are
  unchanged when we instead use the mean $R$ value or the $R$ value
  from the earliest, latest, or closest-to-maximum brightness
  spectrum in our sample.} The dashed vertical line
at $\Delta m_{15}(B) = 1.6$~mag represents a typical cutoff between
normal-declining and fast-declining objects
\citep[e.g.,][]{Ganeshalingam10:phot_paper}. The dotted vertical line
at $\Delta m_{15}(B) = 1.4$~mag is a more conservative fast-declining
cutoff. The horizontal dashed line at $R=1$ 
represents where the pEWs of the HVFs and the PVFs are equal. 

Both $R_\mathrm{\cahk}$ and $R_\mathrm{\cair}$ possibly show an
overall decrease with $\Delta m_{15}(B)$, though the range of observed
$R$ values definitely decreases at higher values of $\Delta
m_{15}(B)$. The overluminous and normal 
luminosity objects ($\Delta m_{15}(B) < 1.6$~mag) exhibit a wide range of
$R$ values, from identically 0 (i.e., no HVFs) to \about7. On the other
hand, the underluminous SNe~Ia ($\Delta m_{15}(B) > 1.6$~mag) almost
all have $R$ values that are 0, and the very few that are
nonzero are all less than 1. A Kolmogorov-Smirnov (KS) test indicates
that $R_\mathrm{\cahk}$ and $R_\mathrm{\cair}$ values for normal and
slow-declining objects are statistically different than those of the
fast-declining objects ($p = 0.007$ and $p = 10^{-5}$ for \cahk\ and
\cair, respectively).

These results still hold true even if the ``fast-declining cutoff'' is
more conservative ($\Delta m_{15}(B) = 1.4$~mag), with KS tests
indicating significant differences in $R_\mathrm{\cahk}$ and
$R_\mathrm{\cair}$ values above and below this cutoff ($p =
10^{-5}$ and $p = 5\times 10^{-8}$, respectively). This is consistent
with what was seen in Section~\ref{ss:ca} when SNID Type was used
instead of light-curve decline rate (i.e., Ia-91bg objects often show
fast-declining light curves). Furthermore, the results presented
here match those of \citet{Maguire12}, \citet{Childress14}, and
\citet{Maguire14}.

\begin{figure*}
\centering$
\begin{array}{c}
\includegraphics[width=5.5in]{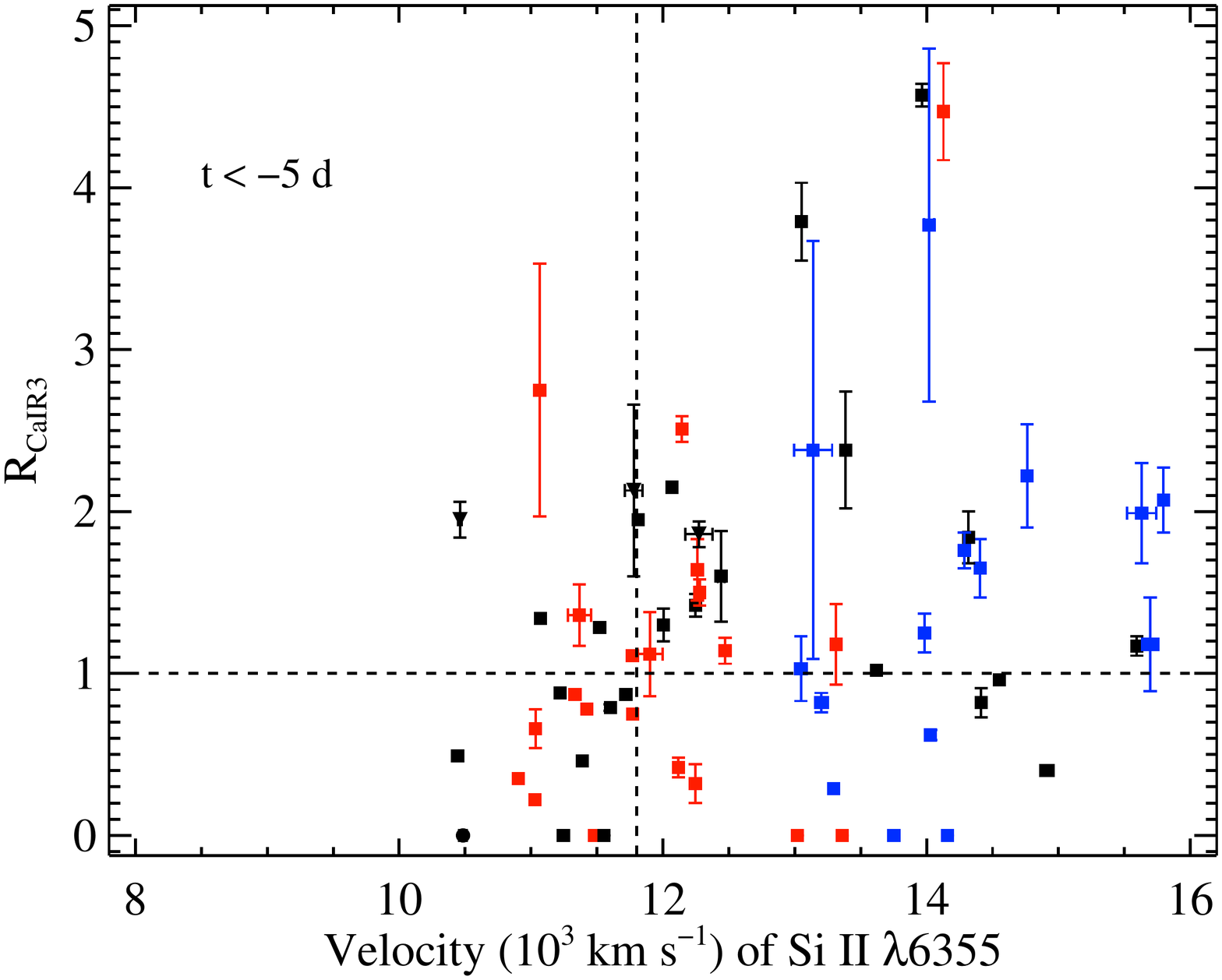} \\
\includegraphics[width=5.5in]{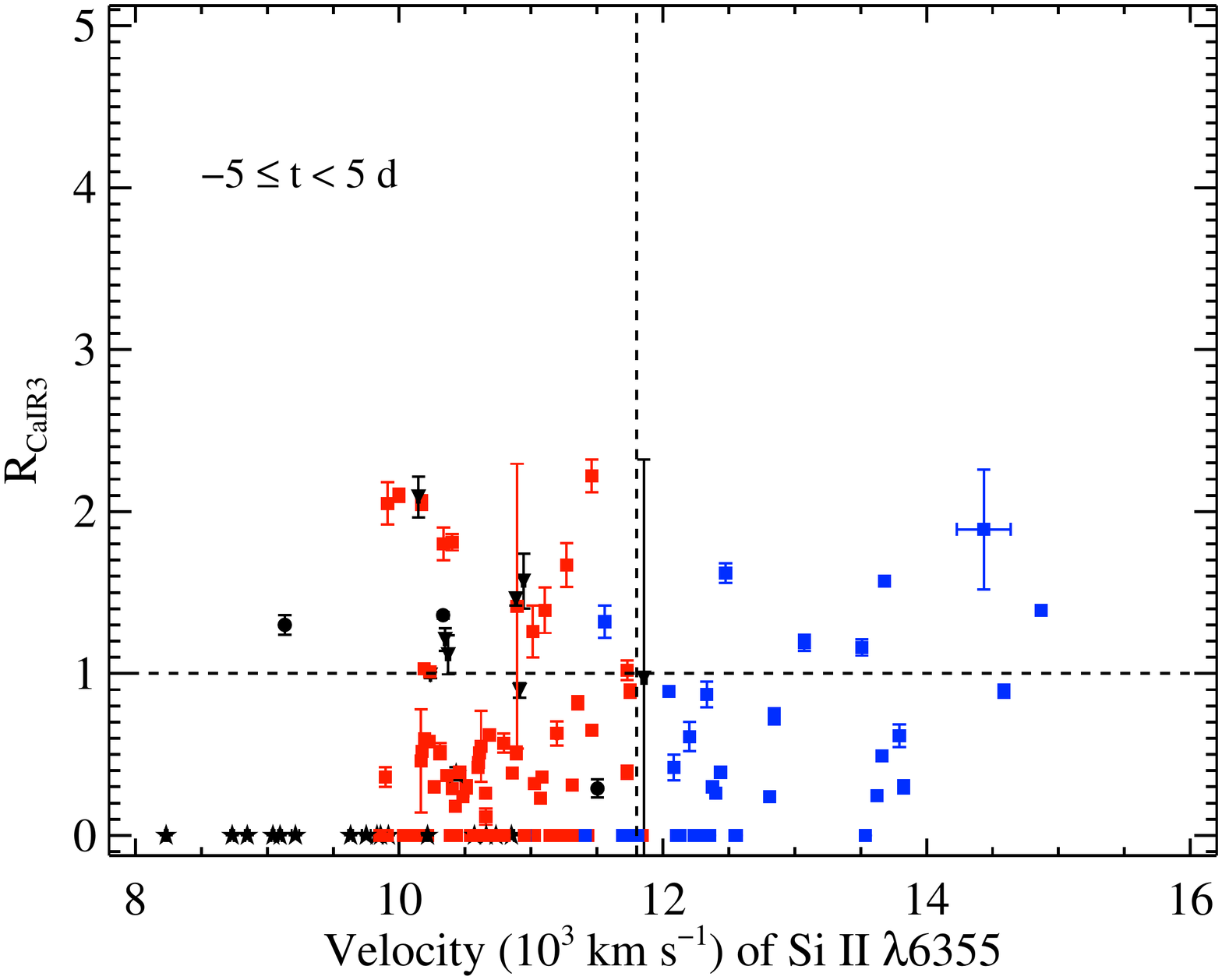} \\
\end{array}$
\caption{$R_\mathrm{\cair}$ versus \siii\ (PVF) velocity for spectra
  obtained earlier than 5~d before maximum brightness ({\it top}) and
  later than 5~d before maximum ({\it bottom}). The median $R$ value
  and velocity of a given SN~Ia are used for objects with multiple
  spectra in each epoch range. The dashed vertical line in each panel
  are the cutoffs between N and HV objects. The horizontal dashed line
  in each panel are where the pEWs of the HVFs and PVFs are
  equal. Colours and shapes of data points are the same as in
  Figure~\ref{f:ew_t_ca}.}\label{f:r_vsi_ca} 
\end{figure*}

Figure~\ref{f:r_vsi_ca} displays the PVF velocity of \siii\ versus
$R_\mathrm{\cair}$ for spectra  obtained earlier than 5~d before
maximum brightness (top) and later than 5~d before maximum (bottom); a
similar plot using $R_\mathrm{\cahk}$ is not shown but is
qualitatively similar, though with fewer data points. Once again, the
median values of both $R_\mathrm{\cair}$ and \siii\ velocity for a
given object are used for SNe~Ia with multiple spectra in each epoch
range. The dashed vertical line at 11,800~\kms\ in each panel
represents the cutoff between N and HV objects while the horizontal
dashed line at $R=1$ represents where the pEWs of the HVFs and the PVFs
are equal.

Aside from the Ia-91bg objects (i.e., the stars in
Figure~\ref{f:r_vsi_ca}), which we have shown almost never contain
HVFs, there is a large range of $R_\mathrm{\cair}$ values at all
\siii\ velocities in both age ranges. Thus, we find no correlation
between these two parameters and no significant difference in
$R_\mathrm{\cair}$ (or $R_\mathrm{\cahk}$) values 
for N versus HV objects. This is inconsistent with \citet{Childress14}
and \citet{Maguire14}, both of which find that HV objects do not show
HVFs. On the contrary, we find many HV objects with relatively strong
HVFs at all epochs, represented by blue points in the upper-right
quadrants of the panels in Figure~\ref{f:r_vsi_ca}. Compare this to
the bottom panel of Figure~5 in \citet{Childress14} and the left panel
of Figure~6 in \citet{Maguire14}, both of which lack objects in the
upper-right quadrant.

Our results are unchanged even if we restrict our sample only to
spectra within 5~d of maximum brightness (top panel of
Figure~\ref{f:r_vsi_ca}), in order to match the epochs studied in the
two aforementioned works. Thus, this discrepancy is 
likely caused by the fact that \citet{Childress14} and \citet{Maguire14}
have too few HV objects in their datasets. As mentioned in
Section~\ref{ss:ca}, \about28~per~cent of the objects studied herein are HV
objects, which matches the overall SN~Ia population
\citep[e.g.,][]{Wang09,Silverman12:BSNIPII}. On the other hand, only
\about13~per~cent of the objects studied by \citet{Childress14} and
\citet{Maguire14} were HV. This difference in sample demographics
likely led to the inconsistency discussed above.

Many other SN~Ia observables were compared to $R_\mathrm{\cahk}$ and
$R_\mathrm{\cair}$, but almost none showed any significant
correlation. For completeness, we list here the parameters
investigated. Some normal SNe~Ia are found to exhibit \ion{C}{II}
absorption features in their early-time spectra
\citep[e.g.,][]{Parrent11,Thomas11,Folatelli12,Silverman12:carbon}. This
C is likely unburned fuel from the progenitor WD. No difference in
$R_\mathrm{\cahk}$ ($R_\mathrm{\cair}$) is found for objects with or
without \ion{C}{II} absorption features when investigating 107 (252)
spectra, consistent with what was found by \citet{Maguire14}.

Narrow \ion{Na}{I}~D absorption features have been found to be
preferentially blueshifted (relative to the host galaxy's rest frame)
in SNe~Ia \citep[e.g.,][]{Sternberg11,Foley12:hires,Maguire13}. When
using 40 spectra that have the shift of these features measured, we
find no difference in $R$ values of objects with blueshifted versus
redshifted \ion{Na}{I}~D lines. The rise time of a SN~Ia light curve
is usually calculated by extrapolating early-time photometry backward
in time to a flux of zero. Using rise times of 74 objects published by
\citet{Ganeshalingam11}, we find no correlation with
$R_\mathrm{\cahk}$ or $R_\mathrm{\cair}$. 

Concentrating now on late-time spectra of SNe~Ia, we compare the
$R$ values calculated herein to nebular velocities --- that is, the average
of the [\ion{Fe}{II}] $\lambda$7155 and [\ion{Ni}{II}] $\lambda$7378
velocities \citep{Maeda10}. The nebular velocity has been
found to correlate with the velocity gradient near maximum
brightness, as well as the near-maximum ejecta velocity (i.e., PVF
velocity) \citep[e.g.,][]{Maeda10,Silverman13:late}, neither of which
were seen 
to correlate with the strengths of HVFs in the current work. Thus, it
is unsurprising that the nebular velocity is unrelated to
$R_\mathrm{\cahk}$ and $R_\mathrm{\cair}$ for 22 and 47 spectra,
respectively.

On the other hand, the full-width at half-maximum intensity
(FWHM) of the [\ion{Fe}{III}] $\lambda$4701 feature, which is detected
in many SN~Ia nebular spectra, is somewhat correlated with
$R_\mathrm{\cahk}$ and $R_\mathrm{\cair}$ (Pearson $r$ value of
\about0.6). This is consistent with previous work which found the FWHM
of this feature to be anticorrelated with $\Delta m_{15}(B)$
\citep[e.g.,][]{Silverman13:late}, and we have shown above that
$R_\mathrm{\cahk}$ and $R_\mathrm{\cair}$ are also anticorrelated
with $\Delta m_{15}(B)$.

For 242 spectra, we compared $R_\mathrm{\cahk}$ and $R_\mathrm{\cair}$
to host-galaxy type, as listed in NED. SNe~Ia in E/S0 hosts are found
to have significantly lower $R$ values than those found in
Sa/Sb/Sc/Sd/Irr hosts. We note that this result was also seen by
\citet{Pan15}. There is a well-established connection between
early-type hosts and underluminous SNe~Ia, and late-type hosts and
normal/overluminous SNe~Ia \citep[e.g.,][]{Hamuy95,Sullivan06,Pan14},
so this finding is completely consistent with the aforementioned
result that underluminous, Ia-91bg objects show relatively weak (or
nonexistent) HVFs.

We searched for correlations between the SN~Ia observables mentioned
above and the velocities of the \ion{Ca}{II} HVFs and PVFs, but found
no significant results. Our analysis indicates that
\ion{Ca}{II} velocities (both the HVFs and PVFs) are uncorrelated with
$\left(B-V\right)_0$, $\Delta m_{15}(B)$, the presence or absence of
\ion{C}{II} absorption, the relative Doppler shift of narrow
\ion{Na}{I}~D absorption, light-curve rise time, nebular velocity,
FWHM of [\ion{Fe}{III}] $\lambda$4701, and host-galaxy type.

\subsection{\ion{Si}{II} $\lambda$3858}\label{ss:si3858}

The \ion{Si}{II} $\lambda$3858 absorption feature has been discussed in 
multiple sections above, but here we summarise our results regarding it. 
\citet{Foley13} claim that \ion{Si}{II} $\lambda$3858 usually 
dominates the \cahk\ profile, but the findings of both 
\citet{Childress14} and \citet{Maguire14} are inconsistent with this
conclusion. We agree with the latter two works, as outlined below.
As discussed in Section~\ref{ss:ambig_hk}, the only extra assumption
we made in order to determine whether \ion{Si}{II} $\lambda$3858 
was present in a given spectrum was as follows: if a
spectrum showed a HVF of \cair, then it should also have a HVF of
\cahk, and vice versa. This broke the degeneracy between
\ion{Si}{II} $\lambda$3858 (PVF) absorption and \cahk\ HVF absorption.  

We tested this assumption by supposing \ion{Si}{II}
$\lambda$3858 was never present. This led to velocity
differences between HVFs of \cahk\ and \cair\ of \about5000~\kms\
(instead of the more typical value of \about500~\kms) and velocity
differences between \cahk\ HVFs and PVFs of
\about5500~\kms\ (as opposed to the median value of
\about9000~\kms). The assumption was further investigated by assuming
that \ion{Si}{II} $\lambda$3858 was {\it always} present (instead of
HVFs of \cahk). This again yielded inconsistent results, namely the
\ion{Si}{II} $\lambda$3858 velocities were \about2600~\kms\ faster
than the \siii\ velocities in a given spectrum \citep[compared to the
more typical value of \about600~\kms,
e.g.,][]{Silverman12:BSNIPII}. Note that this was previously shown
graphically in Figure~\ref{f:cahk_si_vels}. Thus, our assumption seems
to be valid.

\ion{Si}{II} $\lambda$3858 is detected in the \cahk\ profile in
\about24~per~cent (30/126) of the spectra fit, and 
half of these also show evidence for a HVF of \cahk. These
spectra represent \about27~per~cent (23/84) of the SNe~Ia in the current
dataset. When \ion{Si}{II} $\lambda$3858 is detected, its pEW is
\about70~\AA\ smaller than that of \siii\ in the same spectrum, which
represents a factor of \about6 (see Section~\ref{ss:ew_t_si}). This
difference in strength is larger than expected given typical SN~Ia
photospheric temperatures and the $gf$-weights of the two \ion{Si}{II}
lines.\footnote{{\tt\url{http://www.nist.gov/pml/data/asd.cfm}}.}
While weak absorption from \ion{Si}{II} $\lambda$3858 may actually be
present in a higher percentage of the data, our fitting algorithm is
unable to detect such a feature. Furthermore, no evidence of a
HVF of \ion{Si}{II} $\lambda$3858 is detected in any of the
observations. 

\begin{figure*}
\centering
\includegraphics[width=5.5in]{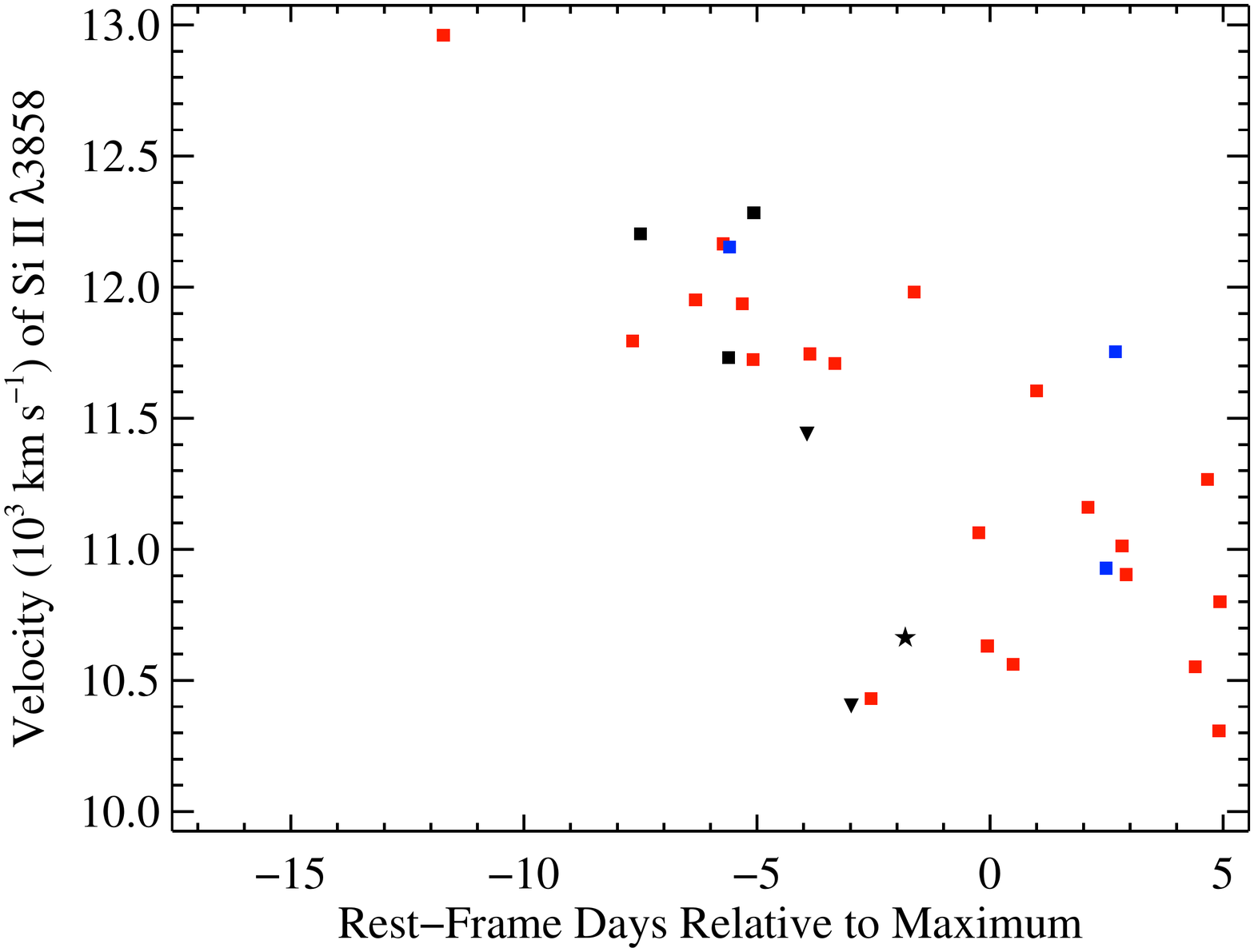}
\caption{The \ion{Si}{II} $\lambda$3858 velocity versus time. Colours
  and shapes of data points are the same as in
  Figure~\ref{f:ew_t_ca}. Measurement uncertainties are comparable to the size of
  the data points.}\label{f:v_t_si3858}
\end{figure*}

Figure~\ref{f:v_t_si3858} shows the temporal evolution of the velocity
of the \ion{Si}{II} $\lambda$3858 absorption feature (using the same
epoch range as previous temporal evolution figures in this work, and
adopting the same data-point colours and shapes as in
Figure~\ref{f:ew_t_ca}). The \ion{Si}{II} $\lambda$3858 feature is
detected mostly at epochs later than 8~d before maximum brightness
(there is one detection at $t \approx -12$~d) and, as expected, the
velocities tend to decrease with time. The detection of \ion{Si}{II}
$\lambda$3858, as well as the strength (i.e., pEW) and velocity of the
feature when detected, are uncorrelated with any of the aforementioned
SN~Ia classification schemes (i.e., SNID Type, Wang Type, and Benetti
Type), as well as any of the other SN~Ia observables mentioned in
Section~\ref{ss:comp_ca} (except, of course, the \siii\ PVF
velocity). Another possible exception is that \ion{Si}{II}
$\lambda$3858 might be detected more frequently in N objects, as
opposed to other subtypes, though the relatively small total number of
detections makes this result rather weak.

\subsection{The Existence of HVFs in  \siii}\label{ss:si}

Using the algorithm described in Section~\ref{s:procedure}, the pEWs 
and expansion velocities were calculated for both HVFs and PVFs of
\siii; the values are listed in Table~\ref{t:siii}. HVFs of \siii\
have rarely been carefully studied previously, with
\citet{Marion13:09ig} representing one of the most detailed works on
the subject. In the present study, \siii\ was fit in a total of 422
spectra of 208 SNe~Ia; 2 of these spectra have only HVFs, 326 have
only PVFs, and 94 have both HVFs and PVFs detected. SN~2012fr is the
only object in our sample with spectra showing only HVFs of \siii,
and this is found for only the two earliest spectra (14.4 and 14.0~d
before maximum brightness). As mentioned in Section~\ref{ss:ca}, this
is consistent with an in-depth study of this object which found HVFs
of \cair\ and \siii\ but no PVFs in these spectra
\citep{Childress13:12fr}. 

Like the \ion{Ca}{II} features, SNe~Ia tend to evolve from having only
\siii\ HVFs, though this is seen only in one object and at very early times,
to having both HVFs and PVFs, to having only PVFs, which is what is
seen in the majority of the spectra, \about77~per~cent. Spectra with only
\siii\ PVFs
are detected as early as 15~d before maximum and as late as 5~d
past maximum (i.e., the oldest spectra in the current study), while
spectra with both HVFs and PVFs are observed at all epochs studied
($-16 < t < 5$~d).

Unlike the \ion{Ca}{II} features, however, HVFs of \siii\ are somewhat
rare in SN~Ia spectra. Only \about16~per~cent of all SNe~Ia studied herein
show evidence for \siii\ HVFs in at least one spectrum (compared to
\about67~per~cent for HVFs of \cahk\ and \cair). In early-time observations
($t \la -5$~d), \about32~per~cent of objects have detectable HVFs of \siii,
much lower than the \about91~per~cent of objects that exhibit HVFs of
\ion{Ca}{II} at these early epochs.

All spectra obtained earlier than 11~d before maximum brightness
contain HVFs of \siii, while they are seen in 21 spectra of 7 SNe~Ia
for $t > -5$~d. This latter result is inconsistent with
\citet{Childress14}, who find no HVFs of \siii\ at these epochs. The
difference is likely caused by the larger dataset used herein, as well as
by differences in the spectral fitting algorithms used. The low detection
rate of HVFs of \siii\ could be explained by an inherent rarity of
HVFs of \siii, the possibility that they disappear at very early
times, or some combination of both.

As when fitting the \ion{Ca}{II} features, we find that \about29~per~cent of
the objects for which we fit the \siii\ feature are HV objects,
again consistent with the overall SN~Ia population
\citep[e.g.,][]{Wang09,Silverman12:BSNIPII}. Furthermore, only 8
objects had their Wang Type changed when using the PVF velocity
calculated in this work as compared to previous work
\citep[e.g.,][]{Silverman12:BSNIPII}, and all of these SNe~Ia had
velocities that were near the cutoff between HV and N objects. For the
SNe~Ia with a Wang Type, 33~per~cent of HV objects contain HVFs of
\siii, while only 6~per~cent of N objects show HVFs of \siii. This
dramatic difference has never been convincingly seen before, although
it was previously suggested by \citet{Tanaka08}, and is significantly
different than the HVFs of \ion{Ca}{II} which are found in similar
percentages of HV and N objects. This result may be surprising since
one might expect PVFs with higher velocities to be blended more
severely with any possible HVF, and thus one might be biased {\it
  against} finding distinct HVFs of \siii\ in HV objects.

Much like \ion{Ca}{II}, however, HVFs of \siii\ are found in similar
numbers of Ia-norm and Ia-91T/99aa objects (14~per~cent and 19~per~cent,
respectively) while no HVFs are seen in Ia-91bg objects. As mentioned
previously, the BSNIP dataset (which represents the bulk of the sample
studied herein) is not well suited to velocity-gradient measurements
or Benetti Type classifications. That being said, we report our results
here for completeness, though note the relatively low numbers of
objects involved. We find no HVFs of \siii\ in FAINT objects,
consistent with Ia-91bg objects having no HVFs. On the other hand,
29~per~cent of LVG objects and 27~per~cent of HVG objects show
evidence for HVFs of \siii. This is somewhat different than the
aforementioned predominance of \siii\ HVFs in HV objects and the
relative lack of them in N objects. However, given the relatively
small number of SNe~Ia for which we can measure a reliable Benetti
Type, these percentages are formally consistent with the results found
when using Wang Types.

\subsection{ \siii\ pEWs}\label{ss:ew_t_si}

\begin{figure*}
\centering
\includegraphics[width=5.5in]{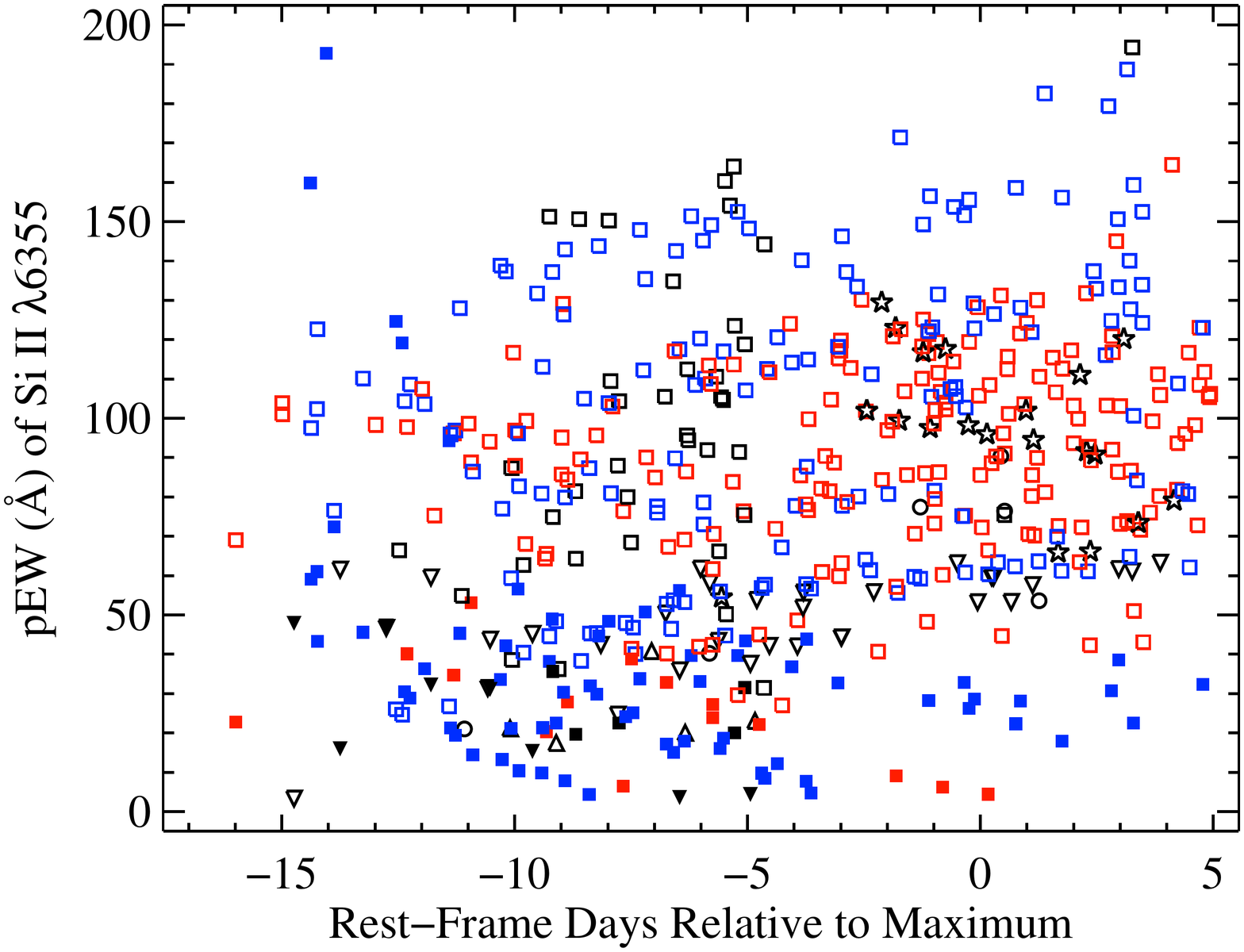}
\caption{The \siii\ pEWs versus time. Colours and shapes of data
  points are the same as in Figure~\ref{f:ew_t_ca}. Measurement uncertainties are
  comparable to the size of the data points.}\label{f:ew_t_si}
\end{figure*}

The pEWs of \siii\ are listed in Table~\ref{t:siii} and their temporal
evolution is displayed in Figure~\ref{f:ew_t_si}. Like the
\ion{Ca}{II} features, there is significant scatter in the
pEWs of the HVFs and PVFs of \siii. In addition, the HVF pEWs tend to
decrease with time while the PVF pEWs typically increase with time. As
discussed above, HVFs of \siii\ disappear at much earlier epochs than
either \ion{Ca}{II} feature. The strengths (i.e., pEWs) of HVFs and
PVFs are seen to be equal for \cahk\ and \cair\ for $-8 < t < 2$~d,
while this is achieved for \siii\ \about11~d before maximum
brightness. In fact, there are only 5 spectra in which the pEW of the
\siii\ HVF is larger than that of the \siii\ PVF, and all of these
were obtained at $t < -11$~d. In spite of this, moderately strong HVFs
of \siii\ are observed in a handful of objects through $t \approx -6$~d
and in some objects through $t \approx 5$~d, and these are almost
exclusively HV objects.

As mentioned in Section~\ref{ss:si}, HV objects tend to show HVFs of
\siii\ significantly more often than N objects, and, according to
Figure~\ref{f:ew_t_si}, on average they exhibit stronger HVFs than
N objects when detected in both subtypes at the same epoch. Similar to
the \ion{Ca}{II} pEWs, though not quite as extreme, Ia-91bg objects
have relatively large pEWs of \siii\ PVFs but no HVFs, and
Ia-91T/99aa objects have relatively low pEWs of both \siii\ PVFs and
HVFs.

Analogous to the \ion{Ca}{II} features, we define $R_\mathrm{Si}$ as
the ratio of pEWs of the HVFs to the PVFs of \siii.\footnote{Note that
  $R_\mathrm{Si}$ defined here is unrelated to the so-called
  ``\ion{Si}{II} ratio,'' $\Re$(\ion{Si}{II}), which was defined by
  \citet{Nugent95} as the ratio of the depth of the \ion{Si}{II}
  $\lambda$5972 feature to the depth of the \siii\ feature.} Once
again, spectra with only PVFs are defined to have a ratio of zero,
while spectra 
with only HVFs have an undefined ratio; the values of $R_\mathrm{Si}$
are listed in Table~\ref{t:siii}. For nearly all spectra studied
herein, $R_\mathrm{Si} < 1$. There are five spectra whose ratio is larger
than unity, and they represent the earliest spectral observations of
SNe~2012cg and 2012fr
\citep[][respectively]{Silverman12:12cg,Childress13:12fr}. Note that
SN~2012fr was also the sole object in this work found to have spectra
showing HVFs of \siii\ {\it without} PVFs.

\subsection{ \siii\ Velocities}\label{ss:v_t_si}

\begin{figure*}
\centering
\includegraphics[width=5.5in]{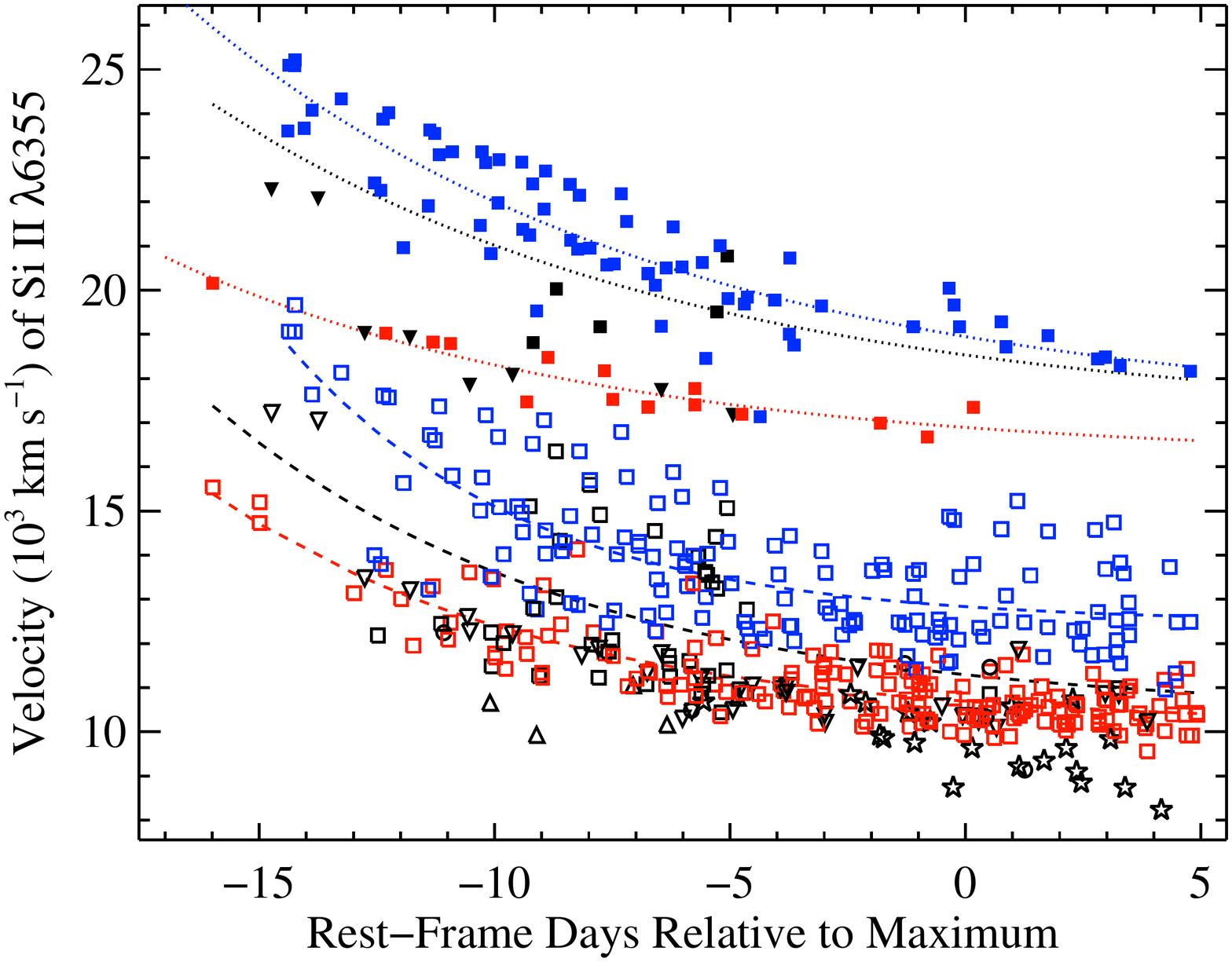}
\caption{The \siii\ velocities versus time. Colours and shapes of data
  points are the same as in Figure~\ref{f:ew_t_ca}. Measurement uncertainties are
  comparable to the size of the data points. Like
  Figure~\ref{f:v_t_ca}, the black, blue, and red dashed lines are
  natural exponential function fits to PVF velocities of all
  objects, HV objects only, and N objects only, respectively. The
  black, blue, and red dotted lines are natural exponential function
  fits to HVF velocities of all objects, HV objects only, and N
  objects only, respectively.}\label{f:v_t_si}
\end{figure*}

The measured PVF and HVF velocities of the \siii\ feature are listed in
Table~\ref{t:siii}, and Figure~\ref{f:v_t_si} shows the temporal
evolution of these velocities. As in Figure~\ref{f:v_t_ca}, the black
dashed line is the best-fitting natural exponential function to all
\siii\ PVF velocities, while the blue and red dashed lines use only HV
and N objects, respectively. The black dotted line is the best-fitting
natural exponential function to all \siii\ HVF velocities, and the
blue and red dotted lines use only HV and N objects, respectively. 

Like the \ion{Ca}{II} features, the measured velocities (both PVFs and
HVFs) of a given object tend to decrease with time. The exponential 
fits in Figure~\ref{f:v_t_si} show that, again as seen in the
\ion{Ca}{II} features, the HVFs (dotted lines) and PVFs (dashed lines)
of HV objects start out with higher velocities than the N objects and
likely decrease their velocity more quickly with time. For \siii,
however, the HVF and PVF velocities of the HV objects are almost
always larger than those of the N objects at a given epoch. This is by
construction, at least for the PVF velocities, since the PVF \siii\
velocity is how a Wang Type is assigned to a given object.

Ia-91bg objects are found to have the lowest \siii\ velocities, with
Ia-91T/99aa objects having slightly larger velocities. N objects have
slightly larger velocities than that and, of course, HV objects exhibit
the highest velocities. This is consistent with previous \siii\
velocity studies \citep[e.g.,][]{Silverman12:BSNIPII}. 

\begin{figure*}
\centering
\includegraphics[width=5.5in]{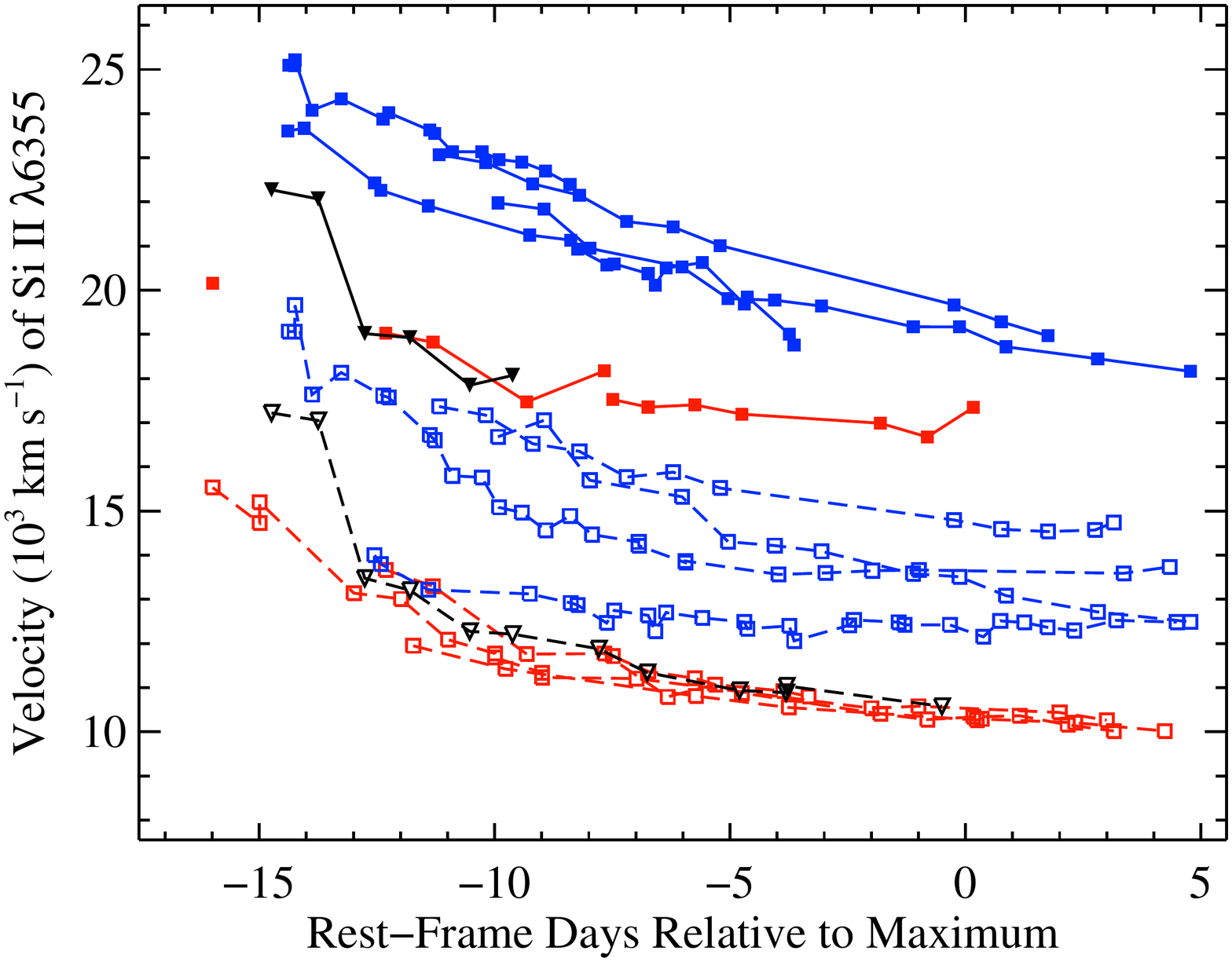}
\caption{The \siii\ velocities versus time for the nine SNe~Ia for which
  we have more than seven spectra (see main text for the list of
  objects). All PVF (HVF) velocities of a given object are connected
  with a dashed (solid) line. Colours and shapes of data points are
  the same as in Figure~\ref{f:ew_t_ca}. Measurement uncertainties are comparable
  to the size of the data points.}\label{f:v_t_si_many}
\end{figure*}

Only \siii\ velocities of objects for which we have more than seven
spectra are shown in Figure~\ref{f:v_t_si_many} (thus, this is a
subset of what is displayed in Figure~\ref{f:v_t_si}). As in
Figure~\ref{f:v_t_ca_many}, the PVF (HVF) velocities of a given object
are connected with a dashed (solid) line. The nine objects plotted in
Figure~\ref{f:v_t_si_many} include the five extremely well-observed
SNe~Ia mentioned above (SNe~2009ig, 2011by, 2011fe, 2012cg --- the lone 
Ia-91T/99aa object in the Figure --- and 2012fr), in addition to
SNe~1994D, 2006X, 2010kg, and 2011ao. The aforementioned conclusions
once again hold for this subset: HV objects have faster
HVFs and PVFs at all epochs, and they likely decrease more
quickly with time than N objects.

As with the \ion{Ca}{II} features, the \siii\ feature shows a distinct
gap between HVF and PVF velocities. This gap is most noticeable for $t
\ga -6$~d. Once again, the difference in velocity between the HVFs and
PVFs in a given spectrum is calculated. While this difference tends to
decrease with time, there is a very large amount of scatter at all
epochs, similar to the \cair\ feature. The typical velocity separation
for \siii\ is \about6000~\kms, slightly less than what was found for
the \ion{Ca}{II} features (i.e., \about9000~\kms). No velocity
differences are found less than 4200~\kms, and most are greater than
5000~\kms. These 
values are somewhat larger than the minimum separation resolvable by
our fitting algorithm (see Section~\ref{sss:synthetic}), so as was 
the case for \cair, the gap between the \siii\ HVFs and PVFs is likely
real. 

As was done for the \ion{Ca}{II} features, the velocity separations
for a given object at all epochs were investigated for 
\siii. Again, each object's maximum PVF velocity, usually from the
earliest spectrum of the object in question, was compared to its
minimum HVF velocity, usually from the latest spectrum of the object
in question. All objects studied herein have all of their HVF
velocities larger than all of their PVF velocities (i.e., the minimum
HVF velocity is larger than the maximum PVF velocity). 

\subsection{HVFs of \siii\ Compared to Other Observables}\label{ss:comp_si}

As with the \ion{Ca}{II} features, we compare the $R_\mathrm{Si}$
values of the \siii\ feature to other SN~Ia 
observables. The SN colour is being characterised in this work using
the observed $B-V$ colour at $B$-band maximum brightness, with
only a correction for MW reddening \citep{Schlegel98} having been
applied. Thus, there 
are many objects in the sample that have large (i.e., red) values
of $\left(B-V\right)_0$ caused by reddening from their host galaxy. When
ignoring highly reddened objects \citep[$\left(B-V\right)_0 >
0.4$~mag; e.g.,][]{Foley11:velb}, the mean 
$R_\mathrm{Si}$ value is more than twice as large for objects with
$\left(B-V\right)_0 > 0$~mag as compared to objects with
$\left(B-V\right)_0 < 0$~mag (\about0.037 and \about0.016~mag,
respectively). Both of these values are quite close to $R_\mathrm{Si}
= 0$, since there are relatively few objects that show HVFs of \siii;
thus, this result is of somewhat low significance. If correct, however,
it indicates that $R_\mathrm{Si}$ is larger for intrinsically redder
SNe~Ia, which are also found to show larger \siii\ velocities
\citep[e.g.,][]{Foley11:velb,Milne14}.

Like the \ion{Ca}{II} features, no HVFs of \siii\ are found in
fast-declining SNe~Ia, while normal- and slow-declining objects show a
range of $R_\mathrm{Si}$ values (from identically 0 to \about0.5). In
fact, no object with $\Delta m_{15}(B) > 1.45$~mag show HVFs of
\siii, and the difference in $R_\mathrm{Si}$ values above and below
this cutoff is statistically significant. These findings are
consistent with the results discussed in Section~\ref{ss:si} where
SNID Type was used instead of light-curve decline rate.

\begin{figure*}
\centering
\includegraphics[width=5.5in]{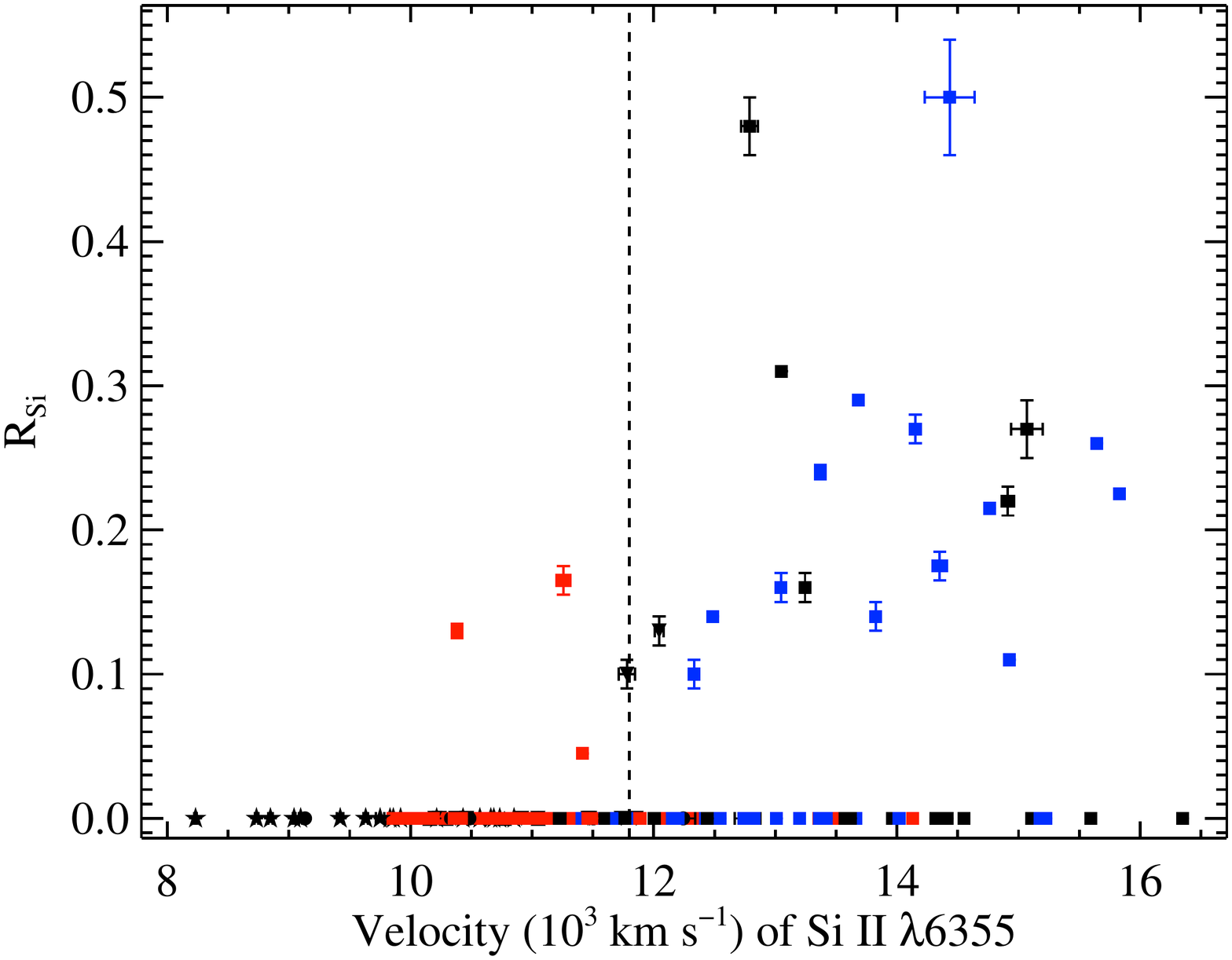}
\caption{$R_\mathrm{Si}$ versus \siii\ (PVF) velocity. The median
  $R$ value and velocity are used for objects with multiple
  spectra. The dashed line is the cutoff between N and HV
  objects. Colours and shapes of data points are the same
  as in Figure~\ref{f:ew_t_ca}.}\label{f:r_vsi_si}
\end{figure*}

As also seen in Section~\ref{ss:si}, the majority of the \siii\ HVFs are
found in HV objects. Another way to present this result is shown in
Figure~\ref{f:r_vsi_si}, which displays the (PVF) velocity of \siii\
versus $R_\mathrm{Si}$. The median values (of both $R_\mathrm{Si}$ and
\siii\ velocity) are used for objects with multiple spectra. The
dashed line at 11,800~\kms\ represents the cutoff between N and
HV objects.

The typical value of $R_\mathrm{Si}$ that we measure increases with
\siii\ velocity, although 
there are only a handful of objects with $R_\mathrm{Si} \ne 0$. Of
these SNe~Ia that show HVFs of \siii, three are N objects, while two
are Ia-91T/99aa and five are unclassified objects. On the other hand,
13 are HV objects, while HV objects make up less than
30~per~cent of the current dataset. Also note that the five unclassified
objects with HVFs of \siii\ are all found to have velocities greater
than 11,800~\kms, and are thus {\it likely} HV objects. About half of
the highest-velocity objects (i.e., \siii\ velocity greater than 14,000~\kms)
show HVFs of \siii; compare this to the entire dataset, in which only
\about16~per~cent of objects show HVFs of \siii.

As mentioned in Section~\ref{ss:comp_ca}, \ion{C}{II} absorption
features are sometimes found in early-time optical spectra of SNe~Ia,
likely coming from unburned progenitor WD material
\citep[e.g.,][]{Parrent11,Thomas11,Folatelli12,Silverman12:carbon}. For
288 spectra, we find a statistically significant difference in values of
$R_\mathrm{Si}$ for objects with versus without \ion{C}{II} absorption
features. The mean $R_\mathrm{Si}$ value for SNe~Ia with detected
\ion{C}{II} is \about0.011, while the mean $R_\mathrm{Si}$ value for
those without \ion{C}{II} is nearly twice as large (\about0.019). This
implies that objects lacking \ion{C}{II} absorption features also show
stronger HVFs. Consistent with this finding, and previous results in
the current work, HV objects have also been shown to lack \ion{C}{II}
absorption \citep[e.g.,][]{Silverman12:carbon}.

Other SN~Ia observables were also compared to $R_\mathrm{Si}$, but
none of them showed any significant correlation. As was done for the
\ion{Ca}{II} feature, we list the observables investigated for
completeness. No difference in $R_\mathrm{Si}$ values is found for objects with
blueshifted versus redshifted narrow \ion{Na}{I}~D absorption lines
when using 47 spectra. Similarly, no correlation is found between
$R_\mathrm{Si}$ and the light-curve rise times of 43 of the objects
published by \citet{Ganeshalingam11}, the nebular velocity (of 17
objects), or the FWHM of the [\ion{Fe}{III}] $\lambda$4701 feature (of
17 objects). Finally, the host-galaxy type (as reported in NED) is
related to the value of $R_\mathrm{Si}$ for 287 spectra in the same
way as $R_\mathrm{\cahk}$ and $R_\mathrm{\cair}$; objects in E/S0
hosts tend to have lower $R$ values than those found in other galaxy
types.

In summary, \siii\ HVFs are never found in slow-declining (i.e.,
Ia-91bg/FAINT) objects, as is the case with HVFs of \cahk\ and \cair. Unlike
the HVFs of \ion{Ca}{II}, however, HVFs of \siii\ are relatively rare
overall, yet there are significantly more HV objects with \siii\ HVFs
as compared to any other SN~Ia subtype. Furthermore, we find that
objects showing strong HVFs of \siii\ also tend to have redder optical
colours at maximum brightness and lack \ion{C}{II} absorption in their
early-time spectra. This connection between photospheric velocity,
ultraviolet (UV)/optical colour near maximum, and \ion{C}{II} absorption have 
all been recognised in previous works mentioned above, but the addition
of the association between HV objects and relatively strong HVFs of 
\siii\ is new and unique to the current work. We also note the
possibility that HV objects preferentially occur in the inner regions
of their host galaxies \citep{Wang13,Pan15}.

As discussed above and in \citet{Wang09}, the Wang Type
classifications, and thus the correlations in the previous paragraph, 
only apply to ``typical'' SNe~Ia. These objects make up \about70~per~cent of
the SN~Ia population \citep{Li11a}, are spectroscopically ``normal''
(e.g., according to SNID), and usually have $\Delta m_{15}(B) = 1.1
\pm 0.3$~mag \citep[e.g.,][]{Ganeshalingam10:phot_paper}. For such
objects, there seems to be an observational dichotomy, or, more likely,
a continuous distribution of multiple observables that are mutually 
correlated. Table~\ref{t:sneia} presents this dichotomy, or rather,
the extremes of this continuous distribution. In essence, we find that
a ``typical'' SN~Ia with a relatively large near-maximum photospheric
velocity, which would lead to classification as a HV object, will
likely lack early-time \ion{C}{II} absorption, tend to have relatively
red UV/optical colours near maximum brightness, will show relatively
strong HVFs of \siii, and may be found in the inner regions of its
host galaxy.

\setlength{\tabcolsep}{8pt}
\begin{table*}
\begin{center}
\caption{Correlated Observables of ``Typical''$^{\rm a}$ SNe~Ia}\label{t:sneia}
\begin{tabular}{lccl}
\hline\hline
Observable & HV objects & N objects & Example Reference(s) \\
\hline
\siii\ (PVF) velocity near maximum & $>$\,11,800~\kms & $<$\,11,800~\kms & \citet{Wang09} \\
Early-time \ion{C}{II} absorption features & No & Yes & \citet{Silverman12:carbon} \\
UV/optical colours near maximum & Red & Blue & \citet{Foley11:velb,Milne14} \\
HVFs of \siii\ & Strong & Weak/None & {\bf This Work} \\
Location within host galaxy & Inner \about70~per~cent? & Everywhere? & \citet{Wang13,Pan15} \\
\hline\hline
\multicolumn{4}{p{6in}}{$^{\rm a}$``Typical'' SNe~Ia are objects
  that are spectroscopically ``normal'' (e.g., according to SNID) and
  usually have $\Delta m_{15}(B) = 1.1 \pm 0.3$~mag
  \citep[e.g.,][]{Ganeshalingam10:phot_paper}; thus, they can be
  assigned a Wang Type.} \\
\hline\hline
\end{tabular}
\end{center}
\end{table*}
\setlength{\tabcolsep}{6pt}


\section{Conclusions}\label{s:conclusions}

We have conducted the most detailed study of HVFs performed to date, 
using a sample of 445 low-resolution optical and NIR spectra 
at epochs up to 5~d past maximum brightness of 210 low-redshift SNe~Ia
that follow the ``Phillips relation.'' By fitting a series of Gaussian
functions, we are able to determine whether a given spectrum
shows evidence for HVFs of \cahk, \siii, or \cair, as well as measure
the velocities and pEWs of the PVFs and HVFs of these three spectral
features. Our measured values are consistent with previous studies of
HVFs \citep[e.g.,][]{Marion13:09ig,Childress14}, and our detection, or
lack thereof, of HVFs also matches spectral fits produced via {\tt
  SYNAPPS}.

In general, SNe~Ia are found to have HVFs with no corresponding PVFs
at the earliest epochs and these features weaken and slow down with
time. PVFs appear later and grow stronger with time, while also
slowing down. HVFs and PVFs of \cahk, \siii, and \cair\ are found (in
at least some objects) at nearly all epochs studied herein. SNe~Ia
with faster PVFs tend to have faster HVFs.

About two-thirds of all objects in the current sample show HVFs of 
\ion{Ca}{II}. For objects with spectra obtained earlier than \about4~d
before maximum brightness, \about91~per~cent show HVFs, and the
remaining \about9~per~cent all seem to be underluminous/Ia-91bg/FAINT
objects. This connection between the relative strength of \ion{Ca}{II}
HVFs and luminosity has also been seen in previous work
\citep{Maguire12,Childress14,Maguire14}. Our analysis further
indicates that \ion{Si}{II} $\lambda$3858 is detectable in the \cahk\
profile of \about24~per~cent of spectra, implying that it does not
usually dominate the spectral profile in this wavelength range. 

We also investigate HVFs of \siii, a
relatively unexplored area of research, but see \citet{Marion13:09ig}
for one of the best previous studies of \siii\ HVFs. As with the
\ion{Ca}{II} features, no HVFs of \siii\ are found in
underluminous/Ia-91bg/FAINT objects. On the other hand, \siii\ HVFs
are much rarer than their \ion{Ca}{II} counterparts, and are detected in
only \about16~per~cent of the objects in the current sample. Even at
early times ($t < -5$~d), HVFs of \siii\ are seen in only \about32~per~cent
of SNe~Ia.

Despite their rarity, \siii\ HVFs are observed about one-third of the time
in HV objects, compared to only 5--10~per~cent of the time in all
other SN~Ia subtypes. We also find that
stronger HVFs of \siii\ are associated with a lack of \ion{C}{II}
absorption at early times and relatively red UV/optical colours near
maximum brightness. These new-found connections, in conjunction with
previous work, led to Table~\ref{t:sneia}, which presents a list of
correlated parameters that likely constitute a continuous distribution
of SN~Ia observables.

Future SN~Ia models should utilise the empirical relations
and observational constraints set forth in this and previous work
regarding HVFs. For example, if HVFs arise purely from an opacity
effect, then a stronger line (i.e., one with a larger pEW) would form
at a larger radius in the SN photosphere. Assuming homologous
expansion, this would correspond to a larger observed velocity for the
HVFs of a stronger line.

Using the measurements of HVFs of \siii\ and \cair\ discussed above,
94~per~cent of all spectra have $R_\mathrm{Si} < R_\mathrm{\cair}$, while the
remaining \about6~per~cent are consistent with equality. Thus, all HVFs of
\cair\ are consistent with being stronger, relative to their PVFs, 
than those of \siii. In \about80~per~cent of observations, HVFs of
\cair\ have larger velocities than HVFs of \siii, with
\about5~per~cent consistent with equality. The remaining
\about15~per~cent of spectra, where \siii\ HVFs are significantly
faster than \cair\ HVFs, include SNe~Ia with some of the fastest
\siii\ velocities ever observed (e.g., SNe~2006X and 2010kg;
\citealt{Wang08}; Silverman et~al., in preparation,
respectively). Thus, \cair\ is stronger and also faster than \siii,
except in these extremely high-velocity objects, and so the HVFs of
\siii\ and \cair\ {\it could} be caused primarily by opacity effects
in most SNe~Ia.

\begin{figure}
\centering
\includegraphics[width=3.5in]{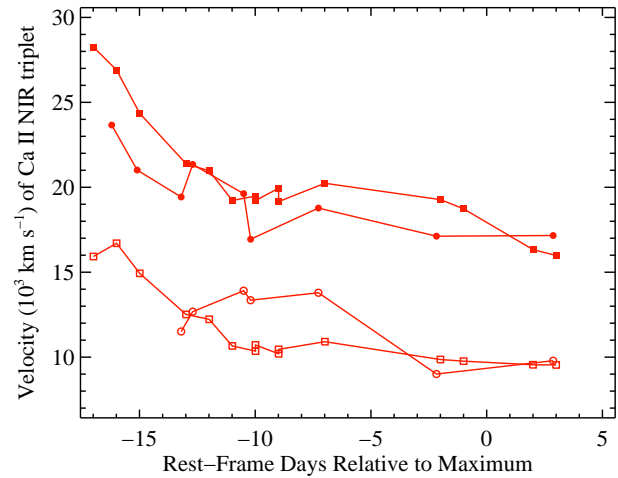}
\caption{The measured \cair\ velocities of SN~2011fe from the current
  work (squares) and model \cair\ velocities from Mulligan \& Wheeler
  (2015, in preparation; circles). Open points are PVFs and filled
  points are HVFs.}\label{f:model_vels} 
\end{figure}

While opacity may play a role, an abundance or density enhancement or
an ionisation change at high velocity (i.e., large radius) is likely
required to produce detectable HVFs. \citet{Gerardy04} showed that a
model in which SN ejecta collide with a circumstellar shell can yield
observed velocities of the \cair\ HVF feature. Mulligan \& Wheeler
(2015, in preparation) show that the evolution in time of the PVF and
HVF profiles, and the nearly constant velocity gap between the two sets
of features, can be reproduced by a model of the interaction between
SN~Ia ejecta and a circumstellar shell of small mass contained within
a few tenths of a solar radius near the exploding
WD. Figure~\ref{f:model_vels} shows the evolution of the \cair\ PVF
and HVF velocities as open and filled circles, respectively, for a
shell with mass 0.005~M$_\odot$. Also shown in the figure (as open and
filled squares, respectively) are the PVF and HVF velocities measured
herein for SN~2011fe.

The model PVF velocities match quite well the measured values from
the current work at early times and near maximum brightness, though in
between these times they are slightly higher than the data indicate. The
model HVF velocities match fairly well at most epochs, but are
sometimes slightly slower than the measured values. Prior to about
13~d before maximum brightness the photosphere is at the contact
discontinuity, below the shell, and thus only the HVF is seen in the
model. At later times, the photosphere moves deeper into the ejecta,
allowing the PVF to be observed, but the CSM shell still has enough
optical depth to yield HVFs. The evolution of the features in the
model is caused by the receding velocity of the photosphere and the
free expansion of the higher-velocity shell material that leads to
weaker absorption in the HVFs (Mulligan \& Wheeler 2015, in
preparation). 

The current work now stands as an observational benchmark against
which theoretical models of SNe~Ia can be compared. Any successful
model of a normal or overluminous SN~Ia that follows the``Phillips
relation'' must naturally produce HVFs of \ion{Ca}{II} since
they are found so ubiquitously and at a range of pre- and 
near-maximum-brightness epochs. On the other hand, models of 
underluminous/Ia-91bg
objects should never produce \ion{Ca}{II} HVFs. Furthermore, based on
spectropolarimetric observations, the HVFs of \ion{Ca}{II} should also
show polarisation, distinct from any polarisation present in the PVFs
\citep[e.g.,][]{Leonard05,Wang03,Wang06,Chornock08,Patat09,Maund13}. Similarly,
HVFs of \siii\ should be occasionally produced in N and
overluminous/Ia-91T/99aa objects, never produced in
underluminous/Ia-91bg objects, and preferentially more often (but
still somewhat rarely) produced in HV objects. It is now clear that the
detection and characterisation of HVFs is yet one more piece of the
SN~Ia puzzle.

\section*{Acknowledgments}

We thank the referee, M.~J.~Childress, for useful comments and
stimulating discussion that helped improve this paper, R.~J.~Foley,
K.~Maguire, A.~A.~Miller, and 
L.~Wang for useful discussions, and the staffs at the Lick, Keck, and
McDonald Observatories for their assistance with the observations. 
The HET is a joint project of the University of Texas at Austin, the
Pennsylvania State University, Stanford University,
Ludwig-Maximilians-Universit\"{a}t M\"{u}nchen, and
Georg-August-Universit\"{a}t G\"{o}ttingen. The HET is named in honor
of its principal benefactors, William P. Hobby and Robert
E. Eberly. The Marcario Low Resolution Spectrograph is named for Mike
Marcario of High Lonesome Optics who fabricated several optics for the
instrument but died before its completion. The LRS is a joint project
of the HET partnership and the Instituto de Astronom\'{i}a de la
Universidad Nacional Aut\'{o}noma de M\'{e}xico. 
Some of the data presented herein were obtained at the W. M. Keck
Observatory, which is operated as a scientific partnership among the
California Institute of Technology, the University of California, and
NASA; the
observatory was made possible by the generous financial support of the
W. M. Keck Foundation. The authors wish to recognise and acknowledge
the very significant cultural role and reverence that the summit of
Mauna Kea has always had within the indigenous Hawaiian community; we
are most fortunate to have the opportunity to conduct observations
from this mountain.
This research has made use of the NASA/IPAC Extragalactic Database
(NED) which is operated by the Jet Propulsion Laboratory, California
Institute of Technology, under contract with NASA.
J.M.S. is supported by an NSF Astronomy and Astrophysics Postdoctoral
Fellowship under award AST-1302771.
J.V. and T.S. are supported by Hungarian OTKA Grants NN 107637 and PD
112325, respectively. 
J.C.W.'s supernova group at UT Austin is supported by NSF Grant AST
11-09801.
A.V.F. is grateful for support from
the Christopher R. Redlich Fund,
the TABASGO Foundation, and NSF grant AST-1211916.
Some work on this paper was done in the hospitable climate of the Aspen
Center for Physics that is supported by NSF Grant PHY-1066293.

\bibliographystyle{mn2e}
\bibliography{/Users/jsilv/astro_refs}

\appendix

\section{Tables of Objects and Spectral Measurements}\label{a:tables}
Table~\ref{t:objects} lists each SN~Ia that we analyse herein and the
(rest-frame) phases of their spectra. Also presented are published
$\Delta m_{15}(B)$ and $\left(B-V\right)_0$ values, as well as
spectral classifications based on various classification schemes.

\onecolumn
\small
\begin{landscape}
\begin{center}
\begin{longtable}{lccccccc}
\caption{Summary of SNe~Ia\label{t:objects}} \\[-2ex]
\hline \hline
SN Name & Phase(s) & $\Delta m_{15}(B)$ & $\left(B-V\right)_0$ & SNID & Benetti & Wang \\
   & (d)$^\textrm{a}$ & (mag)$^\textrm{b}$ & (mag)$^\textrm{c}$ & (Sub)Type$^\textrm{d}$ & Type$^\textrm{e}$ & Type$^\textrm{f}$ \\
\hline
\endfirsthead

\multicolumn{7}{c}{{\tablename} \thetable{} --- Continued} \\
\hline \hline
SN Name & Phase(s) & $\Delta m_{15}(B)$ & $\left(B-V\right)_0$ & SNID & Benetti & Wang \\
   & (d)$^\textrm{a}$ & (mag)$^\textrm{b}$ & (mag)$^\textrm{c}$ & (Sub)Type$^\textrm{d}$ & Type$^\textrm{e}$ & Type$^\textrm{f}$ \\
\hline
\endhead

\hline \hline
\multicolumn{7}{l}{Continued on next page} \\
\endfoot

\hline \hline
\endlastfoot

SN 1989M & $2.5$,$3.5$ & 1.10 (20) & $\cdots$ & Ia-norm & HVG & HV \\
SN 1991T & $-10.1$,$-9.1$ & 0.92 (03) & $\phantom{-}0.090$ (036) & Ia-91T & $\cdots$ & $\cdots$ \\
SN 1991bg & $0.1$,$1.1$ & 1.89 (05) & $\phantom{-}0.714$ (044) & Ia-91bg & FAINT & $\cdots$ \\
SN 1994D & $-12.3$,$-11.3$,$-9.3$,$-7.7$,$-6.3$,$-5.3$,$-3.9$,$-3.3$ & 1.36 (05) & $-0.053$ (035) & Ia-norm & LVG & N \\
SN 1994S & $1.1$ & 0.98 (04) & $-0.090$ (038) & Ia-norm & $\cdots$ & N \\
SN 1995D & $3.8$ & 1.03 (07) & $-0.121$ (033) & Ia-norm & $\cdots$ & N \\
SN 1995E & $-2.5$ & 0.98 (06) & $\phantom{-}0.668$ (040) & Ia-norm & $\cdots$ & N \\
SN 1995ac & $-6.3$ & 0.93 (04) & $-0.124$ (036) & Ia-91T & $\cdots$ & $\cdots$ \\
\end{longtable}
\end{center}
\end{landscape}
\normalsize
\twocolumn

Tables~\ref{t:cahk}, \ref{t:siii}, and \ref{t:cair} list measured
values of the \ion{Ca}{II}~H\&K (\cahk), \siii, and \ion{Ca}{II}
near-IR triplet (\cair) features, respectively. The velocity for each
component is displayed, as well as the pseudo-equivalent width
\citep[pEW; e.g.,][]{Garavini07,Silverman12:BSNIPII}. Also shown for
each feature is $R$, the ratio of the pEW of the HVF to the pEW of the
PVF, as defined by \citet{Childress14}. Fits with no HVF have $R
\equiv 0$ and fits with no PVF have undefined values of $R$. 

\onecolumn
\small
\begin{landscape}
\begin{center}
\begin{longtable}{lrrrrrrrr}
\caption{\cahk\ Fit Results\label{t:cahk}} \\[-2ex]
\hline \hline
SN Name & \multicolumn{1}{c}{Phase (d)$^\textrm{a}$} & \multicolumn{1}{c}{HVF $v$ ($10^{3}$ \kms)} & \multicolumn{1}{c}{HVF pEW (\AA)} & \multicolumn{1}{c}{PVF $v$ ($10^{3}$ \kms)} & \multicolumn{1}{c}{PVF pEW (\AA)} & \multicolumn{1}{c}{$R_\mathrm{\cahk}^\textrm{b}$} & \multicolumn{1}{c}{\ion{Si}{II} $\lambda$3858 $v$ ($10^{3}$ \kms)} & \multicolumn{1}{c}{ \ion{Si}{II}$\lambda$3858 pEW (\AA)} \\
\hline
\endfirsthead

\multicolumn{9}{c}{{\tablename} \thetable{} --- Continued} \\
\hline \hline
SN Name & \multicolumn{1}{c}{Phase (d)$^\textrm{a}$} & \multicolumn{1}{c}{HVF $v$ ($10^{3}$ \kms)} & \multicolumn{1}{c}{HVF pEW (\AA)} & \multicolumn{1}{c}{PVF $v$ ($10^{3}$ \kms)} & \multicolumn{1}{c}{PVF pEW (\AA)} & \multicolumn{1}{c}{$R_\mathrm{\cahk}^\textrm{b}$} & \multicolumn{1}{c}{\ion{Si}{II} $\lambda$3858 $v$ ($10^{3}$ \kms)} & \multicolumn{1}{c}{\ion{Si}{II} $\lambda$3858 pEW (\AA)} \\
\hline
\endhead

\hline \hline
\multicolumn{9}{l}{Continued on next page} \\
\endfoot

\hline \hline
\endlastfoot

SN 1989M & $2.5$ & $\cdots$ & $\cdots$ & 11.19 (0.05) & 81.8 (0.8) & 0 & 10.93 (0.05) & 92.3 (0.4) \\
SN 1991bg & $0.1$ & $\cdots$ & $\cdots$ & 11.19 (0.05) & 233.1 (2.4) & 0 & $\cdots$ & $\cdots$ \\
SN 1994D & $-12.3$ & 26.95 (0.07) & 97.3 (1.2) & 17.53 (0.07) & 85.4 (1.1) & 1.14 (0.02) & $\cdots$ & $\cdots$ \\
SN 1994D & $-7.7$ & 23.56 (0.05) & 120.0 (0.8) & 11.31 (0.05) & 36.7 (0.6) & 3.27 (0.05) & 11.79 (0.05) & 5.2 (0.2) \\
SN 1994D & $-6.3$ & 20.74 (0.12) & 110.0 (1.2) & 9.95 (0.05) & 20.5 (0.9) & 5.36 (0.24) & 11.95 (0.05) & 5.4 (0.3) \\
SN 1994D & $-5.3$ & 19.60 (0.14) & 101.0 (1.4) & 10.16 (0.05) & 18.5 (0.9) & 5.46 (0.28) & 11.94 (0.05) & 8.0 (0.3) \\
SN 1994D & $-3.9$ & 22.87 (0.10) & 22.2 (0.6) & 11.86 (0.05) & 55.6 (0.2) & 0.40 (0.01) & 11.75 (0.05) & 15.9 (0.5) \\
SN 1994D & $-3.3$ & 24.17 (0.09) & 13.1 (0.4) & 11.10 (0.05) & 50.0 (0.2) & 0.26 (0.01) & 11.71 (0.05) & 30.3 (0.4) \\
\end{longtable}
\end{center}
\end{landscape}
\normalsize
\twocolumn

\onecolumn
\small
\begin{center}
\begin{longtable}{lrrrrrr}
\caption{\siii\ Fit Results\label{t:siii}} \\[-2ex]
\hline \hline
SN Name & \multicolumn{1}{c}{Phase (d)$^\textrm{a}$} & \multicolumn{1}{c}{HVF $v$ ($10^{3}$ \kms)} & \multicolumn{1}{c}{HVF pEW (\AA)} & \multicolumn{1}{c}{PVF $v$ ($10^{3}$ \kms)} & \multicolumn{1}{c}{PVF pEW (\AA)} & \multicolumn{1}{c}{$R_\mathrm{Si}^\textrm{b}$} \\
\hline
\endfirsthead

\multicolumn{7}{c}{{\tablename} \thetable{} --- Continued} \\
\hline \hline
SN Name & \multicolumn{1}{c}{Phase (d)$^\textrm{a}$} & \multicolumn{1}{c}{HVF $v$ ($10^{3}$ \kms)} & \multicolumn{1}{c}{HVF pEW (\AA)} & \multicolumn{1}{c}{PVF $v$ ($10^{3}$ \kms)} & \multicolumn{1}{c}{PVF pEW (\AA)} & \multicolumn{1}{c}{$R_\mathrm{Si}^\textrm{b}$} \\
\hline
\endhead

\hline \hline
\multicolumn{7}{l}{Continued on next page} \\
\endfoot

\hline \hline
\endlastfoot

SN 1989M & $2.5$ & $\cdots$ & $\cdots$ & 12.33 (0.05) & 132.9 (0.1) & 0 \\
SN 1989M & $3.5$ & $\cdots$ & $\cdots$ & 12.19 (0.05) & 124.2 (0.1) & 0 \\
SN 1991T & $-10.1$ & $\cdots$ & $\cdots$ & 10.66 (0.06) & 21.1 (0.3) & 0 \\
SN 1991T & $-9.1$ & $\cdots$ & $\cdots$ & 9.93 (0.09) & 17.4 (0.4) & 0 \\
SN 1991bg & $0.1$ & $\cdots$ & $\cdots$ & 9.63 (0.05) & 96.0 (0.6) & 0 \\
SN 1991bg & $1.1$ & $\cdots$ & $\cdots$ & 9.21 (0.05) & 94.5 (0.9) & 0 \\
SN 1994D & $-12.3$ & 19.03 (0.05) & 40.1 (0.9) & 13.67 (0.05) & 97.8 (0.6) & 0.41 (0.01) \\
SN 1994D & $-11.3$ & 18.82 (0.06) & 34.7 (1.5) & 13.30 (0.08) & 96.2 (0.5) & 0.36 (0.02) \\
\end{longtable}

\begin{longtable}{lrrrrrr}
\caption{\cair\ Fit Results\label{t:cair}} \\[-2ex]
\hline \hline
SN Name & \multicolumn{1}{c}{Phase (d)$^\textrm{a}$} & \multicolumn{1}{c}{HVF $v$ ($10^{3}$ \kms)} & \multicolumn{1}{c}{HVF pEW (\AA)} & \multicolumn{1}{c}{PVF $v$ ($10^{3}$ \kms)} & \multicolumn{1}{c}{PVF pEW (\AA)} & \multicolumn{1}{c}{$R_\mathrm{\cair}^\textrm{b}$} \\
\hline
\endfirsthead

\multicolumn{7}{c}{{\tablename} \thetable{} --- Continued} \\
\hline \hline
SN Name & \multicolumn{1}{c}{Phase (d)$^\textrm{a}$} & \multicolumn{1}{c}{HVF $v$ ($10^{3}$ \kms)} & \multicolumn{1}{c}{HVF pEW (\AA)} & \multicolumn{1}{c}{PVF $v$ ($10^{3}$ \kms)} & \multicolumn{1}{c}{PVF pEW (\AA)} & \multicolumn{1}{c}{$R_\mathrm{\cair}^\textrm{b}$} \\
\hline
\endhead

\hline \hline
\multicolumn{7}{l}{Continued on next page} \\
\endfoot

\hline \hline
\endlastfoot

SN 1989M & $2.5$ & $\cdots$ & $\cdots$ & 13.33 (0.05) & 197.1 (0.2) & 0 \\
SN 1989M & $3.5$ & $\cdots$ & $\cdots$ & 13.00 (0.05) & 209.9 (0.2) & 0 \\
SN 1991bg & $1.1$ & $\cdots$ & $\cdots$ & 11.84 (0.05) & 268.8 (1.2) & 0 \\
SN 1994D & $-7.7$ & 20.85 (0.05) & 41.7 (0.2) & 10.01 (0.05) & 37.7 (0.8) & 1.11 (0.02) \\
SN 1994D & $-3.9$ & 20.37 (0.05) & 21.5 (0.2) & 10.22 (0.05) & 48.4 (1.9) & 0.44 (0.02) \\
SN 1994D & $-3.3$ & 20.77 (0.05) & 21.3 (0.2) & 10.51 (0.05) & 64.1 (2.2) & 0.33 (0.01) \\
SN 1994S & $1.1$ & 18.12 (0.06) & 31.3 (1.2) & 10.14 (0.06) & 109.1 (6.9) & 0.29 (0.02) \\
SN 1995D & $3.8$ & 17.73 (0.05) & 55.0 (0.3) & 10.06 (0.05) & 104.8 (2.0) & 0.52 (0.01) \\
\end{longtable}
\label{lastpage}
\end{center}
\normalsize
\twocolumn

\end{document}